%% file: 10may09.tex
\magnification=1200
\hfuzz2pc
\raggedbottom
\overfullrule=0pt
\input gen-dfn
\input epsf
\def\lam{\lambda}
\def\Im{\mathop{\rm Im}}
\def\Re{\mathop{\rm Re}}
\input amssym.def
\def\bbc{{\Bbb C}}
\def\ep{\epsilon}
\def\lam{\lambda}

\let\eno=\sno
\newif\ifproofmode
\newcount\referno
\referno=0
\def\xrefsfilename{turning.xrf}  
\def\myinput#1{\immediate\openin0=#1\relax
   \ifeof0\write16{Cannot input file #1.}
   \else\closein0\input#1\fi}
\def\ifundefined#1{\expandafter\ifx\csname#1\endcsname\relax}
\myinput \xrefsfilename
\immediate\openout1=\xrefsfilename
\def\bibitem#1#2\par{\ifundefined{REFLABEL#1}\relax\else
 \global\advance\referno by 1\relax
 \immediate\write1{\noexpand\expandafter\noexpand\def
 \noexpand\csname REFLABEL#1\endcsname{\the\referno}}
 \global\expandafter\edef\csname REFLABEL#1\endcsname{\the\referno}
 \item{\the\referno.}#2\ifproofmode [#1]\fi\fi}
\def\cite#1{\ifundefined{REFLABEL#1}\ignorespaces
   \global\expandafter\edef\csname REFLABEL#1\endcsname{?}\ignorespaces
   \write16{ ***Undefined reference #1*** }\fi
 \csname REFLABEL#1\endcsname}
\def\nocite#1{\ifundefined{REFLABEL#1}\ignorespaces
   \global\expandafter\edef\csname REFLABEL#1\endcsname{?}\ignorespaces
   \write16{ ***Undefined reference #1*** }\fi}
\vfootnote{*}{This work was supported by NSF grant number DMS-0500641.}

\centerline{\bf WKB and Turning Point Theory for Second Order Difference
Equations:}
\centerline{\bf External Fields and Strong Asymptotics for Orthogonal Polynomials}
\medskip

\centerline{Jeffrey S. Geronimo}
\centerline{School of Mathematics}
\centerline{Georgia Institute of Technology}
\centerline{Atlanta, GA 30332-0160}
\centerline{\tt geronimo@math.gatech.edu}
\bigskip

{\narrower
\noindent{\bf Abstract.} A LG-WKB and Turning point theory is developed for three term recurrence formulas
associated with  monotonic recurrence coefficients. This is used to find strong asymptotics for certain
classical orthogonal polynomials including Wilson polynomials.

\bigskip

\noindent{\bf Key words:} monotonic recurrence coefficients, Airy functions, Wilson polynomials, Turning points, WKB.
\par}

\baselineskip=18pt
\beginsection I. Introduction

There has been much recent interest in so-called $\ep$-difference equations. 
As shown by Deift and and McLaughlin [\cite{DM}] the
recurrence formula satisfied by polynomials orthogonal on the real,
$$a_{n+1} \psi_{n+1} (x)+(b_{n}-x)\psi_n (x)+a_{n} \psi_{n-1} (x)=0
\eq\eno[fdiff]$$
with $a_{n}>0$, and $b_{n}$ real can be recast to satisfy such an equation. To see this we  assume that $a_n$, and $b_n$ are discretizations of the function $a(u)$ and $b(u)$ and in order to use Taylor series
we will write $u={t\over\ep}$ where in this article $t$ will be restricted to a compact
interval of the real line. In the examples considered below an auxiliary
scaling $x=\lambda_{\ep} y+\lambda^1_{\ep}$ first suggest by Nevai and Dehesa~[\cite{ND}] and developed by Van Assche~[\cite{V}] will be performed where
$\lambda_{\ep}$ and $\lambda^1_{\ep}$   are chosen so that $a(t,\ep)=a({t\over\ep})/\lambda_{\ep}$
and $b(t,\ep)=(b({t\over\ep})-\lambda^1_{\ep})/\lambda_{\ep}$ are bounded
functions of $t$ and $\ep$ for $(t,\ep)$ in a compact set say $[0, T]\times[0,\ep_0]$. In most of the cases we will discuss 
$\lambda_{\ep}=a({1\over\ep}+{1\over2})$ and $\lambda^1_{\ep}=0$. This produces the $\ep$ difference equation,
$$
a(t_n +\ep,\ep)\tilde\psi(t_n +\ep,y,\ep)+(b(t_n,\ep)-y)
\tilde\psi(t_n, y,\ep)+a(t_n,\ep) \tilde\psi(t_n -\ep,y,\ep)=0, \eq\eno[epdiff]
$$
where $t_n=n\ep$, and $\tilde \psi(t_n,y,\ep)=\psi_n(\lam_\ep y)$.
Costin and Costin~[\cite{CC}] considered  higher order difference epsilon
equations. (For an application of this theory to knots see
[\cite{GG}].)  In both cases it was assumed that the coefficients in
the difference equation were $C^\infty$ functions of $t$.  In the above
works the authors developed a LG-WKB theory in the region where the roots
of the eikonal  equation (see e.g.~[\cite{DM}]) are distinct.
Furthermore they also produced solutions in terms of Airy functions in a
neighborhood where two roots of the eikonal equation collide at a rate
$(\lam_1(t)-\lam_2(t))\approx\sqrt{(t-t_p)}$.  Here the point of collision
$t_p$ is called a turning point.  The neighborhood of $t_p$ in which the
above solutions are valid was shown to be of order $|t-t_p|\le\ep^{1/2}$  which is 
sufficiently large to allow matching of the
WKB solutions of the ``exterior" region to those of the ``interior" region 
around
$t_p$.  In a later paper Geronimo et al~[\cite{GBVA}] developed a discrete analog
of the Langer transformation used in differential equations.  With this
transformation the authors gave conditions on the recurrence coefficients
that allowed solutions, in terms of Airy functions, valid in an $O(1)$
neighborhood of $t_p$.  They then went on to use these solutions to find
strong asymptotics for Hermite polynomials that resembled the well known
Plancheral-Rotach asymptotics obtained by other means (see Olver~[\cite{O}],
Szeg\"o~[\cite{S}]).  In the process of finding useful approximate
solutions the authors considered the singular initial value problem.
This problem arises when the off-diagonal coefficients in the recurrence
coefficients, $a(t_n,\ep)$, equal or tend to zero with $\ep$ in a neighborhood of $t_n=0$.  Such a
phenomena occurs commonly for instance in the continuum limit of
the Toda lattice [\cite{DM}], in the case where the recurrence coefficients
tend to infinity with $n$ and in many cases in the theory of varying 
recurrence coefficients [\cite{KVA}] and [\cite{AGV}].

The strong asymptotics of polynomials orthogonal with respect to 
exponential weights (i.e., Hermite polynomials, Freud weights, etc.)
has received much attention in recent years, Baik et al~[\cite{BKMM}], Bleher and Its~[\cite{BI}], Deift et al~[\cite{DKMVZ}], Levin and Lubinsky~[\cite{LLone}], McLaughlin and Kriecherbauer~[\cite{KM}], VanLessen~[\cite{V}] (see also Nevai~[\cite{No}, \cite{Nt}], Plancherel and Rotach~[\cite{PR}], and Sheen~[\cite{Sh}]).  In the above studies the main starting point is
in weight function with respect to which the polynomials are orthogonal.
In these studies heavy use is made of the theory of external fields
both constrained (Buyarov and Rakhmanov~[\cite{BR}], Dragnev and Saff~[\cite{DS}], Kuijlaars and Van Assche~[\cite{KVAt}], Lubinsky and Levin~[\cite{LLtwo}], Rakhmanov~[\cite{Raone}, \cite{Rathree}]) and
unconstrained (Gonchar and Rakhmanov~[\cite{GR}], Saff and Totik~[\cite{ST} and references therein]).  Other recent methods used to study
these types of polynomials have been developed by Jin and Wong~[\cite{JWa}], Maejima and Van Assche  ~[\cite{MV}].  Strong asymptotics with error bounds for orthogonal polynomials having
recurrence coefficients that tend in magnitude to infinity have
been obtained by Geronimo, Bruno, and VanAssche~[\cite{GBVA}], Geronimo and
Smith~[\cite{GS}], and Geronimo, Smith and Van Assche~[\cite{GSV}], Geronimo and Van Assche~[\cite{GVA}], Van Assche and Geronimo~[\cite{VG}] and Wang and Wong~[\cite{WW}] (for a heuristic approach see Dominici~[\cite{Do}]). In Wang and
Wong second order difference equations of the form
$$P_{n+1}(x)-(A_nx+B_n)P_n(x)+ P_{n-1}(x)=0$$
are considered where
$$A_n\sim n^{-\theta}\sum^\infty_{b=0} {\alpha_s\over n^s}\quad\hbox{and}
\quad B_n\sim\sum^\infty_{n=0} {\beta_s\over n^s}\qquad\theta\ne 0$$
For real $x$ approximate solutions in terms of Airy functions having a complete asymptotic expansions
are found.  Use of their technique to study the asymptotics of
orthogonal polynomials however requires that the solution of the 
initial value problem be computed by other means. Also in this vein are the uncontracted asymptotics obtained by Janus and Naboko~[\cite{JN}]. It is perhaps worthwhile to emphasize  that the 
theory developed below allows us, in a self contained manner, to obtain the 
strong asymptotics of solutions of {\it initial value problems} associated with 
epsilon difference equations along with error bounds. The main results of this 
paper are to obtain strong asymptotics of continuous dual Hahn, Hermite, Laguerre, Meixner, 
Meixner-Pollaczek and Wilson polynomials $p_n$ with an error of $O(\ep)$ where $\ep=1/n$.

We begin in Section~II by reviewing the 
results of [\cite{GBVA}] and applying them to monotonic recurrence coefficients. 
Then in Section~III we review the theory of external
fields and show the connection between the discrete analog of the Langer transformation
and potentials arising from these fields. These results are applied to investigate monotonic recurrence coefficients that allow an analytic continuation to a wedge. This allows us to simplify formulas for the external field and the Langer Transformation.  In Section IV we discuss the singular initial value problem which allows us to match solutions of the initial value problem with those obtained in terms of Airy functions. The matching problem is considered in Section V where the case of discrete weights is also considered. In Section~VI we consider some
model recurrence coefficients associated with symmetric, antisymmetric, and
discrete Freud weights. In Section VII special cases of these coefficients are chosen so as to
well approximate the recurrence coefficients of the classical
orthogonal polynomials mentioned above. The location of their zeros is also discussed. 
\advance\subsecno by 1\no=0\thmno=0
\beginsection II. Preliminaries

\noindent {\it Turning Point Theory}

\item{(1)} Real case

We begin by reviewing the results of [\cite{GBVA}] that will be needed for the rest of the paper. The objective is to obtain solutions to \epdiff\ in terms of Airy functions.  We suppose that $y$ is a real variable, and in order to exploit the symmetry in equation~\epdiff\ we expand $a(t+\ep,\ep)=a(t+\ep/2,\ep)+a(t+\ep/2,\ep)'\ep/2 + O(\ep^2)$ and $a(t,\ep)=a(t+\ep/2,\ep)-a(t+\ep/2,\ep)'\ep/2 + O(\ep^2)$. Set
$$
\eqalign{
\cosh k(t,y,\epsilon)&={y-b(t,\epsilon)\over 2 a\left(t+{\epsilon\over 2},\epsilon\right)}\quad
{\rm for}\quad t\le t_p\cr
\cos k(t,y,\epsilon)&={y-b(t,\epsilon)\over 2 a\left(t+{\epsilon\over 2},\epsilon\right)}\quad {\rm for}\quad t> t_p,}\eq\eno[eight]
$$
where $t_p$ is such that ${y-b(t_p,\ep)\over2a(t_p+\ep/2)}=1$ and $k/(t_p-t)$ is assumed to be positive in a neighborhood of $t_p$. As in [\cite{GBVA}] we define the Langer transformation for \epdiff\ as
$$\eqalign{
{2\over3}\rho^{3/2} (t,y,\ep)&=
\int_t^{t_p} \cosh^{-1}\left({y-b(u,\epsilon)\over 2 a\left(u+{\epsilon\over 2},\epsilon\right)}\right)du\quad{\rm for }\quad t\le t_p,\cr
{2\over3}(-\rho)^{3/2}(t,y,\ep)
&=\int_{t_p}^t \cos^{-1}\left({y-b(u,\epsilon)\over 2 a\left(u+{\epsilon\over 2},\epsilon\right)}\right)du\qquad{\rm for }\quad t\ge t_p. }\eq\eno[langer]
$$
With
$$g(t)=\left\{\matrix{{\rho (t,y,\epsilon)\overwithdelims ()
a^2\left(t+{\epsilon\over 2},\epsilon\right)\sinh^2 k(t,y,\epsilon)}^{1/4}
& t\le t_p\cr{\rho (t,y,\epsilon)\overwithdelims ()
a^2\left(t+{\epsilon\over 2},\epsilon\right)\sin^2 k(t,y,\epsilon)}^{1/4}
& t> t_p}\right. ,\eq\eno[ninea]$$
we set
$$
\psi_1(t,y,\ep)=g(t,y,\ep){\rm Ai}(\ep^{-{2\over3}}\rho(t,y,\ep)), \eq\eno[psione]
$$
and
$$
\psi_2(t,y,\ep)=g(t,y,\ep){\rm Bi}(\ep^{-{2\over3}}\rho(t,y,\ep)), \eq\eno[psitwo]
$$
where ${\rm Ai}$ and ${\rm Bi}$ are
Airy functions [\cite{O}] and satisfy the differential
equation $$\chi(x)''=x\chi(x).$$
  From \eight\ it follows that,
$$
k^2(t,y,\ep)=\ln^2\left({y-b(t,\ep)\over2a(t+{\ep\over2},\ep)}+
\sqrt{\left({y-b(t,\ep)\over2a(t+{\ep\over2})}\right)^2-1}\,\right), \eq\eno[logform]
$$
and equation \langer\ can be written in the compact form,
$$
{2\over3}\rho^{3\over2}(t,y,\ep)=\int_t^{t_p}k(u,y,\ep) du\,. \eq\eno[langercom]
$$

\remk For real variables the above expression should be considered as a
short  hand notation for \langer. 

\remk The expansion about $t+1/2$ discussed above is motivated by the
special functions we will consider below and it gives a simpler error bound for the approximation problem discussed below.

The rest of this section is devoted to showing that \psione\ and \psitwo\ can be used as useful approximate solution to \epdiff\.
Unless otherwise stated the branch of  $\ln(z)$ which is
positive  for $z>1$ and analytic in a neighborhood of the positive real
axis will be used.  In what is to follow much use will be made of the analytic
properties  of the function,
$$
z(y)=y+\sqrt{y^2-1},\eq\eno[map]
$$
where the branch of the square root function is chosen so that for large 
$y$, $z\sim 2y$.
The above function maps the exterior of $[-1,1]$ to the exterior of the unit circle so that,
$$|z|>1,\quad y\in \bbc\setminus[-1,1],\quad |z|=1, \quad y\in[-1,1].\eq\eno[mag]
$$
For $y$ real the functions $z_+=\lim_{h->0}z(y+ih)$ and
$z_-=\lim_{h->0}z(y-ih)$ are continuous functions of $y$.

We will denote the interior (relative to the complex plane)
of a  set $K$ as ${\rm int}(K)$, $\bbc_+=\{y: \Im y>0\}$, $\bbc_-=\{y: \Im
y<0\}$ and $\bar \bbc_{\pm}=\bbc_{\pm}\cup R$.

We begin with,

\lem logsquare. Let $f(z)=\ln^2(z+\sqrt{z^2-1})$ then $f$ is analytic for $z\in \bbc\backslash(-\infty,-1]$. In this region $f$ has a sole zero which is simple and located at $z=1$.

\pf From the mapping properties of $z+\sqrt{z^2-1}$ we see that
$\ln ^2\left(z+\sqrt{z^2-1}\right)$ is analytic for 
$z\in\bbc\backslash (-\infty,1]$.  Furthermore 
$\Im(\ln^2\left(z+\sqrt{z^2-1}\right))=0$ for $z\in [-1,1]$.
Consequently by the Schwarz reflection principal $\ln^2\left(z+
\sqrt{z^2-1}\right)$ is analytic for $z\in \bbc\backslash (-\infty,-1]$. If we set
$w=z-1$ and use the representation
$$\ln \left(1+w+\sqrt{w(w+2)}\right)=\int^w_0 {dx\over
\sqrt{x(x+2)}},$$
we find
$$\ln \left(1+w+\sqrt{w(w+2)}\right)/\sqrt{w}={1\over\sqrt2}\sum^\infty_{i=0} {-1\overwithdelims () 2}^i
{(1/2)_i\over (1)_i (i+1/2)} \, w^i,\eq\eno[simzero]$$
for $|w|<2$. Thus $f$ has a simple zero at $z=1$. That this is the only zero in $\bbc\backslash (-\infty,1]$ follows from the mapping properties of $z+\sqrt{z^2-1}$. \qed

We will suppose that there exists $T>1$ and $\ep_0$ sufficiently small 
so that
$$a(t,\ep), b(t,\ep)\in C([0,T]\times [0,\ep])\eq\eno[condcoeffone]$$
$${\partial^ia(t,\ep)\over \partial t^i}\,,
{\partial^i b(t,\ep)\over \partial t^i}\in C((0,T]\times [0,\ep_0])\qquad
i=1,\ldots 4\eq\eno[condcoefftwo]$$
$$a(0,0)=0 = b(0,0)\quad\hbox{but }
a(t,\ep)>0\quad (t,\ep)\in (0,T]\times [0,\ep_0].\eq\eno[condcoeffthree]$$
Set 
$$\gamma^{\pm}(t,\ep)=b(t,\ep)\pm 2a(t + \ep/2,\ep),$$
and when they exist
$$t_p^{\pm}(y,\ep)=(\gamma^{\pm})^{-1}(y,\ep).$$
Note that from the assumptions on $a(t,\ep)$ and $b(t,\ep)$,
$\gamma^{\pm}(0,0)=0$. 

We will assume that one of the following cases occur:  
\item{1)} For each $\ep\in[0,\ep_0]$ $\gamma^+(t,\ep)$, and
$-\gamma^-(t,\ep)$ are increasing for $t\in [0,T]$ with nonzero  derivatives in $t$ for $(t,\ep)\in(0,T)\times[0,\ep_0]$.  For every interval 
$[y_1,y_2]\in (0,\gamma^+_0(T))$ there is an $\ep_1$  such that $t_p^+\in
C([y_1,y_2]\times[0,\ep_1])$. 
Likewise for every interval $[y_1,y_2]\in (\gamma^-_0(T),0)$ there is an $\ep_1$ such that 
$t_p^-\in C([y_1,y_2]\times[0,\ep_1])$.
\item{2)} For each $\ep\in[0,\ep_0]$, $\gamma^+(t,\ep)$ is increasing  and 
$\gamma^-(t,\ep)=0$ for $t\in [0,T]$. The derivative of $\gamma^+(t,\ep)$ with respect to $t$ is nonzero for $(t,\ep)\in(0, T)\times[0,\ep_0]$. For every interval $[y_1,y_2]\in (0,\gamma^+_0(T))$ there is an $\ep_1$ such that $t_p^+\in
C([y_1,y_2]\times[0,\ep_1])$.
\item{3)} $\gamma^+(t,\ep)$ and $\gamma^-(t,\ep)$ are increasing for $t\in [0,T]$  with nonzero derivatives in $t$ for $(t,\ep)\in(0,T)\times[0,\ep_0]$. For every interval $[y_1,y_2]\in(0,\gamma^+_0(T))$ there is an $\ep_1$ such that $t_p^+\in
C([y_1,y_2]\times[0,\ep_1])$. Likewise for every interval $[y_1,y_2]\in (0,\gamma^-_0(T))$ there is an $\ep_1$ such that $t_p^+\in
C([y_1,y_2]\times[0,\ep_1])$.

We restrict our attention to the above cases since these are the ones that are relevant for the 
examples we wish to consider.  Other cases for instance $\gamma^+(t,\ep)$
and $\gamma^-(t,\ep)$ both decreasing or $\gamma^+(t,\ep)=0$ and
$\gamma^-(t,\ep)$ decreasing are also important and can be treated by interchanging the roles of $\gamma^+$ and $\gamma^-$. Set $\gamma^+_{\ep}(t)=\gamma^+(t,\ep)$, $\gamma^-_{\ep}(t)=\gamma^-(t,\ep)$, $z^+(t,y,\ep)={y-b(t,\ep)\over2a(t+\ep/2,\ep)}$, and $z^-(t,y,\ep)={b(t,\ep)-y\over2a(t+\ep/2,\ep)}$. Note that  $z^+-1={y-\gamma_{\ep}^+(t)\over2a(t+\ep/2,\ep)}$ and $z^--1={\gamma_{\ep}^-(t)-y\over2a(t+\ep/2,\ep)}$.

\lem contone.  Suppose that $a(t,\ep)$ and $b(t,\ep)$ satisfy
conditions {\rm\condcoeffone--\condcoeffthree}\ and case~$1$, $2$, or $3$ 
holds.
Then for every $[t_{in},t_{fi}]\times[y_1,y_2]\subset(0,T)\times(0,\gamma^+_0(T))$ there is an $\ep_1$ such that 
${\partial^i\over\partial t^i} {z^+(t,y,\ep)-1\over t^+_p(y,\ep)-t}\in C([t_{in},t_{fi}]\times[y_1,y_2]\times[0,\ep_1])$ 
for $i=0,\ldots,3$. In case~$1$ and $y<0$ for every $[t_{in},t_{fi}]\times[y_1,y_2]\subset(0,T)\times(\gamma^-_0(T),0)$ 
there is an $\ep_1$ such that ${\partial^i\over\partial t^i} {z^-(t,y,\ep)-1\over t^-_p(y,\ep)-t}\in C([t_{in},t_{fi}]\times[y_1,y_2]\times[0,\ep_1])$ for $i=0,\ldots,3$. Finally for case~$3$ there exits an $\ep_1$ such that for every interval $[t_{in}, t_{fi}]\times[y_1,y_2]\subset (0,T)\times(0,\gamma^-_0(T))$, ${\partial^i\over\partial t^i} {z^-(t,y,\ep)-1\over t^-_p(y,\ep)-t}\in C([t_{in},t_{fi}]\times[y_1,y_2]\times[0,\ep_1])$ for $i=0,\ldots,3$.

\pf Fix $[y_1,y_2]\in(0,\gamma^+_0(T))$ by enlarging $[t_{in},t_{fi}]$ if need be and using the continuity of $\gamma^+_{0}$ we can assume that $[y_1,y_2]\subset\gamma^+_0((t_{in},t_{fi}))$. Choose $\ep_1$ so that
$t^+_p(y,\ep)\in C([y_1,y_2]\times[0,\ep_1])$ and
$[y_1,y_2]\subset\gamma^+_{\ep}((t_{in},t_{fi}))$ for all $\ep\in[0,\ep_1]$.   
For $(y,t,\ep)\in[t_{in},t_{fi}]\times[y_1,y_2]\times[0,\ep_1]$ and 
$i=0,\ldots, 3$ set, 
$$
{\partial^i\over\partial t^i}f(t,y,\ep)=\left\{\matrix{{\partial^i\over\partial t^i}\left({z^+(t,y,\ep)-1\over t^+_p(y,\ep)-t}\right)\hfill & t\ne t^+_p(y,\ep)\cr -{\partial^{i+1}\over \partial t^{i+1}} {z^+(t,y,\ep)\over i+1} |_{t=t^+_p(y,\ep)} & t=t^+_p(y,\ep)}\right.
$$
To show that $f\in C([t_{in},t_{fi}]\times[y_1,y_2]\times[0,\ep_1])$ it is sufficient to demonstrate that the above function is continuous at $t=t_p^+(y,\ep)$. Other values of $(t,y,\ep)$ follow from the the smoothness conditions imposed on $t_p^+$, $\gamma^+_{\ep}$ 
and $a(t,\ep)$. Fix $(t=t_p^+(y,\ep),y, \ep)$ and 
suppose $(\hat t, \hat y, \hat\ep)$ is a point close by. For $(\hat y, \hat\ep)$ close to $(y,\ep)$, $t_p^+(\hat y, \hat\ep)$ is close 
to $t_p^+(y,\ep)$.  If $\hat t=t_p^+(\hat y, \hat\ep)$ then use the second 
part of the definition of $f$ otherwise expand $\gamma^+(\hat t,\hat \ep)$ in 
a Taylor polynomial with remainder about $t_p^+(\hat y, \hat\ep)$. These formulas give the contiuity of $f$. 
An argument similar to that above shows that ${\partial^i\over\partial t^i}f(t,y,\ep)\in C([t_{in}. t_{fi}]\times[y_1,y_2]\times[0,\ep_1])$,
 $i=0,\ldots,3$. An analogous discussion for case~$1$ with $y<0$ and the 
second part of case three gives the result. \qed

With the above we can now discuss the smoothness properties of $k^2$. 

\lem gplus.  Suppose that $a(t,\ep)$ and $b(t,\ep)$ satisfy
conditions {\rm\condcoeffone--\condcoeffthree}\ and case~$1$ or $2$ holds.
Then for every $[t_{in},t_{fi}]\times[y_1,y_2]\subset(0,T)\times(0,\gamma^+_0(T))$ there is an
$\ep_1$ such that for $k$ given by equation~\logform,
\item{i.} ${\partial^i\over\partial t^i}
{k^2(t,y,\epsilon)\over t^+_p(y,\epsilon)-t}\in 
C([t_{in},t_{fi}]\times[y_1,y_2]\times [0,\epsilon_1])$, 
$i=0,\dots,3$ with ${k^2(t,y,\epsilon)\over t^+_p(y,\epsilon)-t}$ 
strictly positive in $[t_{in},t_{fi}]\times[y_1,y_2]\times[0,\ep_1]$, and
\item{ii.} $|{\sinh^2(k(t,y,\ep))\over t-t^+_p(y,\ep)}|> 0$, $(|{\sin^2(k(t,y,\ep))\over t-t^+_p(y,\ep)}|> 0)$, 
for $(y,t,\ep)\in[t_{in},t_{fi}]\times[y_1,y_2]\times [0,\epsilon_1]$ and $t\le t^+_p(y,\ep)\ (t\ge t^+_p(y,\ep))$.\hfill\break
 For case~$1$ and $y<0$ set
$$k^2(t,y,\ep)=\ln^2\left(
{b(t,\ep)-y\over 2a(t+\ep/2,\ep)} + 
\sqrt{\left({b(t,\ep)-y \over
2a(t+\ep/2,\ep)}\right)^2-1}\right).\eq\eno[kcaseone]$$
Then for every $[t_{in},t_{fi}]\times[y_1,y_2]\subset(0,T)\times (\gamma^-_0(T),0)$ there is an
$\ep_1$ so that i), and ii) hold with $t^+_p$ replaced by $t^-_p$.

\pf Fix $[y_1,y_2]\in(0,\gamma^+_0(T))$ and choose $\ep_1$ so that
$t^+_p(y,\ep)\in C([y_1,y_2]\times[0,\ep_1])$. For $(t,y,\ep)\in[t_{in},t_{fi}]\times[y_1,y_2]\times[0,\ep_1]$ and $i=0,\ldots,3$ set,
$$
{\partial^i\over\partial t^i}h(t,y,\ep)=\left\{\matrix{{\partial^i\over\partial t^i}{k^2(t,y,\ep)\over t^+_p(y,\ep)-t}\hfill& t\ne t^+_p(y,\ep)\cr -{\partial^{i+1}\over\partial t^{i+1}}{k^2(t,y,\ep)\over i+1}|_{t=t^+_p(y,\ep)}& t=t^+_p(y,\ep).}\right. 
$$

Lemmas~\logsquare, \contone, the conditions on $\gamma^+$ and its  derivative with respect to $t$
show that $h$ is positive and ${\partial^i\over\partial t^i}h\in C([t_{in},t_{fi}]\times[y_1,y_2]\times [0,\epsilon_1]),\  i=0,\ldots,3$. Part ii) of the Lemma follows from part i) and equation~\eight. For $y<0$ we see from
equation \kcaseone\ that 
$k^2(t_p^-(y,\ep),y,\ep)=0$ thus the argument above can be taken over
to this case so that the second part of the Lemma follows. \qed

For case~3, $\gamma^-$ becomes an obstacle to the smoothness of
$k^2$. Thus the intervals to be considered will need to be restricted.

\lem gcasethree.  Suppose that $a(t,\ep)$ and $b(t,\ep)$ satisfy
conditions {\rm\condcoeffone--\condcoeffthree}\ and case~$3$ holds.
Then for each $[t_{in},t_{fi}]\times[y_1,y_2]\in(0,
T)\times(\gamma^-_0(T),\gamma^+_0(T))$ there is an $\ep_1$ such that
i), ii) of Lemma~\gplus\ hold. If  $y_1<\gamma^-_0(T)$ then for each
$[t_{in},t_{fi}]\times[y_1,y_2]$ with $[y_1, y_2]\in(0,\gamma^+_0(T))$ and 
$[t_{in},t_{fi}]\subset(0,(\gamma^-_0)^{-1}(y_1))$  there is an $\ep_1$
such that  i), ii) of Lemma~\gplus\ hold. If 
$$k^2(t,y,\ep)=\ln^2 \left({b(t,\ep)-y\over 2a(t+\ep/2,\ep)}
+\sqrt{\left({b(t,\ep)-y \over 2a(t+\ep/2,\ep)}\right)^2-1}\right)\eq\eno[kcasethree]$$
then for 
$[\hat t_{in},\hat t_{fi}]\times[y_1,y_2]$ with $[y_1, y_2]\subset(0,\gamma^-_0(T))$ and  $[\hat t_{in},\hat t_{fi}]\subset ((\gamma^+_0)^{-1}(y_2),T)$
there is an
$\ep_1$ 
such that $t^-_p(y,\ep)=(\gamma_{\ep}^-)^{-1}(y), (y,\ep)\in[y_1,y_2]\times[0,\ep_1]$ exists and  i), ii) of Lemma~{\rm\gplus}\ hold with $t^+_p$ replaced by $t^-_p$ and in  i) ${k^2\over t_p^+-t}$  replaced by ${k^2\over t-t^-_p}$. For $[y_1,y_2]\subset(0,\gamma^-_0(T))$ then it is possible to  $t_{fi}<\hat t_{in}$.

\pf Supppose that $[y_1,y_2]\in(0, \gamma^+_0(T))$ and $\ep_1$ is such
that that $t_p^+(y,\ep)=(\gamma^+_{\ep})^{-1}(y)\in C([y_1,y_2]\times[0,\ep_1])$. 
Likewise for any interval $[y_1,y_2]\in(0,\gamma^-_0(T))$, there is an $\ep_1$ such that $t^-_p(y,\ep)=(\gamma^-_{\ep})^{-1}(y)\in C([y_1,y_2]\times[0,\ep_1])$. From $a(t,\ep)>0$ for $(t,\ep)\in(0,T]\times[0,\ep_0]$ we see that 
$\gamma^+(t,0)>\gamma^-(t,0)$ for $t\in(0,T)$. Thus for $[t_{in},t_{fi}]\times[y_1,y_2]\in(0,T)\times(\gamma^-_0(T),\gamma^+_0(T))$ by \condcoeffone\ we can choose $\ep_1$ sufficiently small so that $y>\gamma^-_{\ep}(t)$ for 
all $(t,y,\ep)\in [t_{in},t_{fi}]\times[y_1,y_2]\times[0,\ep_1]$. That i)
and ii) hold now follows from an argument similar to that given in Lemma~\gplus. Suppose now $y_1<\gamma^-_0(T)$ and $[t_{in},t_{fi}]\times[y_1,y_2]$ is such that $[y_1, y_2]\in(0,\gamma^+_0(T))$ and $([t_{in},t_{fi}]\subset(0,(\gamma^-_0)^{-1}(y_1))$. As above 
 $\ep_1$ may be chosen sufficiently small so that  
$y>\gamma^-_{\ep}(t)$ for all $(t,y,\ep)\in
[t_{in},t_{fi}]\times[y_1,y_2]\times[0,\ep_1]$ so that i), and ii) will be
satisfied. Suppose now that $[\hat t_{in}, \hat t_{fi}]\times[y_1, y_2]$ is such that $[y_1, y_2]\subset(0,\gamma^-(T,0))$ and $[\hat t_{in}, \hat t_{fi}]\subset((\gamma^+_0)^{-1}(y_2),T)$. Then the
uniform continuity of $a(t,\ep)$ and $b(t,\ep)$ given by \condcoeffone\
shows that by choosing $\ep_1$ sufficiently small we can insure that
$y<\gamma^+_{\ep}(t)$ for all $(t,y,\ep)\in[\hat t_{in},\hat
t_{fi}]\times[y_1, y_2]\times[0,\ep_1]$. Thus smoothness part of
condition i) for $k^2$ given by \kcasethree\  follow from
Lemmas~\logsquare\ and \contone\ and an argument analogous to the one used above. Since $k^2>0$ for $y<\gamma^-_\ep(t)$ and has a simple zero at $y=\gamma^-_{\ep}(t)$ the properties of $t^-_p(y,\ep)$ and ${d\over dt}\gamma^-_{\ep}$ imply that ${k^2(t,y,\ep)\over t-t_p^-(y,\ep)}>0$. Property ii) follows by the choice of $[\hat t_{in},\hat t_{fi}]\times[
 y_1, y_2]\times[0,\ep_1]$ and the fact that $\sinh^2
k=({b(t,\ep)-y\over a(t+{\ep\over2},\ep)})^2-1$. The fact that
$\gamma^+_0>\gamma^-_0$ for $t>0$ allows us to choose $t_{fi}<\hat t_{in}$.\qed

For cases~1 and 3 we will have need of
$$\eqalign{
{2\over3}\rho_2^{3/2} (t,y,\ep)&=\int_t^{t^-_p} 
\cosh^{-1}\left({b(u,\epsilon)-y\over 2 a\left(u+{\epsilon\over 2},\epsilon\right)}\right)du\quad{\rm for }\quad t\le t^-_p,\cr
{2\over3}(-\rho_2)^{3/2}(t,y,\ep)&=\int_{t^-_p}^t 
\cos^{-1}\left({b(u,\epsilon)-y\over 2 a\left(u+{\epsilon\over 2},\epsilon\right)}
\right)du\qquad{\rm for }\quad t\ge t^-_p. }\eq\eno[langero]
$$
and
$$\eqalign{
{2\over3}\rho_2^{3/2} (t,y,\ep)&=\int_{t^-_p}^t 
\cosh^{-1}\left({b(u,\epsilon)-y\over 2 a\left(u+{\epsilon\over 2},\epsilon\right)}\right)du\quad{\rm for }\qquad t\ge t^-_p,\cr
{2\over3}(-\rho_2)^{3/2}(t,y,\ep)&=\int_t^{t^-_p} 
\cos^{-1}\left({b(u,\epsilon)-y\over 2 a\left(u+{\epsilon\over 2},\epsilon\right)}
\right)du\qquad{\rm for }\quad t\le t^-_p. }\eq\eno[langert]
$$

With the above results we will be able to obtain approximate solutions to the difference equation.

\remk In the proof of the Theorem below we will use Theorem~3.3 of
[\cite{GBVA}] where it is assumed that $k^2$ is monotonic. While this is true
in our case examination of the proof of Theorem~3.3  shows that only conditions i)--iii) or ia)--iiia) in [\cite{GBVA}]. Note that in ii) and iiia) of [\cite{GBVA}] there is an error, the running index should be $i=0,1,2,3$.

\thm {threethree}. Suppose that $a(t,\ep)$ and $b(t,\ep)$ satisfy
conditions {\rm\condcoeffone--\condcoeffthree}\ and case~$1$, or $2$ holds.
Let $\rho_1$ be given by equation~\langer\  and  $\psi_i(t,y,\epsilon)\ i= 1,2$ be given by equations \psione\ and \psitwo\ respectively. Then for each $[t_{in}, t_{fi}]\times[y_1,y_2]\subset(0,T)\times(0,\gamma^+_0(T))$ there is  an $\ep_1$ such that for all $(t,y,\ep)\in[t_{in},t_{fi}]\times[y_1,y_2]\times(0,\ep_1]$, $\psi_i$ satisfies,
$$ \eqalign{
a(t+\epsilon,\ep)\psi_i(t&+\epsilon,y,\epsilon)+a(t,\ep)\psi_i(t-\epsilon,y,\epsilon)\cr
&-2a(t+\ep/2,\ep)\cosh
k(t,y,\epsilon)\psi_i(t,y,\epsilon)= \beta_i(t,y,\epsilon)\qquad
i =1,2} \eq\eno[ten]
$$
where  $\psi_i$ and $\beta_i\in C([t_{in},t_{fi}]\times[y_1,y_2]\times(0,\ep_1])$. For $(t,y)\in[t_{in},t_{fi}]\times[y_1,y_2]$,  $\beta_i\ i=1,2$ satisfy the inequalities,
$$
|\beta_1(t,y,\epsilon)|\le
c(y)\epsilon^2\sup_{u\in(t-\epsilon,t+\epsilon)}\left[|{\rm
Ai}(\epsilon^{-{2\over3}}\rho_1(u,y,\epsilon))|+\epsilon^{1\over3}|{\rm Ai}'(\epsilon^{-{2\over3}}\rho_1(u,y,\epsilon))|\right],\eq\eno[betaone]
$$
and
$$
|\beta_2(t,y,\epsilon)|\le c(y)\epsilon^2\sup_{u\in(t-\epsilon,t+\epsilon)}
\left[|{\rm Bi}(\epsilon^{-{2\over3}}\rho_1(u,y,\epsilon))|+\epsilon^{1\over3}|{\rm Bi}'(\epsilon^{-{2\over3}}\rho_1(u,y,\epsilon))|\right],\eq\eno[betatwo]
$$
with $c(y)$  uniformly bounded on $[y_1,y_2]$. For case~1 with $[t_{in}, t_{fi}]\times[y_1,y_2]\subset(0,T)\times(\gamma^-_0(T),0)$ let $\rho_2$ be given  
by \langero, $k^2$  by \kcaseone\ 
and $\psi_i(t,y,\epsilon)\ i= 1,2$  by equations \psione\ and \psitwo\ 
respectively with the appropriate substitutions. Then there is  an $\ep_1$ 
such that for all $(t,y,\ep)\in[t_{in},t_{fi}]\times[y_1,y_2]\times(0,\ep_1]$, $\psi_i$ and $\beta_i, i=1,2$ satisfy \ten\ with $\psi_i,\ \beta_i\in C([t_{in},t_{fi}]\times[y_1,y_2]\times(0,\ep_1])$ and $\beta_i$ satisfies the above bound 
with $\rho_1$ replaced by $\rho_2$. Finally suppose case~$3$ holds. Then for 
each $[t_{in},t_{fi}]\times[y_1,y_2]\in(0,T)\times (\gamma^-_0(T),\gamma^+_0(T))$ or if $y_1<\gamma^-_0(T)$ then for each $[t_{in},t_{fi}]\times[y_1,y_2]$ is such that $[y_1, y_2]\in(0,\gamma^+_0(T))$ and $([t_{in},t_{fi}]\subset(0,(\gamma^-_0)^{-1}(y_1))$ there exists 
an $\ep_1$ such  that for all $(t,y,\ep)\in[t_{in},t_{fi}]\times[y_1,y_2]\times(0,\ep_1]$, $\psi_i$ satisfies, \ten\ with $\psi_i$ and $\beta_i\in C([t_{in},t_{fi}]\times[y_
 1,y_2]\times(0,\ep_1])$. For $(t,y)\in[t_{in},t_{fi}]\times[y_1,y_2]$,  $\beta_i\ i=1,2$ satisfy the inequalities \betaone\ or \betatwo\ respectively 
with $c(y)$ uniformly bounded in $[y_1,y_2]$. If  $k^2$ is given by equation~\kcasethree\  and $\rho_2$ by equation \langert,
then for $[\hat t_{in}, \hat t_{fi}]\times[y_1, y_2]$ is such that $[ y_1, y_2]\subset(0,\gamma^-(T,0))$ and $[\hat t_{in}, \hat t_{fi}]\subset((\gamma^+_0)^{-1}(y_2),T)$ there is an $\ep_1$ such  that for all $(t,y,\ep)\in[\hat t_{in},\hat t_{fi}]\times[y_1,y_2]\times(0,\ep_1]$, $\psi_i$ satisfies, \ten\ with $\rho_1$ replaced by $\rho_2$. The functions $\psi_i$ and $\beta_i\in C([\hat t_{in},\hat t_{fi}]\times[y_1,y_2]\times(0,\ep_1])$. Furthermore  for $(t,y)\in[\hat t_{in},\hat t_{fi}]\times[y_1,y_2]$,  $\beta_i\ i=1,2$ satisfy the inequalities \betaone\ or \betatwo\ respectively with $c(y)$ uniformly bounded in $[y_1,y_2]$.

\pf Suppose $y\in(0,\gamma^+(T,0))$ and set
$p(t,y,\ep)=\left({k^2(t,y,\ep)\over t^+_p(y,\ep)-t}\right)^{1/2}$ for
$(t,,y,\ep)\in[t_{in},t_{fi}]\times[y_1,y_2]\times[0,\ep_1]$ where
$[t_{in},t_{fi}]\times[y_1,y_2]\times[0,\ep_1]$ are the sets given in the
first part of Lemma~\gplus. Lemma~\gplus\  shows that 
$p$ is positive and ${\partial^i\over\partial t^i}p(t,y,\ep)\in C([t_{in},t_{fi}]\times[y_1,y_2]\times[0,\ep_1]),\ i=0,\ldots,3$. For $(t,,y,\ep)\in[t_{in},t_{fi}]\times[y_1,y_2]\times[0,\ep_1]$ set
$$
q(t,y,\ep)=\left\{\matrix{{1\over
(t^+_p(y,\ep)-t)^{3/2}}\int_t^{t^+_p(y,\ep)}(t^+_p(y,\ep)-u)^{1/2}p(u,y,\ep)du&
{\rm for}\ t< t_p(y,\ep)\cr{2\over3}p(t^+_p(y,\ep),y,\ep)&{\rm for}\ t=t_p(y,\ep)\cr{1\over(t-t^+_p(y,\ep))^{3/2}}\int^t_{t^+_p(y,\ep)}(u-t^+_p(y,\ep))^{1/2}p(u,y,\ep)du&{\rm for}\ t> t_p(y,\ep)}\right.\ \eq\eno[tinv] 
$$
Again the issue is to show that $q$ is continuous at $(t=t^+_p(y,\ep),y,\ep)$ for fix $(y,\ep)$.
Lemma~\gplus\ (for the definition of $p(t^+_p(y,\ep),y,\ep))$ and the mean value 
theorem for integrals (see\break Lemma~3.1 of Olver [\cite{O}, p.~399]) shows that $q(t,y,\ep)\in C([t_{in},t_{fi}]\times[y_1,y_2]\times[0,\ep_1])$  and is positive. Next set,
$$
{\partial\over\partial t} q(t,y,\ep)=\left\{\matrix{{1\over
(t^+_p(y,\ep)-t)^{5/2}}\int_t^{t^+_p(y,\ep)}(t^+_p(y,\ep)-u)^{3/2}{\partial\over\partial u}p(u,y,\ep)du&
{\rm for}\ t< t_p(y,\ep)\cr{2\over5}{\partial\over\partial t}p(t,y,\ep)|_{t=t^+_p(y,\ep)}&{\rm for}\ t=t_p(y,\ep)\cr{1\over(t-t^+_p(y,\ep))^{5/2}}\int^t_{t^+_p(y,\ep)}(u-t^+_p(y,\ep))^{3/2}{\partial\over\partial u}p(u,y,\ep)du&{\rm for}\ t> t_p(y,\ep).}\right. 
$$
Integration by parts, Lemma~\gplus, and   the mean value theorem for
integrals gives that the above definition is self consistent and ${\partial\over\partial t} q(t,y,\ep)\in C([t_{in},t_{fi}]\times[y_1,y_2]\times[0,\ep_1])$ which is positive. Continuing on in this way gives that ${\partial^i\over\partial t^i} q(t,y,\ep)\in C([t_{in},t_{fi}]\times[y_1,y_2]\times[0,\ep_1])$ for $i=0,\ldots,3$.
Since
${\rho_1(t,y,\ep)\over t^+_p(y,\ep)-t}=({3\over2}q(t,y,\ep))^{2/3}$ we obtain
that ${\rho_1(t,y,\ep)\over t^+_p(y,\ep)-t}$ is positive and  ${\partial^i\over\partial t^i}{\rho_1(t,y,\ep)\over t^+_p(y,\ep)-t}\in C([t_{in},t_{fi}]\times[y_1,y_2]\times[0,\ep_1])$ for $i=0,\ldots,3$.
An analogous argument for the sets in the second part of Lemma~\gplus\ shows
that for case 1 and $y\in(\gamma^-(T,0),0)$, ${\rho_2(t,y,\ep)\over t^-_p(y,\ep)-t}$ is positive and  ${\partial^i\over\partial t^i}{\rho_2(t,y,\ep)\over t^-_p(y,\ep)-t}\in C([t_{in},t_{fi}]\times[y_1,y_2]\times[0,\ep_1])$ for $i=0,\ldots,3$.
Case 3 follows from Lemma~\gcasethree\ and arguments similar to 
those given above.  For $y>0$ that there are solutions well approximated  by 
equations~\psione\ and \psitwo\  follows from  Theorem
3.3 in [\cite{GBVA}].  Case~1 with
$y\in(\gamma^-(T,0),0)$ also
follow from Theorem 3.3 in [\cite{GBVA}] after performing the
substitutions indicated. The proof for case~$3$ holds in a similar manner after the substitutions indicated. \qed

\remk Because of the connection  to the Airy functions $\psi_i$, 
$i= 1,2$  satisfy the differential equation [\cite{O}, p.~396],
$$
\left(\ep^2 {d^2\over dt^2}-(k^2(t)+\ep^2 h(t))\right)w(t)\psi(t)=0
$$
where $k^2$ is given by equation~\logform\ and
$$
w(t)=\left({a^2(t+{\ep\over2})\sinh^2 k(t)\over k(t)^2}\right)^{1\over4}
$$
and
$$
h(t)=\left({k^2\over\rho}\right)^{1/4}{d^2\over dt^2}
\left({\rho\over k^2}\right)^{1/4}.
$$

Set, $u_j(x)=g(x)\tilde u_j(x),\ j=1,2$ where
$$\tilde u_j(x)= {-x\overwithdelims () 3}^{1/2}e^{(-1)^{j+1}i\pi\over6} H^{(j)}_{1/3}
\left({2\over 3}\, (-x)^{3/2}\right), \eq\eno[hank]
$$ 
where $H^{(1)}_{1/3}$ and $H^{(2)}_{1/3}$
are Hankel functions of the 1st and 2nd kind respectively. An important property of these functions is that they do not vanish for real $x$ [\cite{W}].
Consider the difference equation,
$$a_1((n+1)\ep,\ep)f(n+1)+ a_1(n\ep,\ep)f(n-1)-(y-b_1(n\ep,\ep))f(n)=0,
\eq\eno[fourteen]$$
where for $0\le\ep\le\ep_0$,
$$
\sup_{t\in
[t_{in},t_{fi}]}|a(t,\ep)-a_1(t,\ep)|=0(\ep^2)=\sup_{t\in [t_{in},t_{fi}]}|b(t,\ep)-b_1(t,\ep)|, \eq\eno[abep]
$$
for every interval $[t_{in},t_{fi}]\subset (0, T)$,
then the techniques leading to Theorem~3.8 in [\cite{GBVA}] give the following,

\thm {thmnext}.  Suppose that the hypotheses of Theorem~\threethree\  hold,  that
$a_1(t,\ep)$ and $b_1(t,\ep)\in C([0,T]\times [0,\ep_0])$,
satisfy \abep\ and that $a_1(t,\ep)$ is strictly positive on every
compact subset of $(0,T]\times[0,\ep_0]$.
For cases~$1$ and $2$ with $y>0$ or the first part of case~$3$ 
let $[t_{in},t_{fi}]\times[y_1,y_2]\times[0,\ep_1]$ given from 
Theorem~\threethree. 
Then there exist solutions $f_i$, $i = 1,2$ of \fourteen\ such that 
for each  $(y,\ep)\in[y_1,y_2]\times[0,\ep_1]$ and all $n\ep\in[t_{in},t_{fi}]$,
$$
f_i(n)=\psi_i(n)+r_i(n),
$$
where
$$
\Bigl|{r_i(n)\over u_i(n)}\Bigr|=\Bigl|{f_i(n)-\psi_i(n)\over u_i(n)}\Bigr|\le c\ep, \quad \eq\eno[ineqone].
$$
The constant $c$ is uniform on $[t_{in},t_{fi}]\times[y_1,y_2]\times[0,\ep_1]$ and ${r_i(n)\over u_i(n)}\in C([y_1,y_2]\times[0,\ep_1])$. 
For case~$1$ and $y<0$ or the second part of case~$3$
 with $[t_{in},t_{fi}]\times[y_1,y_2]\times[0,\ep_1]$ given in Theorem~\threethree\ there exist solutions $f_i,\ i = 1,2$ of \fourteen\ such that for each  $(y,\ep)\in[y_1,y_2]\times[0,\ep_1]$ and all $n\ep\in[t_{in},t_{fi}]$,
$$
\Bigl|{r_i(n)\over u_i(n)}\Bigr|=\Bigl|{(-1)^n f_i(n)-\psi_i(n)\over u_i(n)}\Bigr|
\le c\ep, \eq\eno[ineqone]$$
where the constant $c$ is uniform on $[t_{in},t_{fi}]\times[y_1,y_2]\times[0,\ep_1]$ and ${r_i(n)\over u_i(n)}\in C([y_1,y_2]\times[0,\ep_1])$.

\pf The result for cases~1--3 and $y>0$ follows from Theorem 3.8 in [\cite{GBVA}].  For case~1
and $y<0$ or the second part of case 3 apply Theorem 3.8 of [\cite{GBVA}]
to $(-1)^n f_i(n)$. \qed

\advance\subsecno by 1\no=0\thmno=0
\beginsection III. External Fields

In order to facilitate the matching problem considered below and to
consider solutions of \epdiff\ with $y$ complex we will find it convenient
to  use the theory of potentials with external fields. 

Let $Q(x)$ be a continuous function on a closed interval $\Sigma$ of the real line. If $\Sigma$ is unbounded we asume that $\lim_{|x|\to\infty}{Q(x)\over\ln{|x|}}=+\infty,\ x\in \Sigma$.
The above assumptions are not the most
general possible (see [\cite{ST}]) but they are sufficient for the problem
we are considering. Let $M(\Sigma)$ be the collection of
positive Borel probability measures on $\Sigma$ with compact support and define
$$I_Q(u)=-\int\int \log |z-t|du(z)du(x)+2\int Q(t)du(t)$$
$u\in M(\Sigma)$, and $E=\inf\{u\in\Sigma : I_Q(u),\
{\rm exists}\}$ then there is a unique equilibrium measure $v$ such that $E=I_Q(v)$
(see Saff-Totik [\cite{ST}], 
Rakhmanov [\cite{Raone}], Gonchar-Rakhmanov [\cite{GR}]).  
Set $V^v(z)=\int\log{1\over |z-t|}dv(t)$
and $F_Q=E-\int Q\,dv$, then $V^v(z)+Q(z)\ge F_Q$ on
$\Sigma$ and $V^v(z)+Q(z) =F_Q$ on
supp$(Q)$. The function $Q$ is called the external field. 

In order to consider discrete measures we need to generalize the above problem to a constrained equlibrium problem. Let $\Sigma$ be a closed bounded intervel and $\sigma$ a positive measure supported on this interval with continuous potential and $\sigma(\Sigma)>1$.
Let $M^{\sigma}(\Sigma)$ be the set of Borel probability measures such that if $\mu\in M^{\sigma}(\Sigma)$ then $\mu\le\sigma$.
It was proved by 
Dragnev and Saff [\cite{DS}, Theorem 2.1] and Rakhmanov [\cite{Rathree}, Theorem 3] (see also [\cite{KR}]), that
for $Q$ as above and constraint $\sigma$ there is a unique Borel probability measure $\nu_Q^{\sigma}\in M^{\sigma}(\Sigma)$ such that for a constant $F^{\sigma}_Q$ we have
$$U^{\sigma}_{\nu^{\sigma}_Q}+Q\le  F^{\sigma}_Q\ {\rm on\
supp}(\nu^{\sigma}_Q),$$
$$U^{\sigma}_{\nu^{\sigma}_Q}+Q= l^{\sigma}_Q\ {\rm on\ supp}(\nu^{\sigma}_Q)\cap{\rm supp}(\sigma-\nu^{\sigma}_Q),$$
$$U^{\sigma}_{\nu^{\sigma}_Q}+Q\ge  l^{\sigma}_Q\  {\rm on\ supp}(\sigma-\nu^{\sigma}_Q).$$

These results have been applied to difference equations by Deift and
McLaughlin [\cite{DM}], Kuijlaars and
VanAssche [\cite{KVA}] and Kuijlaars and Rahkmanov [\cite{KR}] and we have the useful

\thm thma. {\rm[\cite{KR}, Theorem 9.2]} Suppose that $\tilde
a(t)>0$ and $\tilde b(t)$ are continuous for $t\in[0,T]$, 
$\int_0^T |\ln \tilde a(u)|
du<\infty$,
$\tilde\gamma^+(t)=\tilde b(t)+2\tilde a(t)$
has at most one extremum, which if it
exists is a maximum and $\tilde\gamma^- (t)=\tilde b(t)-2\tilde a(t)$ has at most one
extremum, which if it exists is a minimum.  Set $0<c<T$
$$
Q(x)=\int_0^{t_-}\log\left|{x-\tilde b(u)\over 2 \tilde
a(u)}+\sqrt{\left({x-\tilde b(u)\over 2
\tilde a(u)}\right)^2-1}\right|du,
$$
where $t_-(x)=\inf\{0<t: x\in[\tilde\gamma^-(t),\tilde\gamma^+(t)]\}$
and
$$
\sigma=\int_0^T \omega_u du,
$$
with
$$
{d\omega_t(x)\over
dx}={1\over\pi}{1\over\sqrt{(\tilde\gamma^+(t)-x)(x-\tilde\gamma^-(t))}}
=\omega_t '.
$$
Then the constrained equilibrium problem with external field $Q_c={Q\over c}$
and constraint $\sigma_c={\sigma\over c}$ has the solution
$$
\nu_c ={1\over c}\int_0^c\omega_u du. \eq\eno[meas]
$$

As noted by [\cite{LLtwo}] the condition $\int |\ln\tilde a(t)|dt<\infty$
is needed to insure that $Q(x)$ exists. We will assume that 
$\tilde\gamma^+(t)\ne\tilde\gamma^-(t),\ 0<t<T$ which implies that $\tilde
a(t)>0,\ 0<t<T$. Kuijlaars and Van Assche [\cite{KVA}, Remark~1.5] have noted that
for $0<s<c$ the
support of the equilibrium measure is in the interval
$$
\left[\inf_{0<s<c}{\tilde\gamma}^-(s),\sup_{0<s<c}{\tilde\gamma}^+(s)\right].
$$
Furthermore from the definition of the constraint
we see [\cite{KR}] that it is not active if $\tilde\gamma^-$ 
is decreasing on
$[0,c]$ and $\tilde\gamma^+$ is increasing over the same region. However
if this is not the case then the constraint will be active on the
interval
$[{\tilde\gamma}^-_{\min},{\tilde\gamma}^-(c)]$ and
on the interval $[{\tilde\gamma}^+(c), {\tilde\gamma}^+_{\max}]$ where
${\tilde\gamma}^-_{\min}=\min_{t\in[0,c]} \tilde\gamma^-(t)$ 
and 
${\tilde\gamma}^+_{\max}=\max_{t\in[0,c]} \tilde\gamma^+(t)$ .

For $x\in R\backslash{\rm supp}(w_t)$, set
$$
\hat w'_t=|w'_t|. \eq\eno[freemeas]
$$

\remk  Note that equation~\meas\ is to be interpreted as
$$
\nu_c(E)={1\over c}\int_0^c \omega_u(E) du,\ \eq\eno[remeas]
$$
where $E$ is any Borel subset of ${\rm supp}(\nu_c)$ and since ${\rm supp}(\omega_t)\subset {\rm supp}(\nu_c)$ this can be
rewritten as
$$
\int_{{\rm supp}(\nu_c)} f(x) d\nu_c ={1\over c}\int_0^c\int_{{\rm supp}(\omega_u)}f(x)d\omega_u du, \eq\eno[remeasweak]
$$
where $f\in C[{\rm supp}(\nu_c)]$ the set continuous function on ${\rm supp}(\nu_c)$.

If
$$
g^t(z)=\ln\left({z-\tilde b(t)\over2\tilde a(t)}+
\sqrt{\left({z-\tilde b(t)\over2\tilde a(t)}\right)^2-1}\right),\eq\eno[freegreen]
$$
then $g^t$ is the complex Green's function associated with the interval 
$[b(t)-2a(t),b(t)+2a(t)]$, $0<t<\beta$. 
  From the mapping properties of $z(y)$ (see equation \map) we see that $g^t$  is analytic for $z\in \bbc\setminus(-\infty,b(t)+2a(t)]$ and 
$g^t_{\pm}=g^t|_{C_{\pm}}$ is continuous on $\bar C_{\pm}$. Furthermore
for $0<t<\beta$
$$
g^t_+-g^t_-=\cases{2\pi i, & $x\le \tilde\gamma^-(t)$\cr 2\pi i\int_x^{\tilde\gamma^+(t)} d\omega_t, & $x\in {\rm supp}(\omega_t)$\cr 0,& $x>\tilde\gamma^+(t)$.} \eq\eno[gtone]
$$
We also find that,
$$
g^t_++g^t_-=\cases{2\Re g^t, & $x\le \tilde\gamma^-(t)$\cr 
0,& $x\in{\rm supp}(\omega_t)$\cr  
2\Re g^t,& $x>\tilde\gamma^+(t)$.}\eq\eno[gttwo]
$$

\remk It was noticed by [\cite{LLtwo}] that for $x>\tilde\gamma^+(t)$
$$\Re g^t=\int^x_{\tilde\gamma^+(t)} {ds\over
\sqrt{(s-\gamma^+(t))(s-\gamma^-(t))}}=\pi\int^x_{\tilde\gamma^+(t)}\hat w'_t ds, \eq\eno[altgone]$$
and in case 1 for $x<\tilde\gamma^-(t)$
$$\Re g^t=\int^{\gamma^-(t)}_x {ds\over
\sqrt{(\gamma^+(t)-s)(\gamma^-(t)-s)}}=\pi\int^{\gamma^-(t)}_x \hat w'_tds.\eq\eno[altgtwo]$$
These formulas are easily obtained by differentiating $\Re g^t$ with
respect to $x$ then integrating and using equation~\freemeas.

Set
$$
V^c(z)={1\over c}\int_{{\rm supp}{\nu_c}}\ln(z-x)d\nu_c(x), \eq\eno[gc]
$$
for $z\in \bbc\setminus(-\infty,\sup_{0<s<c}\tilde\gamma^+(s)]$

We have  the simple,

\lem repgc. With the hypotheses of Theorem \thma,
$V^c(z)$ is analytic for
$z\in \bbc\setminus(-\infty,\gamma^+_{\max}]$
and
$$
V^c(z)={1\over c}\int_0^c\ln\left({z-\tilde b(u)\over2\tilde a(u)}+
\sqrt{\left({z-\tilde
b(u)\over2\tilde a(u)}\right)^2-1}\right)du + l_c. \eq\eno[newrepgc]
$$
Furthermore $V^c_{\pm}=V^c|_{C_{\pm}}$ have continuous extensions to $\bar
C_{\pm}$ which satisfy
$$
V^c_+(x)-V^c_-(x)=\cases{ 
2\pi i, & $x<\tilde\gamma^-_{\min}$\cr 
2\pi i\int^{\tilde\gamma^+(c)}_x d\nu_c,& $ x\in
{\rm supp}
{\nu_c}$\cr 0, & $x>\tilde\gamma^+_{\max}$}.\eq\eno[gcpm]
$$
Here $l_c={1\over c}\int_0^c \ln \tilde a(t) dt$.

\pf The analyticity properties of $V^c$ follow from its definition. For $z\in \bbc\setminus(-\infty,b+2a], a>0 ,b$ real, the useful integral representation holds [\cite{ST}],
$$
\ln\left({z-b\over2a}+ \sqrt{\left({z-b\over2a}\right)^2-1}\right)+\ln a=
{1\over\pi}\int_{b-2a}^{b+2a}\ln(z-x)
{dx\over\sqrt{4a^2-(x-b)^2}}. \eq\eno[logg]
$$
The integrals from $0<u<c$ of $g_{\pm}$ exist by the hypothesis of
Theorem~\thma\ and give rise to continuous functions. Equation~\newrepgc\ can now be
obtained from \remeasweak\ with $f=\ln(z-x)$ and \logg. The second part of the Theorem is a consequence of the continuity  and
integrability of $g^t_{\pm}$, \gtone, and \meas.\qed

We will now restrict the coefficients to those
cases that arise in the examples to be considered. For $t\in(0,T)$ assume one of the following statements hold:
\item{1)} $\tilde\gamma^-(t)$ is a strictly decreasing function
of $t$ and $\tilde \gamma^+(t)$ is a strictly increasing function of $t$. 
\item{2)} $\tilde\gamma^-(t)$ is a constant and $\tilde \gamma^+(t)$ is
a strictly increasing function of $t$. 
\item{3)} $\tilde\gamma^-(t)$ and $\tilde\gamma^+(t)$ are strictly
increasing functions of $t$.

In all cases we will assume that
$\tilde\gamma^+(0)=\tilde\gamma^-(0)$. With these conditions Theorem \thma\
insures that for cases 1--3 and for each $\tilde\gamma^+(T)>x>\tilde\gamma^+(0)$ there is an external field. Also in case 1, for each $\tilde\gamma^-(0)>x>\tilde\gamma^-(T)$ there is an associated external
field. It is not difficult to see from the definition of $Q$ that $\lim_{x\to\infty}(Q(x)-\ln x)=\infty$ for the above cases and $\lim_{x\to-\infty}(Q(x)-\ln|x|)=\infty$ for case~1. The support of the equilibrium measures for the cases above are
given in Figure 1.
\bigskip
\centerline{\epsfxsize=3in\epsfbox{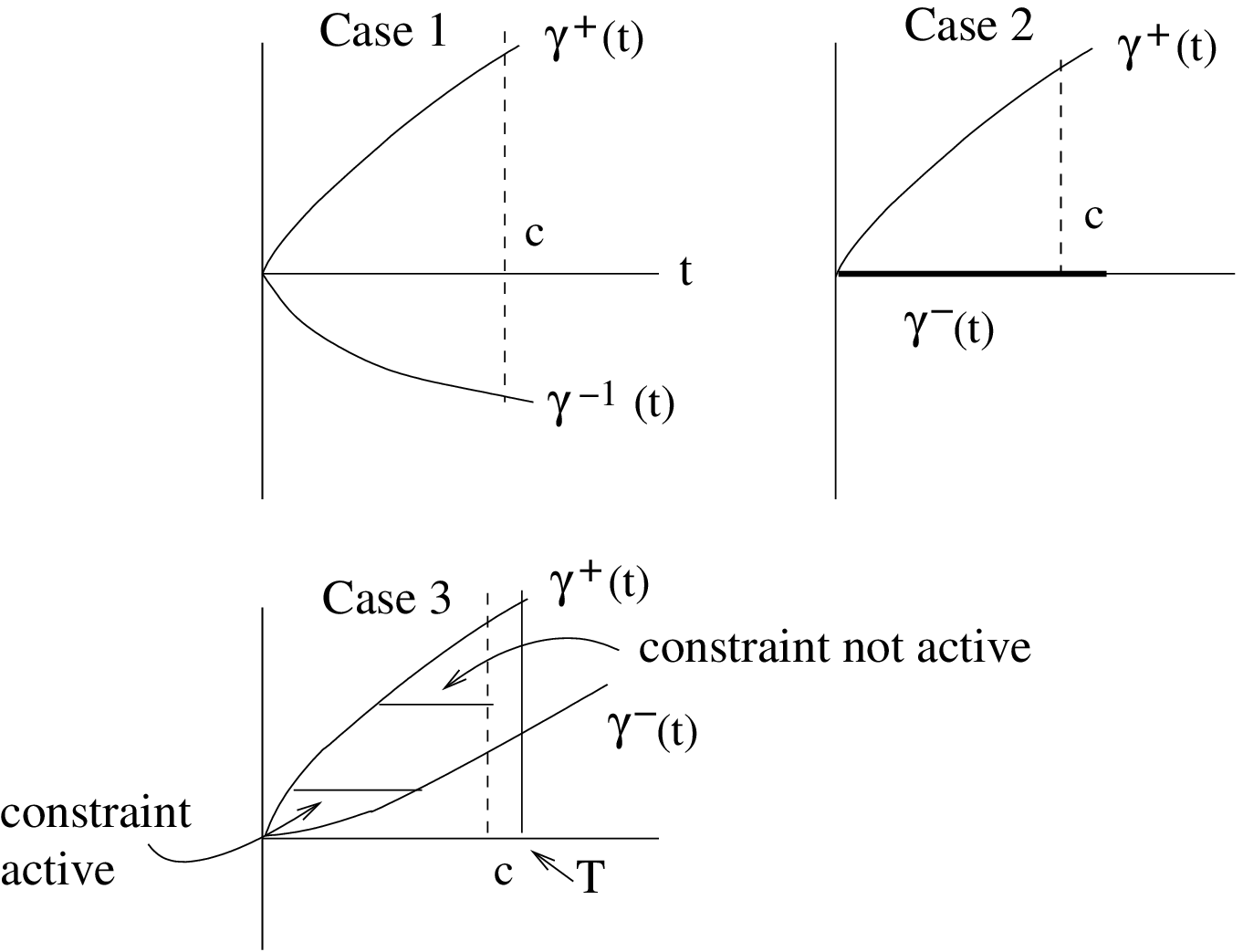}}

\centerline{Figure 1: Support of the equilibrium measures for cases 1--3}

It is not difficult to compute the densities for these measures [\cite{BR}],
[\cite{LLtwo}].

\lem emeas. The density of the equilibrium measure
is given by
$$
{d\nu_c\over dx}={1\over c}\int_{f(x)}^c \omega_t ' dt,\quad
 \tilde\gamma^-(c)< x\le\tilde\gamma^+(c) 
$$
where for case one,
$$f(x)=\cases{(\tilde\gamma^-)^{-1}(x),& 
$\tilde\gamma^-(c)< x\le 0$\phantom{$1\over 2^{1\over 2}$}\cr
(\tilde\gamma^+)^{-1}(x), & $0< x\le \tilde\gamma^+(c)$ } \eq\eno[emeascone]
$$
while for case two
$$f(x)=(\tilde\gamma^+)^{-1}(x),\quad 0< x\le \tilde\gamma^+(c).\eq\eno[emeastwo]$$
For case 3 
$$
{d\nu_c\over dx}={1\over c}\int^{f(x)}_{(\tilde\gamma^+)^{-1}(x)} \omega_t ' dt,
\quad 0< x\le\tilde\gamma^+(c) 
$$
where
$$f(x)=\cases{(\tilde\gamma^-)^{-1}(x), & $0< x\le \tilde\gamma^-(c)$
\phantom{$1\over 2^{1\over 2}$}\cr
c,& $ \tilde\gamma^-(c)< x\le \tilde\gamma^+(c)$}.  \eq\eno[emeascthree]
$$

\pf The proof follows by substituting the above formulas into the left hand
side of \remeasweak\ then interchanging the order of integration
(see Figure~1) we obtain
$$
\int^{\tilde\gamma^t(c)}_d g(x)dv_c=\int^c_0 {1\over c} 
\int^{\tilde\gamma^+(t)}_{\tilde\gamma^-(t)} g(t){dw_t\over dx}\, dx,
$$
where $g\in C[{\rm supp}\; \nu_c]$.  The result now follows from formula
\remeasweak\ and the definition of $\nu_c$.
Let ${d\hat\nu_c\over dx}$ be the extension of 
${dv_c\overwithdelims || dx}$ to $x\in R\backslash{\rm supp}\; \nu_c$
and ${d\tilde \nu_c\over dx}$ be the extension of 
${d(\sigma_c-\nu_c)\overwithdelims || dx}$ to $x\in R\backslash
{\rm supp}\;(\sigma- \nu_c)$. We now have the following,

\lem externalg. Let $V^c_{\pm}$ be as above then, for case 1
$$
V^c_+(x) +V^c_-(x)-2 l_c -2Q_c(x)=\cases{-2\pi\int^{\tilde\gamma^-(c)}_x d\hat\nu_c ,
& $\tilde\gamma^-(T)<x\le\tilde\gamma^-(c)$ \phantom{$1\over 2^{1\over 2}$}\cr
\qquad 0,&
$\tilde\gamma^-(c)<x\le\tilde\gamma^+(c)$\phantom{$1\over 2^{1\over 2}$}\cr
-2\pi\int_{\tilde\gamma^+(c)}^x d\hat\nu_c , &
$\gamma^+(c)<x\le\tilde\gamma^+(T)$} \eq\eno[cgcaseone]
$$ 
while for case 2
$$
V^c_+(x) +V^c_-(x)-2 l_c
-2Q_c(x)=\cases{\qquad 0,
&$\tilde\gamma^-(c)<x\le\tilde\gamma^+(c)$
\phantom{$1\over 2^{1\over 2}$}\cr
-2\pi\int_{\tilde\gamma^+(c)}^x
d\hat\nu_c ,& $\tilde\gamma^+(c)<x\le\tilde\gamma^+(T)$} \eq\eno[cgcasetwo]
$$
and for case 3
$$
V^c_+(x) +V^c_-(x)-2 l_c -2Q_c(x)=\cases{+2\pi\int^{\tilde\gamma^-(c)}_x d\tilde\nu_c ,
& $0< x\le\tilde\gamma^-(c)$\phantom{$1\over 2^{1\over 2}$}\cr
\qquad 0,&
$\tilde\gamma^-(c)<x\le\tilde\gamma^+(c)$\cr-2\pi\int_{\tilde\gamma^+(c)}^x d\hat\nu_c ,&
$\tilde\gamma^+(c)<x\le\tilde\gamma^+(T).$
}\eq\eno[cgcasethree]
$$

\pf The 2nd line of cases~1 and 3 and the first line of case~2 follow
from integrating the 2nd line of equation \gttwo.
For the bottom line of the above cases integrate the last
line of equation \gttwo\ to obtain
$$V^c_+(x)+V^c_-={2\over c} \int^c_0\ln \left(
{x-\tilde b(u)\over 2\tilde a(u)} + 
\sqrt{ {x-\tilde b(u)\overwithdelims () 2\tilde a(u)}^2-1}\right)\, du + 2\ell_c
.
$$
The right-hand side of the above equation exists from the hypothesis of
Theorem~\thma\ and the continuity of $Rg^t$.  Extend the integral to
$(\tilde\gamma^+)^{-1}(x)$ and use the definition of $Q$ to obtain
$$V^c_+ + V^c_-=2Q_c(x)+2\ell_c-
{2\over c}\int^{(\tilde\gamma^+)^{-1}(x)}_c Rg^tdt.$$
Since $\Re g^t=\int^x_{\tilde\gamma (t)} \hat w_t(y)dy$ (see equation~\altgone) the
last line of the above cases follows.  The first line in case~1
may be derived in an analogous manner by integrating the 1st line
in \gttwo.  To obtain the 1st line in case~3 integrate the
first line in \gttwo\ to obtain (see Figure~3)
$$V^c_+ + V^c_- -2\ell_c={2\over c}\left[\int^{(\tilde\gamma^+)^{-1}(x)} _0
+\int^{(\tilde\gamma^-)^{-1}(x)}_{(\tilde\gamma^+)^{-1}(x)} + 
\int^c_{(\tilde\gamma^-)^{-1}(x)} \Re g^tdt\right].$$
The first integral is equal to $Q$ while the second is equal to
zero (since $\Re g^t=0$).  The result now follows from \altgtwo\  and the
definitions of the equilibrium and constraining measures.~\qed

We now consider the continuity in $\ep$ of the above functions. 

\lem contQ. Suppose that $a(t,\ep)$ and $b(t,\ep)$ satisfy
conditions {\rm\condcoeffone--\condcoeffthree}\  and there is an integrable $m(t)$ such that
$$
|\ln(a(t,\ep))|\le m(t),\quad 0<t<T 
$$
for $\ep\in[0,\ep_0]$. Then for cases {\rm 1--3}, $tV_+^t(y,\ep)\in C(\bar
\bbc_+\times[0,T]\times[0,\ep_0])$ while $tV_-^t(y,\ep)\in C(\bar\bbc_-\times[0,T]\times[0,\ep_0])$. Furthermore for every interval
$[y_1,y_2]\subset(0,\gamma^+_0(T))$ there exists an $\ep_1$ such that $Q\in
C([y_1,y_2]\times[0,\ep_1])$. For case $1$, and 
$y\in(\gamma^-_0(T),0)$ the same conclusions hold with $[y_1,y_2]\subset(\gamma^-_0(T),0)$. In case $1$ $\lim_{y->0_+}Q(y,0)=0=\lim_{y->0_-} Q(y,0)$  
while in cases $2$ and $3$ $\lim_{y->0_+}Q(y,0)=0$.

\pf For fixed $\ep, \gamma^{\pm}_{\ep}(t)$, and $a(t,\ep)$ satisfy the hypotheses of Theorem~\thma. If $y>0$ then for cases 1--3 we find from the definition of $t_{-}$ that $t_{-}=t_p^+(y,\ep)$. Consider now,
$$
Q(y,\ep)=\int_0^{t^+_p(y,\ep)}\ln\left|{y-b(u,\ep)\over2a(u+{\ep\over2},\ep)}+\sqrt{\left({y-b(u,\ep)\over2a(u+{\ep\over2},\ep)}\right)^2-1}\right|du.
$$
For $[y_1,y_2]\subset(0,\gamma^+_0(T))$
let $\ep_1$ be such that $t_p^+(y,\ep)\in
C([y_1,y_2]\times[0,\ep_1])$. The integrability of $m$
as well the fact that the integrand is uniformly continuous for
$(t, y,\ep)\in[t_1,t_2]\times[y_1, y_2]\times[0,\ep_1]$ for any $[t_1,t_2]\subset(0,
T)$ gives the continuity property for $Q$. From equation~\newrepgc\ with $\tilde b(t)=b(t,\ep)$ and $\tilde
a(t)=a(t+{\ep\over2},\ep)$ for $z\in\bbc\backslash R$ we find,
$$
tV^t(z)=\int_0^t g^u(z,\ep)du + 
\int_0^t\ln a\left(u+{\ep\over2},\ep\right)du . \eq\eno[newrepgt]
$$
Equation~\freegreen\ and \condcoeffone\ show 
that the function $g^t(z,\ep)$ is uniformly continuous on compact subsets of
$(\bbc\backslash{\rm R})\times(0,T]\times[0,\ep_0]$, while
$g^t_{\pm}(z,\ep)$ are  uniformly continuous on compact subsets of
$\bar\bbc_{\pm}\times(0,T]\times[0,\ep_0]$ respectively. The continuity
properties of $V^t_{\pm}$ now follow from an argument similar to that given 
for $Q$. To show for case 1 that $\lim_{y->0_+}Q(y,0)=0$ note that
for cases 1--3, \condcoeffthree\ implies that $\lim_{y\to0_+}t^+_p(y,0)=0$. The integrability of $g^t$ now gives the result. Case 1 and
$y<0$ follows in a similar manner. \qed

\remk 
With the assumptions on $a(t,\ep)$ and $b(t,\ep)$ above, Lemmas~\repgc\ and \externalg\ are easily extended to the case when $\ep>0$ and we will use the the same numbering.

From equations~\logform\ and \newrepgc\ we have how $\rho$ is related  to the equilibrium measure,

\thm realg. Suppose {\rm\condcoeffone--\condcoeffthree}\ hold then for cases $1$ 
and $2$ and every rectangle $[t_{in},t_{fi}]\times[y_1,y_2]\subset(0,T)\times(0, \gamma^+_0(T))$ or for case $3$ 
with $[t_{in},t_{fi}]\times[y_1,y_2]\subset(0,T)\times(\gamma^-_0(T),\gamma^+_0(T))$ 
or with  $[t_{in},t_{fi}]\times[y_1,y_2]$ such that $[y_1, y_2]\subset[0,\gamma^+_0(T)]$ and $[t_1, t_2]\subset(0, (\gamma^-_0)^{-1}(y_1))$ there exists an
$\ep_1$ such that for all
$(t,y,\ep)\in[t_{in},t_{fi}]\times[y_1,y_2]\times[0,\ep_1]$,
$$\eqalign{
{2\over3}(-\rho_1)^{3/2}(t,y,\ep)&=\pi t
\int_y^{\gamma^+(t,\ep)}d\nu^{\ep}_t\qquad\hbox{\rm for }
 t_p^+(y, \ep)< t\le t_{fi},\cr
{2\over3}(\rho_1)^{3/2}(t,y,\ep)&=\pi t\int^y_{\gamma^+(t,\ep)}d\hat\nu^{\ep}_t
\qquad\,\hbox{\rm ~for }
t_{in}<t\le t^+_p(y,\ep). }\eq\eno[rhoone]
$$

For case~1 and 
every interval $[t_{in}, t_{fi}]\times[y_1,y_2]\in(0,T)\times(\gamma^-_0(T),0)$ there exists an
$\ep_1$ such that for all
$(t,y,\ep)\in[t_{in},t_{fi}]\times[y_1,y_2]\times[0,\ep_1]$,
$$\eqalign{
{2\over3}(-\rho_2)^{3/2}(t,y,\ep)&=\pi t
\int^y_{\gamma^-(t,\ep)}d\nu^{\ep}_t\qquad\,\hbox{\rm ~for } t_p^-(y, \ep)< t\le t_{fi},\cr
{2\over3}(\rho_2)^{3/2}(t,y,\ep)&=\pi t\int_y^{\gamma^-(t,\ep)}d\hat\nu^{\ep}_t
\qquad\hbox{\rm for } t_{in}<t\le t^-_p(y,\ep). }\eq\eno[rhotwocaseone]
$$

Finally for case~3 and every interval $[t_1,t_2]\times[y_1,y_2]$ such that $[y_1, y_2]\in(0, \gamma^-_0(T))$ and $[\hat t_{in},\hat t_{fi}]\subset((\gamma^+_0)^{-1}(y_2),T)$ there exists an
$\ep_1$ such that for all $(t,y,\ep)\in[\hat t_{in},\hat t_{fi}]\times[y_1,y_2]\times[0,\ep_1]$,
$$\eqalign{
{2\over3}(\rho_2)^{3/2}(t,y,\ep)&=\pi t
\int_y^{\gamma^-(t,\ep)}d\tilde\nu^{\ep}_t\qquad\hbox{\rm for } t_p^-(y, \ep)< t\le \hat
t_{fi},\cr
{2\over3}(-\rho_2)^{3/2}(t,y,\ep)&=\pi
t\int^y_{\gamma^-(t,\ep)}d\nu^{\ep}_t
\qquad\,\hbox{\rm ~for } \hat t_{in}<t\le t^-_p(y,\ep). }\eq\eno[rhotwocasetwo]
$$

\pf For cases~1--3, $y\in[y_1,y_2]$ and $t>t^+_p(y,\ep)$ the result follows
by using \altgone\ to rewrite 
$\cosh^{-1}\left({y-b(u,\ep)\over2a(u+\ep/2,\ep)}\right)$ in \langer\ as an integral then
interchanging the order of integration (see equation~\altgone) using Fubini's theorem.  The bottom
line of equation~\rhotwocaseone\ can be obtained in a similar manner (see
\altgtwo). For the top line in \rhoone\
apply the above procedure to the bottom line of \langer. The top lines of
equation~\rhotwocaseone\ and \rhotwocasetwo\  follow in a similar manner
using the bottom lines of \langero\ and \langert. For the top line of
equation~\rhotwocasetwo\ $y< \gamma^-(t,\ep)$, so  recasting
$\cosh^{-1}$ as an integral in the top line of \langert\ and the definition
of $\tilde\nu_t$ gives the result. 
\qed

\item{(2)} Complex case

We now consider complex extensions of the above results. Motivated by the examples discussed earlier we will assume that
$$
a(t,\ep)=a\hat q(t,\ep),\ b(t,\ep)=b\hat q
\left(t+{\ep\over2},\ep\right),\quad  a>0,\quad b\ge0,\eq\eno[ab]
$$
with
$$
\hat q_{\ep}(t)=\hat q(t,\ep)={ \hat q({t\over\ep})\over \hat q({1\over\ep}+{1\over2})}. \eq\eno[qhat]
$$

Set $q_{\ep}(t)=\hat q_{\ep}(t+{\ep\over2})$  then the inverse functions of $\gamma^+_{\ep}$ and if  $b\ne2a$
$\gamma^-_{\ep}$ which we denote respectively as $t^+_p(y,\ep)$ and $t^-_p(y,\ep)$ are given by
$t^+_p(y,\ep)={q_{\ep}}^{-1}({y\over b+2a})$ and
$t^-_p(y,\ep)={q_{\ep}}^{-1}({y\over b-2a})$.
With $\Omega_{\delta}=\{z: |{\rm arg}(z)|<\delta\}$ we will assume that,
\item{ic)} $\hat q(t)$ is nonegative, continuous and strictly increasing for $t\ge0$,
\item{iic)} it has an analytic one-to-one extension to
an  open set $\Omega\subset\bbc$ with $\Omega_{\delta}\subset\Omega$ for some nonzero $\delta$, and there is an $\alpha>0$ such that
$\lim_{(t\to\infty,\ t\in\Omega)}t^{{-1\over\alpha}}\hat q(t)=k,\ k>0$,
\item{iiic)} $\int_0^{t}|\ln \hat q(u)|du <\infty$ for all finite $t>0$.

With the above conditions we find,

\lem hatq. Suppose ic)--iiic) hold then $\hat q_{\ep}\in H(\Omega_{\delta})$
for $\ep\ge0$ and $\hat q_0(t)=t^{1\over\alpha}$. Also for every $\ep_0>0$, $\hat q_{\ep} \in
C^{\infty}(\Omega_{\delta}\times(0,\ep_0])$, ${\partial^i\over\partial
t^t}\hat q_{\ep}\in C(\Omega_{\delta}\times[0,\ep_0])$ for $i=0,\ldots$ and for every $\ep\in[0,\ep_0]$ $\hat q_{\ep}(t)$ is one-to-one on $\Omega_{\delta}$. $\hat q_{\ep}(t)>0$, for
$(t,\ep)\in(0,\infty)\times[0,\ep_0]$ and $\int_0^t|\ln\hat q_{\ep}(u)|du<\infty$ for all finite $t>0$ and $\ep\in[0,\ep_0]$. 

\pf Since $\hat q(t)\in H(\Omega_{\delta})$ we see that $\hat
q({1\over\ep})\in C^{\infty}((0,\ep_0])$ and from iic) that $\lim_{\ep\to
0}\hat q(t,\ep)-t^{1\over\alpha}=0$ uniformly on compact subsets of
$\Omega_{\delta}$. This gives the continuity and analyticity properties of
$\hat q(t,\ep)$ and shows that $\hat q_0(t)=t^{1\over\alpha}$. It follows from iic) that $\hat q_{\ep}(t)>0,$ for $(t,\ep)\in(0,\infty)\times[0,\ep_0]$. That $\hat q_{\ep}(t)$ is conformal for each
$\ep\in[0,\ep_0]$ and increasing for $t>0$ can be deduced  from ic), iic), and the definition of $\hat
q_{\ep}(t)$.  To see the integrability condition write
$$
\eqalign{
&\int_0^t|\ln\hat q_{\ep}(u)|du=\int_0^t\Bigl|\ln
\left({\ep^{1/\alpha}\hat q(u/\ep)\over\ep^{1\over\alpha}\hat q(1/\ep)}\right)
\Bigr|du\cr
&\quad<\ep\int_0^K|\ln(s^{1/\alpha}\hat q(s))|ds+ \ep\int_K^{t/\ep}|\ln(s^{1/\alpha}\hat q(s))|ds+Ct(|\ln t|+1).}\eq\eno[uint]
$$
Property iiic) implies the convergence of the first integral while iic) implies the integrability of the second term. \qed

From the definition of $q_{\ep}(t)$ we find that
$$q^{-1}_{\ep}(y)=\ep\hat q^{-1}\left(\hat
q\left({1\over\ep}+{1\over2}\right)y\right)-{\ep\over2}. \eq\eno[qinv]
$$

Set $S^{\ep}=q_{\ep}(\Omega_{\delta})$. The fact that
$\hat q$ is conformal on $\Omega_{\delta}$ and maps $(0,\infty)$ into
$(0,\infty)$ implies that $S^{\ep}_+=q_{\ep}((\Omega_{\delta})_+)\subset\bbc_+$ and
$S^{\ep}_-=q_{\ep}((\Omega_{\delta})_-) \subset\bbc_-$.

\lem invfuncom. For each fixed $\ep, q^{-1}_{\ep}\in H(S^{\ep})$. If $y\ge r>0$ then there is an $\ep_0$ such that $q^{-1}_{\ep}>0$ $(y,\ep)\in[r,\infty)\times[0,\ep_0]$. If $K$ is a compact subset of $S^0$ then there exists an
$\ep_1$ such that $q^{-1}_{\ep}\in H(K)$ for $\ep\in[0,\ep_1]$,
$q_{\ep}^{-1}\in C^{\infty}(K\times(0,\ep_1])\cap C(K\times[0,\ep_1])$. Finally
${q^{-1}_{\ep}(u)\chi_{[\hat q_{\ep}(0), 1]}(u)\over u}\to u^{\alpha-1}$ in $L([0,1])$, where $\chi_{[\hat q_{\ep}(0), 1]}$ is the characteristic function of the set $[\hat q_{\ep}(0),1]$.

\pf That $q^{-1}_{\ep}\in H(S^{\ep})$ is a conquence of Lemma~\hatq. Since $q^{-1}_0(r)=r^{\alpha}>0$ the continuity of $q^{-1}_{\ep}$ in $\ep$ implies that there is an $\ep_0$ so that $q^{-1}_{\ep}(r)>0$ for all $\ep\in[0,\ep_0]$. The monotonicity of $q^{-1}_{\ep}$ as a function of $y$ shows that $q^{-1}_{\ep}(y)>0$ for all $(y,\ep)\in[r,\infty)\times[0,\ep_0]$. Given $K\subset S_0$ a compact set let $W\subset\Omega_{\delta}$ be such that
$K=q_0(W)$. The continuity of $q_0$ implies that $W$ is compact. The
uniform continuity of $q_{\ep}(t)$ on compact subsets of
$\Omega_{\delta}\times[0,\ep_0]$ implies that there is an $\ep_1>0$ and a
compact set $W_1\subset\Omega_{\delta}$ such that $K\subset q_{\ep}(W_1)$
for all $\ep\in[0, \ep_1]$. Thus $q^{-1}_{\ep}$ is well defined on $K$ for
$\ep\in[0,\ep_1]$. The conformality of  $q_{\ep}$,
Lemma~\hatq, and the inverse function theorem show that
$q^{-1}_{\ep}(t)\in H(K)\cap C(K\times[0,\ep_1])\cap C^{\infty}(K\times(0,\ep_1])$.
To see the $L^1$ convergence we will show that the above family  is  uniformly integrable  and then use Vitali's convergence theorem. Because of the continuity properties of this family only the uniform integrability near zero need be shown. 
Note that
$$
\int_0^t\ln(\hat q_{\ep}(u))du=\int_{\hat q_{\ep}(0)}^{\hat q_{\ep}(t)} \ln(w) q^{-1}_{\ep}(w)'dw
=t\ln(\hat q_{\ep}(t))-\int_{\hat q_{\ep}(0)}^{\hat q_{\ep}(t)}{q^{-1}_{\ep}(w)\over w}dw.
$$
Utilizing the fact that $\hat q_{\ep}(t)\le 1$ for $0\le t\le1$ we find,
$$
0\le\int_0^{\hat q_{\ep}(t)}{\chi_{[\hat q_{\ep}(0),1]}q^{-1}_{\ep}(w)\over w}dw\le\int_0^t|\ln(\hat q_{\ep}(u))|du.
$$
Lemma~\hatq\ shows that $\hat q_{\ep}(t)\to t^{1/\alpha}$ uniformly on compact subsets of R$_+$ so that the uniform integrability follows from \uint. \qed

A limit we will have use of is
$$\lim_{y\to\infty, y\in S^{\ep}} |y^{-\alpha}q^{-1}_{\ep}(y)|=
\ep q\left({1\over\ep}+{1\over2}\right)^{\alpha}. \eq\eno[asqinv]
$$
With the above assumptions we are able to give some representation formulas
for $Q$. Suppose $y>0$ then for $\ep$ sufficiently small $q_{\ep}^{-1}(y)>0$. Thus
$$
\eqalign{
Q(y,\ep)&=\int_0^{t^+_p(y,\ep)}\ln\left({y\over2aq_{\ep}(u)}-{b\over2a}+
\sqrt{\left({y\over2aq_{\ep}(u)}-{b\over2a}\right)^2-1}\,\right)du\cr
&=\int_{q_{\ep}(0)}^{y\over b+2a}
\ln\left({y\over2aw}-{b\over2a}+
\sqrt{\left({y\over2aw}-{b\over2a}\right)^2-1}\right)q^{-1}_{\ep}(w)'dw.}\quad\eq\eno[lnq]
$$
Integration by parts yields,
$$
\eqalign{
Q(y,\ep)&=y\int_{q_{\ep}(0)}^{y\over b+2a}{q^{-1}_{\ep}(w)\over w}{dw\over
\sqrt{(y-bw)^2-(2aw)^2}}\cr&=\int_{q_{\ep}(0)\over y}^{1\over b+2a}{q^{-1}_{\ep}(yu)\over u}{du\over
\sqrt{(1-bu)^2-(2au)^2}}.}\eq\eno[repqygo]
$$ 
Likewise for case~1 and $y<0$ we find
$$
\eqalign{
Q(y,\ep)&=-y\int_{q_{\ep}(0)}^{y\over b-2a}{q^{-1}_{\ep}(w)\over w}{dw\over
\sqrt{(y-bw)^2-(2aw)^2}}\cr&=-\int_{q_{\ep}(0)\over -y}^{1\over 2a-b}{q^{-1}_{\ep}(-yu)\over u}{du\over
\sqrt{(1+bu)^2-(2au)^2}}.}\eq\eno[repqylo]
$$
The definition for $V^t(y,\ep)$ is
$$
\eqalign{
V^t(y,\ep)=&{1\over
t}\int_0^t\ln\left({y\over2aq_{\ep}(u)}-{b\over2a}+
\sqrt{\left({y\over2aq_{\ep}(u)}-{b\over2a}\right)^2-1}\,\right)du\cr&+{1\over
t}\int_0^t\ln aq_{\ep}(u) du,} \eq\eno[greeneq]
$$
 and the density for the equilibrium measure also simplifies to
$$\eqalign{
{d\nu^{\ep}_t\over dy} &={1\over\pi t}\int^t_{(\gamma^+_\ep)^{-1}(y)}
{dp\over\sqrt{(\gamma^+_\ep (p)-y)(y-\gamma^-_\ep (p))}}\cr
&={1\over t\pi} \int^{2a+b}_{y(2a+b)\over \gamma^+_\ep (t)}
{q^{-1}_\ep (y/w)'dw\over
w\sqrt{4a^2-(w-b)^2}}}\eq\eno[eqqmeasure]
$$
for cases~1--3 and $\max \{(b+2a)q_{\ep}(0),\gamma^-_\ep (t)\}<y< \gamma^+_\ep (t)$.
For case~1 and $\gamma^-_\ep (t)<y<(b-2a)q_{\ep}(0)$
$$\eqalign{
{d\nu^{\ep}_t\over dy} &={1\over  t\pi}\int^t_{(\gamma^-_\ep)^{-1} (y)}
{dp\over\sqrt{(\gamma^+_{\ep}(p)-y)(y-\gamma^-_{\ep}(p))}}\cr
&={1\over t\pi}\int^{2a-b}_{-y(b-2a)\over\gamma^-_\ep (t)}
{q^{-1}_\ep (-y/w)'dw\over
w\sqrt{4a^2-(b+w)^2}}}\eq\eno[eqqmeatwo]$$
while for case~3 and $(b+2a)q_{\ep}(0)<y\le \gamma^-_\ep (t)$
$$\eqalign{
{d\nu^\ep_t\over dy} &= {1\over t\pi}
\int^{(\gamma^-_\ep)^{-1}(y)}_{(\gamma^+_{\ep} )^{-1}(y)}
{dp\over\sqrt{(\gamma^+_\ep (p)-y)(y-\gamma^-_{\ep}(p))}}\cr
&={1\over t\pi}\int^{b+2a}_{b-2a} 
{q^{-1}_\ep (y/w)'dw\over w\sqrt{4a^2-(b-w)^2}}\,.}
\eq\eno[eqqmeathree]$$
The above representations are obtained from Lemma~\emeas\ by setting 
$u=q_\ep (p)$, $u=yv$ or $-yv$ and then $w=1/v$.

\thm qgcomone. Suppose ic)--iiic) hold and let $r>0$. Then there exists an $\ep_0$
such that for all $\ep\le\ep_0,\ Q(y,\ep)$ exists for $y\ge r$. Furthermore
$Q(y,\ep)$ has an analytic extension to $S^{\ep}$. If $K\subset S^0$ is  a compact set, then there
exists an $\ep_1$ such that $Q(y,\ep)\in H(K)$ for $\ep\in[0,\ep_1]$, and $Q\in C^{\infty}(K\times(0,\ep_1])\cap
C(K\times[0,\ep_1])$. For case~$1$ and $y<-r$ there
exists an  $\ep_0$
such that for all $\ep\le\ep_0$, $Q(y,\ep)$ exists for $y\le -r$. 
$Q(-y,\ep)$ has an analytic extension to $S^{\ep}$ and if $K\subset S^0$ is  a compact set, then there
exists an $\ep_1$ such that $Q(-y,\ep)\in H(K)$ for all
$\ep\in[0,\ep_1]$. Also $Q(-y,\ep)\in C^{\infty}(K\times(0,\ep_1])\cap
C(K\times[0,\ep_1])$.
For each $t\in(0,t_{fi}]$ and $\ep\ge0,\ V^t(y,\ep)\in
H(\bbc\setminus(-\infty,\gamma^+(t_{fi},\ep)))$, $V^t(y,\ep)\in
 C^{\infty}(\bbc_{\pm}\times(0,t_{fi}]\times(0,\ep_0])$ for $\ep_0>0$. For
each compact set $K\in \bbc\setminus(-\infty,\gamma^+(t_{fi},0))$ there is
an $\ep_1$ such that $tG\in C^{\infty}(K\times(0,t_{fi}]\times(0,\ep_1])$ and ${\partial^i\over \partial t^i}tV^t\in C(K\times(0,t_{fi}]\times[0,\ep_1]),\ i\ge0$.  Finally $tV^t_+$ and $tV^t_-$ have
extensions to $\bar\bbc_+$ and $\bar\bbc_-$ respectively that are in
$C(\bar\bbc_{\pm}\times(0,t_{fi}]\times[0,\ep_0])$ for $\ep_0>0$.

\pf From Lemma~\invfuncom\ we see that for $y=r>0$
there is an $\ep_0$ such that $q^{-1}_{\ep}(y)>0$ for all
$\ep\in[0,\ep_0]$. Since for fixed $\ep,\ q^{-1}_{\ep}(y)$ is an 
increasing function of $y$, $q^{-1}_{\ep}(y)>0$ for all $y\ge r$. Thus
Theorem~\thma\ and Lemma~\hatq\ show that $Q(y,\ep)$ is well defined for each
$\ep\in[0,\ep_0]$ and $y\ge r$ and can be represented by \repqygo\
. Lemma~\invfuncom\ and  Morera's Theorem, applied to the second integral in equation\repqygo\ gives an
extension of  $Q(y,\ep)$ that is  analytic on $S^{\ep}\setminus[0,q_{\ep}(0)(b+2a)]$ for $\ep\in[0,\ep_0]$.  Let $K\subset S^0$ be a compact set. Then
the above argument and Lemma~\invfuncom\ show that there is an $\ep_1$ such
that $Q(\cdot,\ep) \in H(K)$ for all $\ep\in[0,\ep_1]$ and $Q\in C^{\infty}(K\times(0,\ep_1])$. 
Using the convexity of $S^0$ chose  $\hat q_{\ep}(0)<t$  so that the branch of the square root is respected and the line segment $[t,{y\over b+2a}]\in S^{\ep}, \ep\in[0, \ep_1]$ so  the second integral in \repqygo\ may be written as 
$$
\int_{q_{\ep}(0)\over y}^{1\over b+2a}{q^{-1}_{\ep}(yu)\over u}{du\over
\sqrt{(1-bu)^2-(2au)^2}}=\int_{q_{\ep}(0)\over y}^{t\over y}+\int_{t\over y}^{1\over b+2a}{q^{-1}_{\ep}(yu)\over u}{du\over
\sqrt{(1-bu)^2-(2au)^2}}.
$$
Lemma~\invfuncom\ now shows that $Q\in C(K\times[0,\ep_1])$. An analogous argument applies for
case~1 and $y<0$ using \repqylo. Since $\gamma^+_{\ep}(t)$ is an increasing function of $t$,  Lemmas \repgc\ and \hatq\ imply that $V^t(y,\ep)\in
H(\bbc\setminus(-\infty,\gamma^+(t_{fi},\ep)))$ for $\ep\ge0$ and $tV^t_{\pm}$ have
extensions to $\bar C_{\pm}$ that are in
$C(\bar\bbc_{\pm}\times(0,t_{fi}]\times[0,\ep_0])$ for $\ep_0>0$. From the
definitions of $q_{\ep}(t)$, and $V^t$ and Lemma~\hatq\ we find that for
each compact set $K\in \bbc\setminus(-\infty,\gamma^+(t_{fi},0))$ there is
an $\ep_1$ such that $tV^t\in C^{\infty}(K\times(0,t_{fi}]\times(0,\ep_1])$ and ${\partial^i\over\partial t^i}tV^t\in C(K\times(0,t_{fi}]\times[0,\ep_1])$ for $i>0$. \qed

With the above results we are able consider extensions of the Langer
transformation into the complex plane. We use the principal branch of the $\log$.

\lem anafreudpl. Suppose ic)--iiic) hold and $b\le2a$. For $y\ge r>0$ and
$t\in(0,t_{fi}]$, $t_{fi}<\infty$ set
$$
\zeta_1(t,y,\ep)={1\over(t^+_p(y,\ep)-t)^{3/2}}\int_t^{t^+_p(y,\ep)}
\ln\left(z(u,y,\ep)
+\sqrt{z(u,y,\ep)^2-1}\right)du\eq\eno[frrho]$$
with $z(y,t,\ep)={y\over 2aq_{\ep}(t)}-{b\over2a}$. Then there
exists an $\ep_0$ such that $\zeta_1\in
C((0,t_{fi}]\times[r,\infty)\times[0,\ep_0])$, is positive for
$(t,y,\ep)\in(0,t_{fi}]\times[r,\infty)\times[0,\ep_0]$ and
${\partial^i\over\partial t^i}\zeta_1\in
C((0,t_{fi}]\times[r,\infty)\times[0,\ep_0])$ for all $i>0$. For fixed $(t,\ep)\in(0,t_{fi}]\times[0,\ep_0]$,
$\zeta_1(t,y,\ep)$ can be extended so that $\zeta_1(t,\cdot, \ep)\in
H(S^\ep)$ with $\zeta_1\in C((0,t_{fi}]\times
S^{\ep})$. For fixed $y\in S^{\ep}$, $\Re((t^+_p(y,\ep)-t)^{3/2}\zeta_1(t,y,\ep))$ is a decreasing function of $t\in(0,t_{fi}]$. For $0\le\ep\le\ep_0$ there exists a $\delta_1(\ep)>0$ such that  $\zeta_1$ is nonvanishing for $(t,y)\in [t_{in},t_{fi}]\times L_1(\ep)$ where $L_1(\ep)=\{y: y\in S^{\ep},\ |\Im y|<\delta_1\}$.  If $K\subset S^0$ is compact there exists an $\ep_K$ such that ${\partial^i\over\partial t^i}\zeta_1\in C((0,t_{fi}]\times K\times[0,\ep_K])$, $i\ge0$.
\hfill\break
\indent If $b<2a$ and $-y\le r>0$ set,
$$
\zeta_2(t,y,\ep)={1\over(t^-_p(y,\ep)-t)^{3/2}} \int_t^{t^-_p(y,\ep)}\ln(z(u,y,\ep)
+\sqrt{z(u,y,\ep)^2-1})du.\eq\eno[frmrho]$$
with $z(t,y,\ep)={b\over2a}-{y\over 2aq_{\ep}(t)}$. Then there
exists an $\ep_0$ such that $\zeta_2\in
C((0,t_{fi}]\times(-\infty,-r]\times[0,\ep_0])$, is positive for
$(t,y,\ep)\in(0,t_{fi}]\times(-\infty,-r]\times[0,\ep_0]$ and ${\partial^i\over\partial t^i}\zeta_2\in
C((0,t_{fi}]\times(-\infty,-r]\times[0,\ep_0])$ for all $i>0$. For fixed
$(t,\ep)\in(0,t_{fi}]\times[0,\ep_0],\ 
\zeta_2(t,y,\ep)$ can be extended so that $\zeta_2(t,\cdot, \ep)\in
H(-S^\ep)$ with $\zeta_2\in C((0,t_{fi}]\times -
S^{\ep})$.  For fixed $-y\in S^{\ep}$, $\Re (t^-_p(y,\ep)-t)^{3/2}\zeta_2(t,y,\ep)$ is a decreasing function of $t\in(0,t_{fi}]$.   For $0\le\ep\le\ep_0$ there exists a $\delta_1(\ep)>0$ such that  $\zeta_1$ is nonvanishing for $(t,y)\in [t_{in},t_{fi}]\times L_2(\ep)$ where $L_2(\ep)=\{y: y\in -S^{\ep},\ |\Im y|<\delta_1\}$. If $K\subset -S^0$ is compact there exists an $\ep_K$ such that ${\partial^i\over\partial t^i}\zeta_2\in C((0,t_{fi}]\times K\times[0,\ep_K])$, $i\ge0$.

\pf Set $k^2=\ln^2\left(z(t,y,\ep)+\sqrt{z(t,y,\ep)^2-1}\right)$ with
$z(y,t,\ep)={y\over 2aq(t,\ep)}-{b\over2a}$. For $y\ge r>0$ we see from
Lemma~\invfuncom\ that there is an $\ep_0$ so that $t^+_p(y,\ep)>0$ for all
$(y,\ep)\in[r,\infty)\times[0,\ep_0]$. 
 Lemmas~\logsquare, \hatq, and \invfuncom\ imply that
$p(t,y,\ep)=({k^2(t,y,\ep)\over(t^+_p(y,\ep)-t)})^{1\over2}\in
C((0,t_{fi}]\times[r,\infty)\times[0,\ep_0])$ 
is positive in this region. From Lemmas \hatq\ and \invfuncom\ we see that
$$
z(t,y,\ep)=1+\sum_{j=1}^{\infty}c_j(y,\ep)(t_p^+(y,\ep)-t)^j,
$$
where the sum is uniformly convergent for $t$ in an interval about
$t^+_p(y,\ep)$
with $c_j(y,\ep)\in C([r,\infty)\times[0,\ep_0])\cap C^{\infty}([r,\infty)\times(0,\ep_0])$
for $\ep_0$ sufficiently small. Thus ${\partial^i\over\partial
t^i}p(t,y,\ep)\in
C((0,t_{fi}]\times[r,\infty)\times[0,\ep_0])$. For
$(t,y,\ep)\in(0,t_{fi}]\times[r,\infty)\times[0,\ep_0]$ set
$$
f(t,y,\ep)=\left\{\matrix{{1\over
(t^+_p(y,\ep)-t)^{3/2}}\int_t^{t^+_p(y,\ep)}(t^+_p(y,\ep)-u)^{1/2}p(u,y,\ep)du&
{\rm for}\ t< t^+_p(y,\ep)\cr{2\over3}p(t^+_p(y,\ep),y,\ep)&{\rm for}\
t=t^+_p(y,\ep)\cr{1\over(t-t^+_p(y,\ep))^{3/2}}\int^t_{t^+_p(y,\ep)}(u-t^+_p(y,\ep))^{1/2}p(u,y,\ep)du&{\rm
for}\ t> t^+_p(y,\ep)}\right.,
$$
and for $(t,y,\ep)\in(0,t_{fi}]\times[r,\infty)\times[0,\ep_0]$ set
$$
{\partial^i\over\partial t^i}f(t,y,\ep)=\left\{\matrix{{1\over
(t^+_p(y,\ep)-t)^{3/2+i}}\int_t^{t^+_p(y,\ep)}(t^+_p(y,\ep)-u)^{1/2+i}{\partial^i\over\partial
t^i}p(u,y,\ep)du&
{\rm for}\ t< t^+_p(y,\ep)\cr{2\over3+2i}{\partial^i\over\partial t^i}p(t^+_p(y,\ep),y,\ep)&{\rm for}\
t=t^+_p(y,\ep)\cr{1\over(t-t^+_p(y,\ep))^{3/2+i}}\int^t_{t^+_p(y,\ep)}(u-t^+_p(y,\ep))^{1/2+i}{\partial^i\over\partial
t^i}p(u,y,\ep)du&{\rm
for}\ t> t^+_p(y,\ep)}\right..
$$
Lemma~\logsquare, integration by parts and the mean value theorem for integrals shows that
the definition of $f$ is self consistent, $f\in
C((0,t_{fi}]\times[r,\infty)\times[0,\ep_0])$, is positive in that
region, and ${\partial^i\over\partial
t^i}f\in
C((0,t_{fi}]\times[r,\infty)\times[0,\ep_0])$.
Now fix $(t,\ep)$ such that $\gamma^+_{\ep}(t)>r$ and let
$y>\gamma^+_{\ep}(t)$. We find from Theorem~\thma\ and equation~\greeneq\ that
$$
(t^+_p(y,\ep)-t)^{3/2}\zeta_1(t,y,\ep)=Q(y,\ep)-t(V^t(y,\ep)-l_t). \eq\eno[qgrone]
$$
Since $(b+2a)q_{\ep}(0)\le (b+2a)q_{\ep}(t)=\gamma^+(t,\ep)$ we find
from Theorem~\qgcomone\ that $(t^+_p(y,\ep)-t)^{3/2}\zeta_1(t,y,\ep)$ has an
extension that is analytic in $S^{\ep}\setminus[0,\gamma^+(t,\ep)]$ for
each $(t,\ep)\in(0,t_{fi}]\times[0,\ep_0]$. Lemma~\invfuncom\ and the Schwarz reflection
principal show that $\zeta_1(t,\cdot,\ep)\in H(S^{\ep})$ for
$(t,\ep)\in(0,t_{fi})\times[0,\ep_0]$ and this combined with the above argument and Theorem~\qgcomone\ give the remaining continuity and differentiability  properties of $\zeta_1$. Taking the real part of the above equation then
differentiating with respect to $t$ yields
$-\ln|z(t,y,\ep)+\sqrt{z(t,y,\ep)^2-1}|\le0$ which shows that for fixed
$y\in S^{\ep}$, $\Re((t_p^+(y,\ep)-t)^{3/2}\zeta_1)$ is a
decreasing function of $t$.  The dominated convergence theorem applied to the second integral in \repqygo\ shows that $\lim_{y\to\infty y\in S^{\ep}} |y^{-\alpha}Q(y,\ep)|>0$
for $0\le\ep\le\ep_0$ where equation~\asqinv\ has been used. This coupled with \qgrone\ shows that there is a $\delta_1(\ep)$ such that $\zeta_1$ is nonvanishing on $[t_{in},t_{fi}]\times L_1(\ep)$.

For
$b<2a$ and $y<0$ if we replace $k^2$ above by
$k^2=\ln^2\left(z(t,y,\ep)+\sqrt{z(t,y,\ep)^2-1}\right)$ with
$z(t,y,\ep)={b\over2a}-{y\over2aq_{\ep}(t)}$ arguments analogous to those above can be used to show that $\zeta_2\in
C((0,t_{fi}]\times(-\infty,-r]\times[0,\ep_0])$, is positive for
$(t,y,\ep)\in(0,t_{fi}]\times(-\infty,-r)]\times[0,\ep_0]$ and
${\partial^i\over\partial t^i}\zeta_2\in C((0,t_{fi}]\times(-\infty,
-r]\times[0,\ep_0])$. We now extend $\zeta_2$ to complex $y$. Fix $(t,\ep)$
such that $\gamma^-_{\ep}(t)<-r$ and let $y<\gamma^-_{\ep}(t)$. Set $\hat
G_{\pm}(y,\ep)=G_{\pm}^t(y,\ep)-l_t\mp i\pi$. Then the extension of
Lemma~\repgc\ to the case when $\ep>0$ (or by direct computation) we find that $\hat G_+(y,\ep)=\hat G_-(y,\ep)$ for $y<\gamma^-(t,\ep)$. Consequently there is a function $\hat V^t\in H(\bbc\setminus[\gamma^-(t,\ep),\infty))$ such that $\hat V^t =\hat G_{\pm}$ for $y\in C_{\pm}$ and
$$
(t^-_p(y,\ep)-t)^{3/2}\zeta_2(t,y,\ep)=Q(y,\ep)-t\hat V^t(y,\ep). \eq\eno[qgrtwo]
$$ 
Theorem~\qgcomone\ now shows that $(t^-_p(y,\ep)-t)^{3/2}\zeta_2(t,y,\ep)$ 
has an extension that is analytic in $-S^{\ep}$ 
for each $(t,\ep)\in(0,t_{fi}]\times[0,\ep_0]$. The rest of the proof follows 
as above.  \qed
 
We now consider case~3. Again $\gamma^-_{\ep}$ becomes an obstacle to the smoothness 
of $k^2$.

\lem anafreudpla. Suppose ic)--iiic) hold and $b>2a$. With $y\ge r>0$ and
$t\in(0,t_{fi}]$, let $\zeta_1$ be as in equation~\frrho.
For every $(t_{fi},r),\ 0<t_{fi}<(\gamma_0^-)^{-1}(r)),\ 0<r<\infty$, there exists an $\ep_0$ such that $\zeta_1\in C((0,t_{fi}]\times[r,\infty)\times[0,\ep_0])$ is positive for
$(t,y,\ep)\in(0,t_{fi}]\times[r,\infty)\times[0,\ep_0]$ and
${\partial^i\over\partial t^i}\zeta_1\in
C((0,t_{fi}]\times[r,\infty)\times[0,\ep_0])$ for all $i>0$. 
For fixed $(t,\ep)\in(0,t_{fi}]\times[0,\ep_0]$,
$\zeta_1(t,y,\ep)$ can be extended so that $\zeta_1(t,\cdot, \ep)\in
H(S^\ep\setminus[0,\gamma^-_{\ep}(t_{fi})])$ with $\zeta_1\in C((0,t_{fi}]\times
(S^{\ep}\setminus[0,\gamma^-_{\ep}(t_{fi})]))$. For fixed $y\in S^{\ep}\setminus[0,\gamma^-_{\ep}(t_{fi})]$,
$\Re((t^+_p(y,\ep)-t)^{3/2}\zeta_1(t,y,\ep))$ is a decreasing function of
$t\in(0,t_{fi}]$.  For $0\le\ep\le\ep_0$ there exists a $\delta_1(\ep)>0$
such that  $\zeta_1$ is nonvanishing for $(t,y)\in [t_{in},t_{fi}]\times
L_1(\ep)$ where $L_1(\ep)=\{y: y\in S^{\ep}\setminus[0,\gamma^-_{\ep}(t_{fi})],\ |\Im
y|<\delta_1\}$. If $K\subset S^0\setminus[0,\gamma^-_{0}(t_{fi})]$ is compact there exists an $\ep_K$ such that $\zeta_1\in C((0,t_{fi}]\times K\times[0,\ep_K])$ and ${\partial^i\over\partial t^i}\zeta_1\in C((0,t_{fi}]\times K\times[0,\ep_K])$, $i>0$. Consider 
$$
\zeta_2(t,y,\ep)={1\over(t-t^-_p(y,\ep))^{3/2}} \int^t_{t^-_p(y,\ep)}\ln(z(u,y,\ep)
+\sqrt{z(u,y,\ep)^2-1})du\eq\eno[frmarho]$$
with $z(t,y,\ep)={b\over2a}-{y\over 2aq_{\ep}(u)}$. 
Let $r>0$ then for every
$[t_{in},t_{fi}]\times[y_1,y_2]\ y_1<y_2$ with $[y_1, y_2]\in[r,\infty)$ and $[t_{in},t_{fi}]\in((\gamma_0^+)^{-1}(y_2),
\infty)$
there exists an $\ep_0$ such that $\zeta_2$ is positive for
$(t,y,\ep)\in[t_{in},t_{fi}]\times[y_1,y_2]\times[0,\ep_0]$, and ${\partial^i\over\partial t^i}\zeta_2\in
C([t_{in},t_{fi}]\times[y_1,y_2]\times[0,\ep_0])$ for all $i\ge0$. For fixed
$(t,\ep),\ \zeta_2$ can be extended so that $\zeta_2(t,\cdot,\ep)\in H(
(S^{\ep}\setminus[\gamma^+_{\ep}(t),\infty)))$ with
$\zeta_2\in
C([t_{in},t_{fi}]\times
(S^{\ep}\setminus[\gamma^+_{\ep}(t_{in}),\infty)))$. For each fixed
$y\in S^{\ep}\setminus[\gamma^+_{\ep}(t_{in}),\infty)$, $\Re(t-t^-_p(y,\ep))^{3/2}\zeta_2(t,y,\ep)$
is an increasing function of $t\in[t_{in},t_{fi}]$ . For $0\le\ep\le\ep_0$ there exists a $\delta_2(\ep)>0$ such that $\zeta_2$ is nonvanishing on $[t_{in},t_{fi}]\times L_2(\ep)$
where $L_2(\ep)=\{y: y\in S^{\ep},\ 0<\Re(y)<\gamma^+_{\ep}(t_{in}), |\Im(y)|<\delta_2\}$.  If $K\subset S^0\setminus[\gamma^+_0(t_{in}),\infty)$ is compact there exists an $\ep_K$ such that  ${\partial^i\over\partial t^i}\zeta_2\in C([t_{in},t_{fi}]\times K\times[0,\ep_K])$, $i\ge0$.

\pf The first part of the above Lemma follows as the first part 
of Lemma~\anafreudpl\ noting that the restriction on $t_{fi}$ and
$S^{\ep}\setminus[0,\gamma^-_{\ep}(t_{fi})]$ is necessary
to avoid the singularity in $k^2$ due to $\gamma^-_{\ep}$.
To show the second part of the Lemma fix $r>0$ then from Lemma~\invfuncom\ 
there exists an $\ep_0$ such that $t^-_p(y,\ep)$ exists and is greater than
zero for all $(y,\ep)\in[r,\infty)\times[0,\ep_0]$. Set $k^2= \ln^2(z(t,y,\ep)
+\sqrt{z(t,y,\ep)^2-1})$  with $z(t,y,\ep)={b\over2a}-{y\over2aq_{\ep}(t)}$.
 It may be that $t> t_p^-(y,\ep)$ for all
$(t,y)\in[t_{in},t_{fi}]\times[y_1,y_2]$ in this case we increase $y_2$ as
large as possible consistent with the restriction 
on $t_{in}$. Lemmas~\logsquare , \hatq , and \invfuncom\ imply that
$p(t,y,\ep)=({k^2(t,y,\ep)\over t-t^-_p(y,\ep)})^{1\over2}\in
C([t_{in},t_{fi}]\times[y_1,y_2]\times[0,\ep_0])$ 
is positive in this region.  From Lemmas \hatq\ and \invfuncom, $z$ has
the expansion
$$
z(t,y,\ep)=1+\sum_{j=1}^{\infty}c_j(y,\ep)(t-t_p^-(y,\ep))^j,
$$
where the sum is uniformly convergent for $t$ in an interval about
$t^-_p(y,\ep)$
with $c_j(y,\ep)\in C([r,\infty)\times[0,\ep_0])\cap
C^{\infty}([r,\infty)\times(0,\ep_0])$. This implies that
${\partial^i\over\partial t^i}p(t,y,\ep)\in
C([t_{in},t_{fi}]\times[y_1,y_2]\times[0,\ep_0])$ for all $i\ge0$.  For
$(t,y,\ep)\in[t_{in},t_{fi}]\times[y_1,y_2]\times[0,\ep_0]$ and all $i\ge0$
set
$$
{\partial^i\over\partial t^i}f(t,y,\ep)=\left\{\matrix{{1\over
(t-t^-_p(y,\ep))^{3/2+i}}\int^t_{t^-_p(y,\ep)}(u-t^-_p(y,\ep))^{1/2+i}{\partial^i\over\partial
u^i}p(u,y,\ep)du&
{\rm for}\ t> t^-_p(y,\ep)\cr{2\over3+2i}{\partial^i\over\partial
t^i}p(t^-_p(y,\ep),y,\ep)&{\rm for}\
t=t^-_p(y,\ep)\cr{1\over(t^-_p(y,\ep)-t)^{3/2+i}}\int_t^{t^-_p(y,\ep)}(t^-_p(y,\ep)-u)^{1/2+i}{\partial^i\over\partial
u^i}p(u,y,\ep)du&{\rm
for}\ t< t^-_p(y,\ep)}\right..
$$
 Lemma~\logsquare, integration by parts and the mean value theorem for integrals show that
the definition of $f$ is self consistent, $f$ is positive for
$(t,y,\ep)\in[t_{in},t_{fi}]\times[y_1,y_2]\times[0,\ep_0]$ 
and ${\partial^i\over\partial
t^i}f\in
C([t_{in},t_{fi}]\times[y_1,y_2]\times[0,\ep_0])$ for all $i\ge0$. To
 extend 
$\zeta_2$
to the complex plane fix $(t,\ep)$. For $r\le y<\gamma^-_{\ep}(t)$ set 
$$\hat G_{\pm}(y,\ep)=V^t_{\pm}(y,\ep)-l_t\mp i\pi. \eq\eno[linka] $$
Then from Lemma~\repgc\ we find that $\hat G_+(y,\ep)=\hat G_-(y,\ep)$ 
for $y<\gamma^-_{\ep}(t)$. Consequently there is a function 
$\hat V^t\in H(\bbc\setminus[\gamma^-(t,\ep),\infty))$ such 
that $\hat V^t =\hat G_{\pm}$ for $y\in C_{\pm}$ and
$$
(t-t^-_p(y,\ep))^{3/2}\zeta_2(t,y,\ep)=t\hat V^t(y,\ep)-Q(y,\ep). \eq\eno[qgrthree]$$ 
Thus the extension of  Lemma~\repgc\ to $\ep>0$ or direct computation shows that $(t^-_p(y,\ep)-t)^{3/2}\zeta_2(t,y,\ep)$ has
an extension to
$S^{\ep}\setminus[\gamma^{-}_{\ep}(t),\infty)$. Lemma~\invfuncom\ and the
Schwarz reflection principle shows that $\zeta_2\in
H(S^{\ep}\setminus[\gamma^+_{\ep}(t),\infty))$ for each
$(t,\ep)\in[t_{in},t_{fi}]\times[0,\ep_0]$. 
The arguments above and Theorem~\qgcomone\ show that 
${\partial^i\over\partial t^i}\zeta_2\in C([t_{in},t_{fi}]\times
(S^{\ep}\setminus[\gamma^+_{\ep}(t_{in}),\infty))\times[0,\ep_0])$ for all
$i\ge0$. 
Taking the real part of equation~\qgrthree\ then differentiating with 
respect to $t$ yields $\ln|z(t,y,\ep)+\sqrt{z(t,y,\ep)^2-1}|\ge0$ which 
shows that for fixed $(y,\ep)\in
S^{\ep}\setminus[\gamma^+_{\ep}(t_{in},\infty)$, $\Re((t_p^--t)^{3/2}\zeta_2)$
is an increasing
function of $t$. The remaining part of the Lemma follows from the continuity and positivity of $\zeta_2$.\qed

\remk We note that away from the turning point the above $\zeta$ function are in fact $C^{\infty}([t_{in},t_{fi}]\times K\times(0,\ep_1])$

\bigskip

A consequence of the above lemmas is,
\thm freudtur.  Suppose ii)--iiic) hold and $b-2a\le0$. For $0<t_{in}\le t\le t_{fi}$ set
$$
{\rho_1(t,y,\ep)\over t_p^+(y,\ep)-t}=({3\over2}\zeta^1(t,y,\ep))^{2/3}.\eq\eno[comdeqone]
$$
Let $L^1(\ep)$ be as in the Lemmas~\anafreudpl\ or \anafreudpla. Then $\rho_1\in H(L_1(\ep))$ for fixed $(t,\ep)\in[t_{in},t_{fi}]\times [0,\ep_1]$ with $\rho_1\in C((0, t_{fi}]\times L_1(\ep))$ . If $K$ is a compact set in $L_1(0)$ then there is an $\ep_1$ such that ${\partial^i\over\partial t^i}{\rho_1\over t_p^+-t}\in C([t_{in},t_{fi}]\times K\times [0,\ep_1])$ for all $i\ge0$. Also for $(t,y,\ep)\in[t_{in},t_{fi}]\times K\times [0,\ep_1]$, ${d\hat\nu_t^{\ep}\over dx}$ has an analytic extension so that,
$$
\rho_1(t,y,\ep)=({3\over2}\pi t\int^y_{\gamma^+_{\ep}(t)}d\hat\nu_t^{\ep})^{2\over3}.\eq\eno[compeqone]
$$
Likewise for $b-2a<0$ and $L^2(\ep)$  as in the Lemma~\anafreudpl\  set
$$
{\rho_2(t,y,\ep)\over t_p^-(y,\ep)-t}=({3\over2}\zeta^2(t,y,\ep))^{2/3}.\eq\eno[comdeqtwo]
$$
Then $\rho_2\in H(L_2(\ep))$ for fixed $(t,\ep)\in[t_{in},t_{fi}]\times [0,\ep_1]$ with $\rho_2\in C((0, t_{fi}]\times L_2(\ep))$ . If $K$ is a compact set in $L_2(0)$ then there is an $\ep_1$ such that ${\partial^i\over\partial t^i}{\rho_2\over t_p^+-t}\in C([t_{in},t_{fi}]\times K\times [0,\ep_1])$ for all $i\ge0$. For $(t,y,\ep)\in[t_{in},t_{fi}]\times K\times [0,\ep_1]$, ${d\hat\nu_t^{\ep}\over dx}$ has an analytic extension so that,
$$
\rho_2(t,y,\ep)=({3\over2}\pi t\int_y^{\gamma^-_{\ep}(t)}d\hat\nu_t^{\ep})^{2\over3}.\eq\eno[compeqtwo]
$$
If $b-2a>0$ then for $0<\hat t_{in}\le t\le \hat t_{fi}$ let $L^2(\ep)$ be as in Lemma~\anafreudpla\ and set
$$
{\rho_2(t,y,\ep)\over t-t_p^-(y,\ep)}=({3\over2}\zeta^2(t,y,\ep))^{2/3}.\eq\eno[comdeqtwoa]
$$
Then $\rho_2\in H(L_2(\ep))$ for fixed $(t,\ep)\in[\hat t_{in}, \hat t_{fi}]\times[0,\ep_1]$
with $\rho_2\in C([\hat t_{in},\hat t_{fi}]\times L_2(\ep)$. If $K$ is a compact set in $L_2(0)$ then there is an $\ep_1$ such that ${\partial^i\over\partial t^i}{\rho_2\over t-t_p^+}\in C([\hat t_{in}, \hat t_{fi}]\times K\times [0,\ep_1])$ for all $i\ge0$. For
$(t,y,\ep)\in [t_{in},t_{fi}]\times
K\times[0,\ep_1]$, ${d\tilde\nu_t^{\ep}\over dx}$ has an analytic extension so that,
$$
\rho_2(t,y,\ep) =({3\over2}i\pi\int_y^{\gamma^-_{\ep}(t)}d\tilde\nu_t^{\ep})^{2\over3}.\eq\eno[compeqtwoa]
$$
In the above equations $d\hat\nu_t^{\ep}$ and $d\tilde\nu_t^{\ep}$ are
given by Lemma~\externalg.   

\pf The continuity analyticity, and smoothness properties of $\rho_i$, $i=1,2$ follow
from the continuity, analyticity, smoothness, and nonvanishing properties of $\zeta_{i}$,
$i=1,2$ discussed in Lemmas~\anafreudpl\ and \anafreudpla\ and the fact that $L_i(0)\ i=1,2$ are simply connected regions in the plane. In order to
show equation~\compeqone\ note that for fixed $t$, and $x>{\rm
max}\{0,\gamma^-_{\ep}(t)\}$ equation~\eqqmeasure\ shows that
${d\hat\nu_t\over dx}$ has an analytic extension to
$S^{\ep}\setminus[0,\gamma^+_{\ep}(t)]$. Equation~\compeqone\ now follows
from \qgrone\ and the extension of Lemma~\externalg\ to $\ep>0$ . A similar argument for $x<0$ follows using
equation~\eqqmeatwo. Case 3 follows analogously from
\qgrthree\ and the extension of Lemma~\externalg\ to $\ep>0$ .
\qed

For the extension of  Theorem~\thmnext\ to the complex plane we will use  the solutions to the Airy differential equation given by,
${\rm Ai}_{0}= {\rm Ai}$, and
$$
{\rm Ai}_{\pm1}(z)= {\rm Ai}(ze^{\mp2i\pi/3}),\eq\eno[otherairy]
$$
 and the regions $S_0=\{z: |\arg z|\le{\pi\over3}\}$ and,
$$
S_{\pm 1}=e^{\pm2\pi/3}S_0.
$$
Since the above functions and Bi all satisfy the same differential
equation we find, [\cite{O}, p.~414],
$$
{\rm Ai}(ze^{\pm2\pi/3})={1\over2}e^{\pm i\pi/3}({\rm Ai}(z)\pm i {\rm
Bi}(z)).\eq\eno[compairy]
$$
Then it follows from the asymptotic expansions of  Ai
[\cite{O}, p.~413] that ${\rm Ai}_{j}$ is recessive in $S_j$ and
dominant in $S_{j+1}$, and $S_{j-1}$ where the suffix $j$ is
enumerated mod~3. Furthermore since the zeros of Ai are all real and negative 
[\cite{O}, p.~418],\ ${\rm Ai}_1$ is non zero in $S_0\cup S_1$. 
Two other solutions of the Airy equation that are non zero in the region of interest
are $\tilde w^{(i)}$ which for complex values of
$z$ are defined by
$$
\tilde w^{(j)}(y)= {1\over2}
\left({y\over3}\right)^{{1\over2}}e^{(-1)^{j+1}i{\pi\over6}}
H_{1/3}^{(j)}
\left({2\over3}y^{{2\over3}}e^{{i\pi\over2}}\right), \qquad j=1,2 \eq\eno[ucomplex]
$$
where we take the branch of the square root so that
Re$(z^{3/2})\ge0$ for $z\in S_0$ and  Re$(z^{3/2})\le0$
for $z\in S_1$ which is the principal branch of $z^{3/2}$. It follows from the properties of Hankel functions ([\cite{O}, p.~238], 
[\cite{GBVA}]) that $\tilde w^{(1)}$ is recessive in $S_0$ and dominant in $S_1$ while $\tilde w^{(2)}$ is dominant in $S_0$
and $S_1$, furthermore neither vanish in $S_0\cup S_1$ ([\cite{O}], [\cite{GBVA}])

We rewrite $g$ as
$$g(t,y,\ep)={\rho (t,y,\epsilon)\overwithdelims ()
a^2\left(t+{\epsilon\over 2},\epsilon\right)\sinh^2
(\rho^{1\over2}\rho')(t,y,\ep)}^{1/4},$$ and set,
$$
\psi_1(t,y,\ep)=g{\rm Ai}_0(\ep^{-{2/3}}\rho(t,y,\ep)), \eq\eno[cpsione]
$$
$$
\psi_2(t,y,\ep)=g{\rm Ai}_1(\ep^{-{2/3}}\rho(t,y,\ep)), \eq\eno[cpsitwo]
$$
$$
u^{(1)}(t,y,\ep)=g\tilde w^{(1)}(\ep^{-{2/3}}\rho(t,y,\ep)), \quad
u^{(2)}(t,y,\ep)=g{\rm Ai}_1(\ep^{-{2/3}}\rho(t,y,\ep)), \eq\eno[cpsithree]
$$
$$\hat u^{(1)}(t,y,\ep)= e^{{2\over3}\rho(t,y,\ep)^{3\over2}}u^{(1)}(t,y,\ep), 
$$ and
$$
\hat u^{(2)}(t,y,\ep)=e^{-{2\over3}\rho(t,y,\ep)^{3\over2}}u^{(2)}(t,y,\ep).
$$
Set $S=S_0\cup S_1$, and let $\Omega$ a region in the complex $y$ plane. 
It follows from the asymptotic expansions of $u^i\  i=1,2$ [\cite{O}] that
$\hat u^i$ are bounded functions. Furthermore
both are non zero in $S_0\cup S_1$ [\cite{GBVA}, \cite{O}, p.~254]. With the above notation we now have,

\thm cthmnextone. Suppose \ab\ and ic)--iiic) hold. Let $a_1(t,\ep), b_1(t,\ep)\in C([0,\infty)\times[0,\ep_0])$
satisfy \abep\ 
with $b_1(t,\ep)$ real and $a_1(t,\ep)$ strictly positive on every
compact subset of $(0,\infty]\times[0,\ep_0]$. In equations 
\cpsione--\cpsithree\ let $\rho=\rho_1=(t_p^+-t)(\zeta_1)^{3/2}$ and let $L_1(\ep)$
be given as in Lemmas~\anafreudpl\ or \anafreudpla\ and
$L_1^+(\ep)=L_1(\ep)\cap \bar C_+$. Suppose $K\subset L^+_1(0),\ K$ compact, then there is an $\ep_K$ such that for each $(y,\ep)\in K\times(0,\ep_K]$ and all $n: n\ep\in[t_{in},t_{fi}]$ there exists solutions $f_{i}\ i=1,2$ of
equation~\fourteen\ such that 
$$
f_i(n)=\psi_i(n)+r_i(n), \eq\eno[aieqone]
$$
where
$$
\Bigl|{r_i(n)\over u^{(i)}(n)}\Bigr|=
\Bigl|{f_i(n)-\psi_i(n)\over u^{(i)}(n)}\Bigr|\le d(K)\ep
,\ i=1,2. \quad\eq\eno[cineqone]
$$ 
Furthermore for fixed $\ep,\ {r_i(n)\over u^{(i)}(n)}\in H(K\cap C_+)$ and ${r_i(n)\over u^{(i)}(n)}\in C(K\times[0,\ep_K])$. If $b-2a<0$ let $\rho=\rho_2=(t_p^--t)(\zeta_2)^{3/2}$ equations \cpsione--\cpsithree. If $b-2a>0$ let $\rho=\rho_2=(t-t_p^-)(\zeta_2)^{3/2}$. For both cases let  $L_2(\ep)$ be given as in Lemmas~\anafreudpl\ or \anafreudpla\ and $L_2^+(\ep)=L_2(\ep)\cap\bar C_+$ .  Suppose $K\subset L^+_2(0),\ K$ compact, then there is an $\ep_K$ such that for each $(y,\ep)\in K\times(0,\ep_K]$   and all $n: n\ep\in[t_{in},t_{fi}]$ there exists solutions $f_{i}\ i=1,2$ of
equation~\fourteen\ such that 
$$
(-)^nf_i(n)=\psi_i(n)+r_i(n), \eq\eno[aieqtwo]
$$
where
$$
|{r_i(n)\over u^{(i)}(n)}|=|{(-1)^nf_i(n)-\psi_i(n)\over u^{(i)}(n)}|\le d(K)\ep,
\ i=1,2. \quad\eq\eno[cineqtwo]
$$
Furthermore for fixed $\ep,\ {r_i(n)\over u^{(i)}(n)}\in H(K\cap C_+)$ and ${r_i(n)\over u^{(i)}(n)}\in C(K\times[0,\ep_K])$.

The proofs of the above Theorem closely follow the proof of
Theorem~4.4 in [\cite{GBVA}]. However since the hypotheses are different a proof is sketched in Appendix A.

We now have the important

\lem wronskian. With $f_1$ and $f_2$ above,
$$
a_1((n+1)\ep,\ep)\left[f_1(t_{n+1})f_2(t_n)-f_1(t_n)f_2(t_{n+1})\right]=-{i\ep^{1\over3}
e^{-i{\pi\over3}}\over2\pi}(1+O(\ep)).
\eq\eno[eqwr]
$$

\pf The Wronskian of Ai and Bi [\cite{O}] is
$$
W[{\rm Ai}, {\rm Bi}] = {1\over\pi}.
$$
This coupled with the definition of $g$, Theorem~\cthmnextone,\ equation~\compairy,\ and Lemma~3.5 in
[\cite{GBVA}] (with
$u^{(1)}$ and $u^{(2)}$ replaced by $f_1$ and $f_2$ respectively) give the result. \qed

\advance\subsecno by 1\no=0\thmno=0
\beginsection IV. Singular Initial Value problem

The previous results allow us to find solutions to the difference equation
of a prescribed form that are valid uniformly in the neighborhood of a turning 
point. In order to obtain asymptotics for special functions we will need to 
find approximate solutions that satisfy the initial value problem. 
Unfortunately in the case
when $a(n)$ and $|b(n)|$ tend to infinity, in general the partial derivatives 
of $a(t,\ep)$ with respect to $t$ will not be bounded in a neighborhood 
of $t=0$. This is because if $a(t,\ep)$ and $b(t,\ep)$ are to be bounded 
then $\lambda_{\ep}$ must increase to infinity. (See section 6 for an example.) 
A similar problem arises in the case 
of varying recurrence coefficients [\cite{DM}], [\cite{KVA}] which are 
connected to the continuum limit of the Toda lattice.
While the 
main results, Theorem (4.4) below can be obtained from Theorem~(5.5) 
of [\cite{GBVA}] the conditions on $q$ allow simpler proofs which will be 
presented.  We begin by switching to the 
equation satisfied by the
polynomials $\hat p_n =2^np_n/k_n$, where $k(n)$ the leading coefficient 
of $p_n$. Thus,
$$
\hat p_{n+1}(x)+2(b_n-x)\hat p_n(x) + 4a^2_n \hat p_{n-1}(x)=0. \eq\eno[posdif]
$$
If the scaling indicated in the introduction is performed then writing $\tilde p_n=\hat p_n/\lambda_{\ep}^n$ we arrive at the difference equation,
$$
\tilde p_{n+1}(y) +2(b(n\ep,\ep)-y)\tilde p_n(y)
+4a^2(n\ep,\ep)\tilde p_{n-1}(y)=0.\eq\eno[initpos]
$$
The associated $\ep$-difference equation is,
$$
\psi(t+\ep, y,\ep) +2(b(t,\ep)-y)\psi(t,y,\ep)
+4a^2(t,\ep)\psi(t-\ep,y,\ep)=0.\eq\eno[epdifft]
$$
Following Deift and McLaughlin [\cite{DM}], 
Costin and Costin [\cite{CC}], and [\cite{GBVA}] we look for approximate solutions of \epdiff\ of the form
$$
\psi(t)=e^{{1\over\ep}s_0(t)+s_1(t)}.
$$
If we expand  $\psi(t\pm\ep)$ in powers of $\ep$ then equate like powers the eikonal ($\ep^0$) equation gives (see section 6),
$$
e^{s'_0(t)}+2(b(t,\ep)-y)+4a(t,\ep)^2 e^{-s'_0(t)}=0, \eq\eno[eikonalone]
$$
while the geometrical optics equation ($\ep^1$) gives
$$
s_1(t,\ep)'= -{(e^{s_0(t,\ep)'}+4a(t,\ep)^2e^{-s_0(t,\ep)'})s_0''(t,\ep)\over2(e^{s_0(t,\ep)'}-a(t,\ep)^2e^{-s_0(t,\ep)'})}. \eq\eno[geomone]
$$
Combining
$$
(e^{s_0(t))'}-4a(t)^2e^{-s_0(t)'})'=(e^{s_0(t)'}+4a(t)^2e^{-s_0(t)'})s_0''(t)-8a(t)a'(t)e^{-s_0(t)},
$$
with the derivative of \eikonalone\ yields, 
$$
s_1(t,\ep)'= -{(e^{s_0(t,\ep)'}-4a(t,\ep)^2e^{-s_0(t,\ep)'})'\over2(e^{s_0(t,\ep)'}-a(t,\ep)^2e^{-s_0(t,\ep)'})}
+ {b(t,\ep)'\over e^{s_0(t,\ep)')}-a(t,\ep)^2e^{-s_0(t,\ep)'}}+{s_0(t,\ep)''\over2}.\eq\eno[geomone]
$$

Solutions to the above equations are given by,
$$
s^{\pm}_0(t,\ep)=\int^t
\ln\left(y-b(u,\ep)\pm\sqrt{(y-b(u,\ep))^2-4a(u,\ep)^2}\right)du, \eq\eno[szeropm]
$$
and
$$
\eqalign{
s^{\pm}_1(t,\ep)=&-{1\over4}\ln ((y-b(t,\ep))^2-4a(t,\ep)^2)\cr
&+{1\over2}\ln(y-b(t,\ep)\pm\sqrt{(y-b(t,\ep))^2-4a(t,\ep)^2})\cr
&\pm{1\over 2}\int^t
{b'(u,\ep)du\over\sqrt{(y-b(u,\ep))^2-4a(u,\ep)^2}}.} \eq\eno[sonepm]
$$
A formula which we will have use of later is,
$$
s^{\pm}_0(t)''=-{b'(t)\over h_1}{\mp}{2a^2(t)'\over h_2^{\pm}(t)h_1},\eq\eno[sodir]
$$
where
$$
h_1(t, y, \ep)=\sqrt{(y-b(t,\ep))^2-4a(t,\ep)^2},\eq\eno[hatfone]
$$
and
$$
h_2^{\pm}(t,y,\ep)=y-b(t,\ep)\pm\sqrt{(y-b(t,\ep))^2-4a(t,\ep)^2}\eq\eno[hatftwo]
$$

We will take as our approximate solutions
$$
\eqalign{
\psi_1^+(t,y,\ep)&={h^+_2(t,y,\ep)^{1/2}\over
h_1(t,y,\ep)^{1/2}}
\times\exp\left(\int^{t}_0
{b'(u,\ep)du\over2 h_1(u,y,\ep)}\right)\cr&\times
\exp\left({1/\ep}\int^{t}_0\ln h^+_2(u,y,\ep)du\right),}\eq\eno[fplusone]
$$
and,

$$
\eqalign{
\psi_1^-(t,y,\ep)&=
{h_2^-(t,y,\ep)^{1/2}\over
h_1(t,y,\ep)^{1/2}}\times\exp\left(-\int^t_{t_{in}}
{b'(u,\ep)du\over2h_1(u,y,\ep)}\right)\cr
&\quad \times\exp\left({1/\ep}\int^{t}_{t_{in}}\ln h_2^-(u,y,\ep)du\right).}
\eq\eno[fminusone]
$$

Let,
$$
\tilde\gamma^{\pm}(t,\ep)=b(t,\ep)\pm2a(t,\ep),
$$
$$
A(\tilde t,\ep)=\inf_{[0,\tilde t]}\tilde\gamma^-(t,\ep),\ B(\tilde t,\ep)=
\sup_{[0,\tilde t]}\tilde\gamma^+(t,\ep),
$$
$$
I_B(\tilde t,\ep)=(-\infty, B(\tilde t,\ep)], \ {\rm and}\ I(\tilde t,\ep)=[A(\tilde t,\ep),B(\tilde t,\ep)].
$$

We now examine the analytic properties of the above functions. 

\lem wkmone. Suppose $a(t,\ep)$ and $b(t,\ep)$ are given by equation~\ab\
 and conditions ic) and iic) hold. Then for each $(t,\ep)\in[0,t_{fi}]\times[0,\ep_0],\ \ln h^+_2$ is 
nonvanishing and analytic on $\bbc\setminus(-\infty,\tilde\gamma^+(t,\ep)]$ while  $h^+_2$ and $h_1$ are nonvanishing and analytic on $\bbc\setminus[\tilde\gamma^-(t,\ep),\tilde\gamma^+(t,\ep)]$. For each $(t,\ep)\in[t_{in},t_{fi}]\times[0,\ep_0],\ \ln h^-_2$ is analytic and nonvanishing on $\bbc\setminus(-\infty,\tilde\gamma^+(t,\ep)]$ and $h^-_2$ is analytic and nonvanishing on $\bbc\setminus[\tilde\gamma^-(t,\ep),\tilde\gamma^+(t,\ep)]$.
For any compact set $K\subset \bbc\setminus
I_B(t_{fi},0)$   there is an $\ep_K$ such that 
$\ln h^+_2\in C([0,t_{fi}]\times K\times[0,\ep_K])$ and $\ln h^-_2\in C([t_{in},t_{fi}]\times K\times[0,\ep_K])$. For any compact set $\tilde K\subset \bbc\setminus I(t_{fi},0)$ there is an $\ep_{\tilde K}$ such that $h^+_2,\ h_1\in C([0,t_{fi}]\times\tilde K\times[0,\ep_{\tilde K}])$
and ${\partial^i\over\partial t^i}h_1$, ${\partial^i\over\partial t^i}h^+_2$, ${\partial^i\over\partial t^i}\ln h^+_2\in C((0,t_{fi}]\times\tilde K\times[0,\ep_{\tilde K}])$, $i>0$. 
Likewise there 
is an $\ep_{\tilde K}$ such that ${\partial^i\over\partial t^i}h^-_2$, 
and ${\partial^{i+1}\over\partial t^{i+1}}\ln h^-_2\in C([t_{in},t_{fi}]\times\tilde K\times[0,\ep_{\tilde K}])$, $i\ge0$.

\pf The continuity, differentiability and analyticity properties 
of $\ln h^+_2$  follow from equation~\logg,  the fact that $\ln(z-x), x\in
[\tilde\gamma^-(t,\ep), \tilde\gamma^+(t,\ep)]$ is 
analytic for $z\in \bbc\setminus(-\infty, \tilde\gamma^+(t,\ep))$ and the differentiability part of 
Lemma~\hatq. The continuity, 
differentiability, analyticity and nonvanishing properties 
of $h^+_2$ and $h_1$ follow from their definitions (see equations~\map\ and \mag). 
The fact that $a(t,\ep)>0$ for $t>0$ and the formula $h^-_2={4a^2\over
h^+_2}$, imply the results for this function and its log. \qed 

In order to simplify the analysis below we make the  assumptions
that there exist constants $C_1>0$ and $C_2\ge0$ such that, 
\item{ivc)} $|{d^j\over dt^j} \hat q(t)^i|\le C_1(t+C_2)^{p(i,j)}$, 
$j=0,\ldots 2$, $i=1,2$, $t>0$ where
$$
p(i,j)=\left\{\matrix{{i\over\alpha}-j& {\rm if}\ {i\over\alpha}-j\ne-1\cr 0 &{\rm if}\ {i\over\alpha}-j=-1}\right.
$$

Set
$$
J^{\pm}_{s^+_0,s^+_1}(t)={1\over2\ep}\int_t^{t\pm\ep}
 s_0^+(u)'''(t\pm\ep-u)^2 du+\int_t^{t\pm\ep} s^+_1(u)''(t\pm\ep-u) du,
$$
and
$$
L^{\pm}_{s^+_0,s^+_1}(t)={1\over\ep}\int_t^{t\pm\ep} s_0^+(u)''(t\pm\ep-u) du+\int_t^{t\pm\ep} s^+_1(u)' du.
$$

\lem ainitvalue. {\rm [Lemma~5.2, \cite{GBVA} errata]}
Suppose $a(t,\ep)$ and $b(t,\ep)$ are given by equation~\ab,  
conditions ic), iic) and ivc) hold, and $0<\ep_0\ll t_{fi}$. Then for  $0<\ep\le\ep_0$ and  $(t,y)\in[0,t_{fi}]\times \bbc\setminus I_B(t_{fi},\ep)$, $\psi^+_1$ is nonzero  and for fixed $(y,\ep)$ satisfies the difference equation,
$$
\eqalign{
\psi^+_1(t+\ep) &+2(b(t,\ep)-y)\psi^+_1(t)\cr
& + 4a^2(t,\ep)\psi^+_1(t-\ep)=\eta^+_1(t)\psi^+_1(t).
\quad\ep\le t\le t_{fi}-\ep}\eq\eno[eqfplus]
$$
For $\ep_0$ sufficiently small, $\psi^+_1\in C([0,t_{fi}]\times(0,\ep_0])\times H(\bbc\setminus I_B(t_{fi},\ep))$ 
and $\eta^+_1\in C(W^+_{\ep_0})\times H(\bbc\setminus I_B(t_{fi},\ep)),\ W^+_{\ep_0}=\{(t,\ep):\ep\in[0,\ep_0],\ \ep\le t\le t_{fi}-\ep\}$. For each compact set $K\subset \bbc\setminus I_B( t_{fi},0)$ there exists 
an $\ep_K$ such that $\psi^+_1\in C([0,t_{fi}]\times K\times(0,\ep_K])$, $\eta^+_1\in C(W^+_{\ep_K}\times K)$ and 
$$
\sum_{i:i\ep\in[\ep,t_{fi}]}|\eta^+_1(i\ep,y,\ep)|<c\max(b\ep^{1/\alpha},\ep^{2/\alpha},\ep) \eq\eno[zetaopbound].
$$
Here $c$ depends on $K$ and $t_{fi}$. If $t=t_n=n\ep$ then the sets $I_B(t_{fi},\ep)$ and $I_B(t_{fi},0)$ in the above statements maybe replaced by the sets $I(t_{fi},\ep)$ and $I(t_{fi},0)$ respectively.
Likewise for each $0\le\ep\le\ep_0$ and  $(t,y)\in[t_{in},t_{fi}]\times \bbc\setminus I_B(t_{fi},\ep)$, $\psi^-_1$ is nonzero  and for fixed $(y,\ep)$ satisfies the  above difference equation for $t_{in}+\ep\le t\le t_{fi}-\ep$ with $\eta^+_1$ replaced by, $\eta^-_1$. $\psi^-_1\in C([t_{in},t_{fi}]\times(0,\ep_0])\times H(C\backslash I_B(t_{fi},\ep))$
and $\eta^+_1\in C(W^-_{\ep_0})\times H(\bbc\setminus I_B(t_{fi},\ep)),\ W^-_{\ep_0}=\{(t,\ep):\ep\in[0,\ep_0],\ t_{in}+\ep\le t\le t_{fi}-\ep\}$. For each compact 
set $K\subset \bbc\setminus I_B(t_{fi},0)$ there exists 
an $\ep_K$ such that $\psi^-_1\in C([t_{in},t_{fi}]\times K\times(0,\ep_K])$, $\eta^-_1\in C(W^-_{\ep_K}\times K)$ and 
$$
\sum_{i: i\ep\in[t_{in}+\ep,t_{fi}-\ep]}|\eta^-_1(i\ep,y,\ep)|<c\ep, \eq\eno[zetaombound]  
$$
with $c$ depending on $K$ and the interval $[t_{in},t_{fi}]$.

\pf The conditions on $\hat q$ imply $\sup_{\ep\in[0,\ep_0]}\int_0^{t_{fi}}|{d\over d t}\hat q_{\ep}(t)|dt<\infty$. 
Thus Morera's theorem, equations~\hatfone, \hatftwo\ and Lemma~\wkmone\ show that
$\psi^{\pm}_1$ exist and have the claimed continuity and analyticity properties.  Using Taylor's theorem with remainder, equations~\eikonalone\ and \geomone\ we find
(suppressing all variables but $t$),
$$
\eqalign{
\eta^+_1(t)&=e^{s_0^+(t)'}(J^+_{s^+_0,s^+_1}(t)
+R^+_{s^+_0,s^+_1}(t))\cr
&\qquad +4a(t,\ep)^2 e^{-s_0^+(t)'}(J^-_{s^+_0,s^+_1}(t)
+R^-_{s^+_0,s^+_1}(t)).}\eq\eno[etaplus]
$$
where
$$
R^\pm_{s^+_0,s^+_1}(t))=L^{\pm}_{s^+_0,s^+_1}(t)^2\int_0^1 \exp(L^{\pm}_{s^+_0,s^+_1}(t)u)(1-u)du,
$$
The properties of $\eta^+_1$ except equation~\zetaopbound\ follow from Lemmas~\wkmone \ and condition ivc). If $t=t_n=n\ep$ then for $y< A(t_{fi},\ep)$ we find 
$${1\over\ep}\int_0^{t_n}\ln h^+_2(t,y\pm,\ep) dt ={1\over\ep}\int_0^{t_n}\ln |h^+_2(t,y,\ep)|dt\pm in\pi. \eq\eno[htwo]$$ 
Thus $\psi^+_1(t_n,y+,\ep)=\psi^+_1(t_n,y-,\ep)$ which along with Lemma~\wkmone\ gives the stated properties with  $I_B(t_{fi},\ep)$ and $I_B(t_{fi},0)$ above replaced by $I(t_{fi},\ep)$ and $I(t_{fi},0)$ respectively. To obtain \zetaopbound\ note that 
$$\sum_{i:i\ep\in[\ep,t_{fi}-\ep]}|\int_{i\ep}^{(i+1)\ep}{d^l\over d u^l} \hat q_{\ep}(u)^i (i+\ep-u)^jdu|<k{\ep^{-p(i,l)+j-l}\over\hat q({1\over\ep}+{1\over2})^i}\int_{\ep}^{t_{fi}}(u+C_2\ep)^{p(i,l)}du,
$$
which when combined with the definition of $J^{\pm}_{s^+_0,s^+_1}$, $R^{\pm}_{s^+_0,s^+_1}$
and Lemma~\wkmone\ give \zetaopbound. Since
$$
\eta^-_1(t)=e^{s_0^-(t)'}(J^+_{s^-_0,s^-_1}(t)+R^+_{s^-_0,s^-_1}(t))
+4a(t,\ep)^2e^{-s_0^-(t)'}(J^-_{s^-_0,s^-_1}(t)+R^-_{s^-_0,s^-_1}(t)).\eq\eno[etaminus]
$$ 
an argument similar to the one above gives the result for $\psi^-_1$ and $\eta^-_1$.
Equation~\zetaombound\ follows since $t_{in}>0$. \qed

As we will use perturbation theory we suppose for each $\ep\in[0,\ep_0]$,
$$\sup_{t\in[0,t_{fi}]}|b(t,\ep)-b_1(t,\ep)|=O(\ep^2)=\sup_{t\in[0,t_{fi}]}|a^2(t,\ep)-a^2_1(t,\ep)|.\eq\eno[abepw]$$

With the above we obtain solutions to the initial value problem.

\thm thminit. {\rm [\cite{GBVA}, Theorem~5.5]}
 Suppose $a(t,\ep)$ and $b(t,\ep)$ are given by equation~\ab, and ic), 
iic), and ivc) hold.  
Let $a_1(t,\ep)$, $b_1(t,\ep)\in C([0,\infty)\times[0,\ep_0])$
satisfy \abepw\ 
with $b_1(t,\ep)$ real and $a_1(t,\ep)$ strictly positive on every
compact subset of $(0,\infty)\times[0,\ep_0]$.
Let 
$y\in C\backslash I(t_{fi},0)$ and  $\tilde p_n(y,\ep)$ be a solution of
$$
\eqalign{
\tilde p_{n+1}(y,\ep)&+2(b_1(n\ep,\ep)-y)
\tilde p_n(y,\ep)+4a_1(n\ep,\ep)^2\tilde p_{n-1}(y,\ep)=0,\cr
&
\tilde p_0(y,\ep)=1,\ \tilde p_1(y,\ep)=2(y-b_1(0,\ep)).}\eq\eno[inival]
$$
Then there exists an $\ep_1$ such that $y\in \bbc\setminus I(t_{fi},\ep)$,
$ 0\le\ep\le\ep_1$. For each $\ep$, $0<\ep\le\ep_1$ and all $n: n\ep\in[0,t_{fi}]$
$$
\tilde p_n(y,\ep)=\psi^+_1(n,y,\ep)/\psi^+_1(0,y,\ep)(1+\phi_1(n,y,\ep)),\eq\eno[formp]
$$
with $\psi^+_1$ given by equation~\fplusone\ and 
$$
|\phi_1(n,y,\ep)|<d\max(b\ep^{1/\alpha},a\ep^{2/\alpha},\ep).\ \eq\eno[errwkb]
$$
If $K$ is a compact set in $\bbc\setminus I(t_{fi},0)$ and $y\in K$ then there is an $\ep_K$ such that \errwkb\ holds uniformly on $K\times[0,\ep_K]$ and $\phi_1(n)\in H(K)\cap C(K\times[0,\ep_K])$. 

\pf From condition ivc) and  equation~\abepw\ we find that $a(0,\ep)^2\le c \ep^{2\over\alpha}$, $|b_1(i\ep,\ep)-b(i\ep,\ep)|\le c_1\ep^2$ and $|a^2_1(i\ep,\ep)-a^2(i\ep,\ep)|\le c_2\ep^2$. This coupled with equations~\zetaopbound\
and \zetaombound\ and equation~(5.28) of [\cite{GBVA}] give equation~\errwkb\ 
(note that in equation~(5.28) $L(1,n-1)$ should be replaced by $S(n-1)$, 
see [\cite{GBVAe}]). 
The smoothness properties for $\phi_1$ follow from Lemma~\ainitvalue\ and equation~(5.30) in [\cite{GBVA}] with the correspondence $f^+\to \psi^+_1$.

\advance\subsecno by 1\no=0\thmno=0
\beginsection  V. Matching

The previous results can be combined to obtain a uniform asymptotics  for
solutions of the initial value problem. This is accomplished by matching
solutions in various overlapping regions. Throughout this section we will
suppose $a_1(t,\ep), b_1(t,\ep)\in C([0,\infty)\times[0,\ep_0])$ with $b_1(t,\ep)$ real and $a_1(t,\ep)$ strictly positive on $(0,T)\times[0,\ep_0]\, T>1$.

\lem ratio. Suppose that conditions iic) and ivc) hold then
$$
\kappa(n)={\exp({1\over\ep}\int_0^{n\ep}\ln\hat q({u\over\ep}+{1\over2})du)\over\prod_{i=1}^n \hat q(i)}=\kappa(1+c_{n}),
$$
where
$$|c_{n}|=O({1\over n}),
$$
and
$$\kappa=\int_{1/2}^1 \ln\left({\hat q(u)\over\hat q(1)}\right)du -\int_1^{\infty}(B_2-B_2(u-[u])){d^2\over du^2}\ln\hat q(u) du \eq\eno[kap]
$$
Here $B_2$ and $B_2(x)$ are the second Bernoulli number and polynomial respectively. Also if equation~\abep\ holds then for
$n$ such that $t_{in}<t_n<t_{fi}$
$$
\kappa_1(n)=\prod_1^n{a(i\ep,\ep)\over a_1(i\ep,\ep)}=\kappa_1(1+O(\ep)),\  \eq\eno[kaponen]
$$
where
$$\kappa_1=\prod_{i=1}^{[t_{in}/\ep]}{a(i\ep,\ep)\over a_1(i\ep,\ep)}. \eq\eno[kapone]
$$

\pf  Conditions ic) and ivc) show that the second integral in $\kappa$ converges. The result follows by using Euler-Maclaurin formula [\cite{O}, p.~285].  From \abep\ we see that for $t_i>t_{in},\  \ln{a(t_i,\ep)\over a_1(t_i,\ep)}=O(\ep^2)$ which implies \kaponen. \qed

With the above we now prove,
\thm uniformbound. Suppose $a(t,\ep)$ and $b(t,\ep)$ are given by equation~\ab\ and conditions ic)--ivc)  hold. Let $a_1(t,\ep), b_1(t,\ep)$ satisfy equations~\abep\ and \abepw. Suppose $p_n(y,\ep)$ is such that $\tilde p_n$ satisfies the initial value problem \inival. If $y\in\bar\bbc_+\setminus\{0\}$ then there exists an $\ep_1$ and a $t_B\in (0,t_{fi}]$ such that for each $0<\ep\le\ep_1$ and all $n:n\ep\in[0,t_B]$,
$$
\eqalign{
&p_n(y,\ep)=\cr
&\kappa(n)\kappa_1(n)\left({y^2\over
(y-b(t_n,\ep))^2-4a^2(t_n+{\ep\over2},\ep)}\right)^{1\over4}e^{{1\over\ep}\int_0^{t_n} \ln( z_1(u,y,\ep)+\sqrt{z_1(u,y,\ep)^2-1})du}\cr
&\qquad \times (1+\tilde\phi_1(n,y,\ep)),} \eq\eno[asmext]
$$
where $\kappa(n)$ and $\kappa_1(n)$ are given in Lemma~\ratio,\ $t_n=n\ep$, $z_1(u,y,\ep)={y-b(u,\ep)\over2a(u+{\ep\over2},\ep)}$,  
and
$$|\tilde\phi_1(n,y,\ep)|\le d
 \max(|b|\ep^{1\over\alpha},\ep^{2\over\alpha},\ep).
\eq\eno[errortone]
$$
If $K$ is a compact set of $\bar\bbc_+\setminus\{0\}$ with $y\in K$ then there exists an $\ep_K$ so that the  above error can be made uniform on $K\times[0,\ep_K]$ for all $n:n\ep\in[0, t_B]$. If $y\in L_1^+(0)$ given in Lemmas~ \anafreudpl\ or \anafreudpla\ then there exists a $t_{in}<t_{B}$ and an $\ep_2\le\ep_1$ such that for $0\le\ep\le\ep_2$ and all $n:n\ep\in[t_{in},t_{fi}]$
$$
{p_n(y,\ep)\over e^{{1\over\ep}Q(y, \ep)}}=\kappa\kappa_1 2\sqrt{\pi y}\hat g_1(n\ep,y,\ep)({\rm Ai}(\ep^{-{2\over3}}\rho_1(n\ep,y,\ep))+r_1(n,y,\ep)),\eq\eno[asmoneone]
$$
where $Q(y, \ep)$ is given in equation~\repqygo, $\kappa$ by \kap, $\kappa_1$ 
by \kapone,\ and $\rho_1$ by equation~\comdeqone,  and
$$
\hat g_1(t_n,y,\ep)=
\left({\ep^{-{2\over3}}\rho_1(t_n,y,\ep)\over((y-b(t_n,\ep))^2-4a(t_n+{\ep\over2},\ep)^2)}\right)^{1\over4}.
\eq\eno[goneone]
$$
The error term satisfies 
$$
|r_1(n,y,\ep)e^{{2\over3\ep}\Re(\rho_1^{3\over2}(t_n,y,\ep))}|\le d\max(|b|\ep^{1\over\alpha},\ep^{2\over\alpha},\ep).
\eq\eno[errortwo]
$$
If $K$ is a compact set in $L_1^+(0)$ then there exists an $\ep_K$ such that the above bound is uniform on $K\times[0,\ep_K]$ for all $n:n\ep\in[t_{in},t_{fi}]$.
\hfill\break\indent
If $b-2a<0$ and $y\in L_2^+(0)$ given by Lemma~\anafreudpl,\ then there exists a $t_{in}<t_B$ and an $\ep_2\le\ep_1$ such that for $0\le\ep\le\ep_2$ and all $n:n\ep\in[t_{in},t_{fi}]$
$$
{p_n(y,\ep)\over e^{Q(y, \ep)}}=(-1)^n\kappa\kappa_1 2\sqrt{\pi(-y)} 
\hat g_2(n\ep,y,\ep)
\left({\rm Ai}(\ep^{-{2\over3}}\rho_2(n\ep,y,\ep))+r_2(n,y,\ep)\right),\eq\eno[asmonetwo]
$$
where $Q(y, \ep)$ is given in equation~\repqylo\ and
$$
\hat g_2(t_n,y,\ep)=
\left({\ep^{-{2\over3}}\rho_2(t_n,y,\ep)\over((y-b(t_n,\ep))^2-4a(t_n+{\ep\over2},\ep)^2)}\right)^{1\over4}.
\eq\eno[gonetwo]
$$
The error term satisfies 
$$
|r_2(n,y,\ep)e^{{2\over3\ep}\Re(\rho_2^{3\over2}(t_n),y,\ep)}|\le d\max(|b|\ep^{1\over\alpha},\ep^{2\over\alpha},\ep).
\eq\eno[errortwo]
$$
If $K$ is a compact set in $L_2^+(0)$ then there exists an $\ep_K$ such that the above bound is uniform on $K\times[0,\ep_K]$ for all $n:n\ep\in[t_{in},t_{fi}]$.

\pf See Figure~2 for a visualization of the above times. Fix $y$ in 
$\bar\bbc_+\setminus\{0\}$. Since $\gamma^+_0(0)=0$ we see from the definition of $I(t,\ep)$ that there exists a $t_B\in(0,t_{fi}]$ such that for all $(t,\ep)\in[0,t_B]$,\ $y\in \bar\bbc_+\setminus I(t_B,0)$. Thus Theorem~\thminit\ says there exits an $\ep_1$ so that for each $\ep,\ 0<\ep\le\ep_1$ and all $n:n\ep\in[0,t_B]$, 
$$
\tilde p_n(y,\ep)=e^{{1\over\ep}\int_0^{t_n}(s^+_0(u,y,\ep)'+s^+_1(u,y,\ep)')du}(1+\phi(n,y,\ep)),\eq\eno[asymptoone]
$$
where $t_n=n\ep$ and equations~\szeropm--\fplusone\ have been used.
We find  using Taylor's Theorem that,
$$\eqalign{
&{1\over\ep}\int_0^{t_n}\ln\left(y-b(u,\ep)+
\sqrt{(y-b(u,\ep))^2-4a\left(u+{\ep\over2},\ep\right)^2}\right)du
\cr&={1\over\ep}\int_0^{t_n}\ln h^+_2(u,y,\ep)du-{1\over2}\int_0^{t_n}{4a(u,\ep)a'(u,\ep)\over h^+_2(u,y,\ep)h_1(u,y,\ep)}du+{1\over\ep}\int_0^{t_n}e_1(u,y,\ep)du,}\eq\eno[atoaonehalf]
$$
and
$$\ln((y-b(t_n,\ep))^2-4a^2(t_n+{\ep\over2},\ep))=\ln h_1(t_n,y,\ep)+e_2(t_n,y,\ep). \eq\eno[soneto]
$$
Condition ivc) shows that,
$$
|e_1|\le c(y)\max(b\ep^{1/\alpha},a\ep^{2\over\alpha},\ep),\eq\eno[eonebound]
$$
and condition iic) shows that for $t_n$ strictly greater than zero, 
$$
|e_2(t_n,y,\ep)|\le c(y)\ep. \eq\eno[etwobound]
$$
It also follows from conditions ivc) that
$$h_1(0,y,\ep)=y(1+e_3(0,y,\ep)),\eq\eno[honezero]
$$
with
$$
|e_3(0,y,\ep)|\le c(y)\max(b\ep^{1\over\alpha},a\ep^{2\over\alpha}).\eq\eno[ethreebound]
$$
Substituting \sodir\ into equation~\geomone\ for ${s_1^+}'$ then using equations~\atoaonehalf,\ \soneto,\ and \honezero\  in \asymptoone\ yields,
$$\eqalign{
p_n(y,\ep)=&\kappa(n)\kappa_1(n)
\left({y^2\over(y-b(t_n,\ep))^2-4a(t_n+{\ep\over2},\ep)^2}\right)^{1\over4}\cr&\times e^{{1\over\ep}\int_0^{t_n}\ln(z_1(u,y,\ep)+\sqrt{z_1(u,y,\ep)^2-1})du}(1+\phi_1(n,y,\ep)),}\eq\eno[intermed]
$$
where $\phi_1$ has the same properties as $\phi$ and thus satisfies the
bound \errortone\ by Theorem~\thminit. This yields equation~\asmext.
 If $y\in L_1^+(0)$ then Theorem~\freudtur\ shows that there exist a $t_{in}<t_B$ and an $\ep_2$ so that $y\in L_1^+(\ep)$ for all $0\le\ep\le\ep_1$ and $\rho_1$ is well defined for $(t,\ep)\in[t_{in},t_{fi}]\times[0,\ep_2]$. For fixed $\ep$ the function on the right hand side of equation~\asmext\
satisfies the same difference equation \fourteen\ as $p_n(y,\ep)$ so that
for all $n:n\ep\in[t_{in},t_{fi}]$ Theorem~\cthmnextone\ implies that it can be written as
$c_1 f_1+c_2 f_2$. In order to compute these coefficients restrict
$n:n\ep\in[t_{in},t_B]$ in which case equation~\asmext\ can be used for
$p_n$. For $n$ in this region Lemma~\ratio\ shows that
$\kappa(n)=\kappa(1+O(\ep))$ , $\kappa_1(n)=\kappa_1(1+O(\ep))$ and from equations~\qgrone\ and \comdeqone\ we find that
$$
{p_n(y,\ep)\over e^{{1\over\ep}Q(y,\ep)}}=\kappa\kappa_1
\left({y^2\over(y-b(t_n,\ep))^2-4a(t_n+{\ep\over2},\ep)^2}\right)^{1\over4}\times e^
{-{2\over3}{1\over\ep}\rho_1(t_n)^{3\over2}}(1+\phi_1(n,y,\ep))
$$
Since there is no turning point in $[t_{in},t_B]$ we may use the asymptotic expansion of the Airy functions,
$$
{\rm Ai}(\ep^{-{2\over3}}\rho(t_n))={\ep^{1\over6}\over 2\pi^{1\over2}\rho(t_n)^{1\over4}}e^{-{2\over3}{1\over\ep}\rho(t_n)^{3\over2}}(1+0(\ep)),
$$
and
$$
{\rm Bi}(\ep^{-{2\over3}}\rho(t_n))={\ep^{1\over6}\over \pi^{1\over2}\rho(t_n)^{1\over4}}e^{{2\over3}{1\over\ep}\rho(t_n)^{3\over2}}(1+0(\ep)),
$$
in $f_1$ and $f_2$.
Using the Casorati determinant Ca,
$$
{\rm Ca}[u_1,u_2](n)=a_1(n\ep,\ep)(u_1(n)u_2(n-1)-u_1(n-1)u_2(n)), \eq\eno[wi]
$$
we find $c_1={\rm Ca}[p/e^{Q/\ep},f_2](n)\over\rm{Ca}[f_1,f_2](n)$ and
$c_2={\rm Ca}[p/e^{Q/\ep},f_1](n)\over{\rm Ca}[f_2,f_1](n)$.
The above equations coupled with equations~\cineqone\ and \eqwr,
show that
$$
c_1={2\kappa_1\kappa\sqrt{\pi y}\over\ep^{1\over6}}(1+\hat r^1(n_m,y,\ep)), \eq\eno[eqcone]
$$
and,
$$
|c_2|<{c(y)\over\ep^{1\over6}}e^{-{2\over3}{1\over\ep}{\bf Re}(\rho_1^{3/2}(n_m\ep,y,\ep)+\rho_1^{3/2}((n_m+1)\ep,y,\ep))}|\hat r^2(n_m,y,\ep)|
$$
where $n_m$ is the smallest value of $n$ such that $n\ep\in[t_{in},t_B]$ and the fact that Ca is independent of $n$ has been used to obtain the above equations. Theorems \thminit\ and \cthmnextone\ show that 
$$
|\hat r^i|<c(y)\max(|b|\ep^{1\over\alpha},\ep^{2\over\alpha},\ep),\quad i=1,2
$$ 
Since the real part of $\rho_1(t)^{3/2}$ is a decreasing function of $t$ we find,
$$
|c_2 f_2|< c(y)e^{-{2\over3}{1\over\ep}{\bf
Re}(\rho_1^{3/2}((n_m+1)\ep,y,\ep))}|\hat r^2(n_m,y,\ep)|, \eq\eno[eqctwo]
$$
which is exponentially small. This gives equation~\asmoneone. The uniformity of the error terms follows from Theorems~ \cthmnextone\ and \thminit. 

If $b-2a<0$ and $y\in L_2^+(0)$ where $L_2(0)$
is given by Lemma~\anafreudpl,\ there exists  an $\ep_2$ such that $y\in L^+_2(\ep)$ for all $\ep\in[0, \ep_2]$ so that $\rho_2$ is well defined for $(t,\ep)\in[t_{in},t_{fi}]\times[0,\ep_2]$. For $t_n\in[0,t_B]$ we find from equation~\htwo,
$$
\eqalign{
&p_n(y,\ep)=\cr
&(-1)^n\kappa(n)\kappa_1(n)({y^2\over
(b(t_n,\ep)-y)^2-4a^2(t_n+{\ep\over2},\ep)})^{1\over4}e^{{1\over\ep}\int_0^{t_n} \ln( z_2(u,y,\ep)+\sqrt{z_2(u,y,\ep)^2-1})du}\cr
&\qquad\times (1+\tilde\phi_2(n,y,\ep)),} \eq\eno[asmextlo]
$$
From Lemma~\ratio, equations~\qgrtwo\ and \comdeqtwo\ we find for fixed $(y,\ep)$ and all $n:n\ep\in[t_{in},t_{fi}]$
$$
{(-1)^n p_n(y,\ep)\over e^{{1\over\ep}Q(y,\ep)}}=
\kappa\kappa_1\left({y^2\over((y-b(t_n,\ep))^2-4a(t_n+{\ep\over2},\ep)^2,\ep)}\right)^{1\over4}\times e^
{-{2\over3}{1\over\ep}\rho_2(t_n)^{3\over2}}(1+\phi_2(n,y,\ep)),
$$
where $Q(y,\ep)$ is given by equation~\repqylo. In this region we also have from the second part on Theorem~\cthmnextone\ that
$$
{(-1)^n p_n(y,\ep)\over e^{{1\over\ep}Q(y,\ep)}}=c_1 f_1(n)+c_2 f_2(n)
$$
where $f_1(t_n)=g_2(t_n)(A_0(\ep^{-2/3}\rho_2(t_n))+r^1(t_n))$ and
$f_2(t_n)=g_2(t_n)(A_1(\ep^{-2/3}\rho_2(t_n))+r^2(t_n))$ with $g_2$
described in equation~\gonetwo. Since there is no turning point in $[t_{in},t_A]$
the asymptotic expansion of the Airy functions can be used so that 
the coefficients $c_1$ and $c_2$ may now be computed using the Casorati determinant as above. The uniformity of the error terms follows from the second part of Theorem~\cthmnextone\  and Theorem~\thminit.   \qed 

\bigskip

\centerline{\epsfxsize=3in\epsfbox{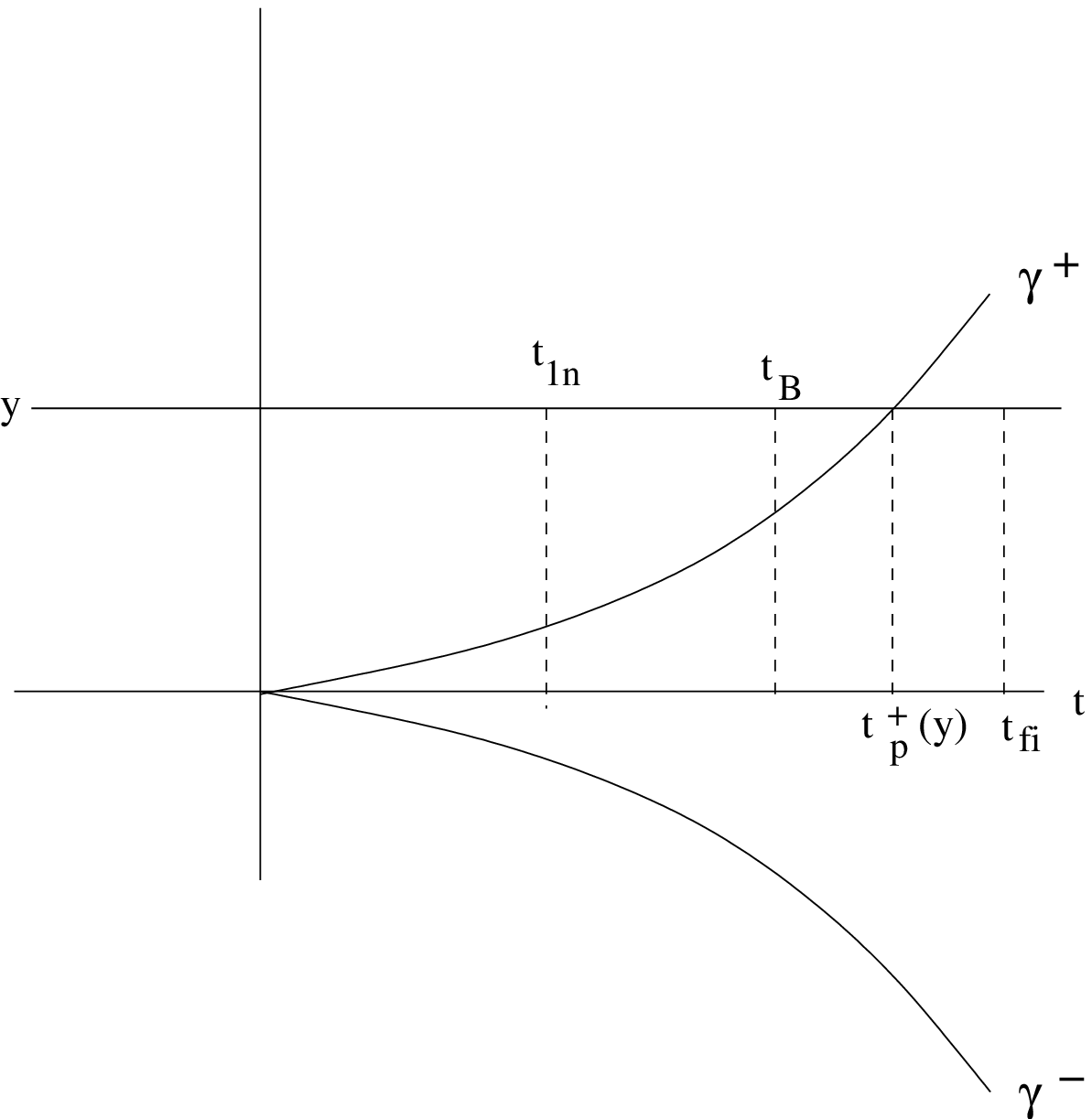}}

\centerline{Figure 2: Various times used in Theorem~\uniformbound}

A case where the above results can be applied  is when $b=0$ and let $T_{2l}(x)$ be a monic even polynomial in $x$ with real coefficients and let $T_{2l}(x_0)=0$ and $T_{2l}'(x)>0$ for all $x>x_0$. Let $\hat q(t)=T^{-1}_{2l}(t)$ be the inverse branch of $T_{2l}$ that is positive for all $t$ large enough. From the assumptions on $T_{2l}$ we see that $q(t)$ is monotonically increasing for $t>0$ and  since $x=\infty$ is a critical point of order $2l$ the open mapping theorem (see Rudin[\cite{R}] p), shows that in a neighborhood of infinity $q(t)$ has the piuseux series representation
$$
\hat q(t)=\sum_{n=-1}^{\infty}q_n t^{-{n\over2l}}. \eq\eno[pius]
$$
Furthermore $q(t)$ has an extension to a wedge $\{t :\arg t <\delta\}$ where $\delta$ is the argument of the closest non zero critical value  to the positive real axis. Because of the conditions on $T_{2l}$ and $T'_{2l}$ there are no critical values on the positive real axis. If $t=0$ is a non analytic point of $\hat q(t)$ the the piuseux series about $t=0$ of form $\sum_{n=1}^{\infty} c_n t^{n/p}$ with $p\ge2l$. From the above argument we see that.

\lem pius. Suppose $\hat q(t)$ is the inverse branch of an even monic polynomial with the properties
described above. Then $\hat q(t)$ will satisfy ic-ivc) with $\alpha=2l$.

We now consider the case when $b-2a>0$ and $0<y<\gamma^-_0(t_{fi})$ then we will be in the two turning point case and  besides $t_B$ above we will need some other times in order to separate these points 
(see Figure~3). We will restrict ourselves to the real line since at this time we are unable to control the error terms for the matching problem beyond the first turning point in the complex plane.

\thm uniformbounda. Suppose $a(t,\ep)$ and $b(t,\ep)$ are given by equation~\ab\ with $b-2a>0$ and conditions ic)--ivc)  hold. Let $a_1(t,\ep), b_1(t,\ep)$ satisfy equations~\abep\ and \abepw\ and $0<y<\gamma^-_0(t_{fi})$. Then there exist $t_B,\ t_{in},\ \hat t_{in},\ \hat t_{fi}$ and an $\ep_1$ satisfying the inequalities $0<t_{in}<t_B<\hat t_{in}<\hat t_{fi}<t_{fi}$ such that $t^+_p( y,\ep)\notin[0,t_B]$, $t^+_p(y,\ep)\in(t_{in},\hat t_{fi})$, $t^-_p(y,\ep)\notin[t_{in},\hat t_{fi}]$, $t^-_p(y,\ep)\in(\hat t_{in}, t_{fi})$ and $t^+_p(y,\ep)\notin[\hat t_{in}, t_{fi}]$ for all $0\le\ep\le\ep_1$. Suppose $p_n(y,\ep)$ is such that $\tilde p_n$ satisfies the initial value problem \inival. Then for $0<\ep\le\ep_1$, $\ep_1$ sufficiently small and all $n:n\ep\in[0,t_B],\ p_n(y,\ep)$ satisfies 
$$
\eqalign{
&p_n(y,\ep)=\cr
&\kappa(n)\kappa_1(n)\left({y^2\over
(y-b(t_n,\ep))^2-4a^2(t_n+{\ep\over2},\ep)}\right)^{1\over4}e^{{1\over\ep}\int_0^{t_n} \ln( z_1(u,y,\ep)+\sqrt{z_1(u,y,\ep)^2-1})du}\cr
&\qquad \times (1+\tilde\phi_1(n,y,\ep)),} \eq\eno[asmexta]
$$
where $\kappa(n)$ and $\kappa_1(n)$ are given in Lemma~\ratio, $t_n=n\ep$, $z_1(u,y,\ep)={y-b(u,\ep)\over2a(u+{\ep\over2},\ep)}$,  
and
$$|\tilde\phi_1(n,y,\ep)|\le d
 \max(|b|\ep^{1\over\alpha},\ep^{2\over\alpha},\ep).
\eq\eno[errortonea]
$$
For all $n:n\ep\in[t_{in},\hat t_{fi}]$
$$
{p_n(y,\ep)\over e^{Q(y, \ep)}}=\kappa\kappa_1 2\sqrt{\pi y}\hat g_1(n\ep,y,\ep)({\rm Ai}(\ep^{-{2\over3}}\rho_1(n\ep,y,\ep))+r_1(n,y,\ep)),\eq\eno[asmoneonea]
$$
where $Q(y, \ep)$ is given in equation~\repqygo, $\kappa$ by \kap, $\rho_1$ by equation~\comdeqone,   $g_1$ by equation~\goneone,  and the error term satisfies 
$$
|r_1(n,y,\ep)e^{{2\over3\ep}\Re(\rho_1^{3\over2}(t_n,y,\ep))}|\le d\max(|b|\ep^{1\over\alpha},\ep^{2\over\alpha},\ep).
\eq\eno[errortwoa]
$$
For all $n:n\ep\in[\hat t_{in},t_{fi}]$
$$\eqalign{
&{p_n(y,\ep)\over e^{{1\over\ep}Q(y,\ep)}}\cr
&\quad = (-1)^n2\kappa\kappa_1\sqrt{\pi y}\,\hat g_2(n\ep,y,\ep)
\Biggl[\left(\sin\left({1\over\ep}\Gamma_{\ep}(y)\right)+R_1(y,\ep)\right)
\Bigl({\rm Ai}(\ep^{-2/3}\rho_2(t_n,y,\ep))\cr
&\qquad +\hat r_1(t_n,y,\ep)\Bigr)
+2\left(\cos\left({1\over\ep}\Gamma_{\ep}(y)\right)+R_2(y,\ep)\right){\rm Bi}(\ep^{-2/3}\rho_2(t_n,y,\ep))+\hat r_2(t_n,y,\ep))\Biggr],}$$
where 
$$
{1\over\ep}\Gamma_{\ep}(y)=\int_{b-2a}^{b+2a} {\hat q}^{-1}
\left({\hat q}\left({1\over\ep}+{1\over2}\right){y\over u}\right){du\over\sqrt{4a^2-(u-b)^2}}. \eq\eno[gammay]
$$
Here 
$$
\eqalign{&
|\hat r_1(n\ep,y,\ep)e^{{2\over3\ep}\Re(\rho_2^{3\over2}(t_n,y,\ep))}|,\  
|\hat r_2(n\ep,y,\ep)e^{-{2\over3\ep}\Re(\rho_2^{3\over2}(t_n,y,\ep))}|,|R_1(y,\ep)|,\|R_2(y,\ep)|\cr&\le d
 \max(|b|\ep^{1\over\alpha},\ep^{2\over\alpha},\ep).}\eq\eno[errt] $$
The error terms in equation \errortwoa\ and \errt\ are uniform on closed intervals of $[0,\infty)$.

\pf See Figure 3 for a visualization of above times. Equations~\asmexta\ and \errortonea\ follow from equations~\asmext\ and \errortone\ of Theorem~\uniformbound. Likewise if in Lemma~\anafreudpla\ we replace $t_{fi}$ in $L_1(\ep)$ with $\hat t_{fi}$, equations~\asmoneonea\
and \errortwoa\ also follow from Theorem~\uniformbound.\ We now consider the
interval $[\hat t_{in},t_{fi}]$. In this interval we have by Theorems~\thmnext\ and \cthmnextone\ that
$$
(-1)^n{p_n(y,\ep)\over e^{{1\over\ep}Q(y,\ep)}}=c_1 f_1
+c_2 f_2,\eq\eno[satone]
$$
with $f_1=g_2(t_n)({\rm Ai}(\ep^{-2/3}\rho_2(t_n))+r_1(t_n))$ and
$f_2(t_n)=g_2(t_n)({\rm Bi}(\ep^{-2/3}\rho_2(t_n))+r_2(t_n))$.
By selection the interval $[\hat t_{in}, \hat t_{fi}]$ is void of turning points so for small enough $\ep_1$ the
asymptotics of the Airy functions on the negative real axis, may 
be used [\cite{O}, p.~392]. Thus with $n:n\ep\in[\hat t_{in}, \hat t_{fi}]$  equation~\asmoneonea\ can be recast as
$$
\eqalign{
&{p_n(y,\ep)\over e^{{1\over\ep}Q(y, \ep)}}\cr
&=2\kappa\kappa_1\left({y\over2a(t_n+{\ep\over2},\ep)\sin z_1(t_n,y,\ep)}\right)^{1\over2}
\left(\cos\left({2\over3\ep}(-\rho_1(n\ep,y,\ep))^{3\over2}-{\pi\over4}\right)+\tilde r_1(n,y,\ep)\right).}\eq\eno[asmoneonea]
$$
Equations~\eight\ and \langert\ show that
$$
{2\over3}(-\rho_1(t_n))^{3\over2}=\int_{t_p^+(y)}^{t_p^-(y)}\cos^{-1}
\left({y-b(u)\over2a(u+{\ep\over2})}\right)du +\pi(t_n-t_p^-(y))+{2\over3}(-\rho_2(t_n))^{3\over2}.
$$
Using equations~\ab, \qhat\ and relations between $q^{-1}_{\ep}$ and $t_p^{\pm}$, we find
$$
\eqalign{
\int_{t_p^+(y)}^{t_p^-(y)}&\cos^{-1}
\left({y-b(u)\over2a(u+{\ep\over2})}\right)du
=\int_{y\over b+2a}^{y\over b-2a}q^{-1}_{\ep}(w)'\cos^{-1}
\left({y-bw\over2aw}\right)dw\cr
&=\pi t_p^-(y)+\ep\int_{b-2a}^{b+2a}\hat q^{-1}
\left(\hat q\left({1\over\ep}+{1\over2}\right){y\over u}\right){du\over\sqrt{4a^2-(u-b)^2}}-{\pi\over2},}
$$
where integration by parts and \qinv\ have been used to obtain the last equality in the above equation. Since $t_n=n\ep$ we find
$$\cos\left({2\over3\ep}\left(-\rho_1(n\ep,y,\ep)\right)^{3\over2}-{\pi\over4}\right)=
(-1)^n\sin
\left({2\over3\ep}\left(-\rho_2(t_n)\right)^{3\over2}+{1\over\ep}\Gamma_{\ep}(y)-{\pi\over4}\right). \eq\eno[linkthree]
$$
where $\Gamma_{\ep}$ is given by equation~\gammay.
Therefore
$$
\eqalign{
&(-1)^n{p_n(y,\ep)\over e^{{1\over\ep}Q(y,\ep)}}\cr&=2\kappa\kappa_1\left({y\over2a(t_n+{\ep\over2},\ep)\sin z_1(t_n,y,\ep)}\right)^{1\over2}\left(
\sin\left({2\over3\ep}(-\rho_2(t_n))^{3\over2}+{1\over\ep}\Gamma_{\ep}(y)-{\pi\over4}\right)+\tilde r_1(n,y,\ep)\right),}\eq\eno[sattwo]
$$
Likewise in this interval
$$f_1(t_n)={\ep^{1/6}\over (\pi2a(t_n+{\ep\over2},\ep)\sin z_2(t_n,y,\ep))^{1/2}}
\left(\cos\left({2\over 3\ep}(-\rho_2(t_n))^{3/2}
-{\pi\over 4}\right)+\hat r_1(t_n)\right),$$
and
$$f_2(t_n)=-{\ep^{1/6}\over 2(\pi2a(t_n+{\ep\over2},\ep)\sin z_2(t_n,y,\ep))^{1/2}}
\left(\sin\left({2\over 3\ep}(-\rho_2(t_n))^{3/2}-
{\pi\over 4}\right) + \hat r_2(t_n)\right).$$
The error term $\tilde r_1(t_n)$ satisfies equation~\errortwo\
and $\hat r_1(t_n)$ and $\hat r_2(t_n)$ satisfy a similar
bound with $\Re(\rho_1^{3\over2})$ replaced by $\mp\Re(\rho_2^{3\over2})$ respectively.  

Comparison of the coefficients in equations~\satone\ and \sattwo\ (or using the Casorati determinant) for $c_1$ and $c_2$ and utilizing the fact that $\sin z_1=\sin z_2$
yields
$$c_1=2\kappa\kappa_1{\sqrt{\pi y}\over\ep^{1/6}}
\left(\sin\left({1\over\ep}\Gamma_{\ep}(y)\right) + R_1\right),
$$
and
$$
c_2=-4\kappa\kappa_1{\sqrt{\pi y}\over\ep^{1/6}}
\left(\cos\left({1\over\ep}\Gamma_{\ep}(y)\right) + R_2\right),
$$
where $|R_i|<d\max(b\ep^{1\over\alpha},\ep^{2\over\alpha},\ep), i=1,2$ which follows since we are in the oscillatory region. The result now follows.\qed

\bigskip

\centerline{\epsfxsize=3in\epsfbox{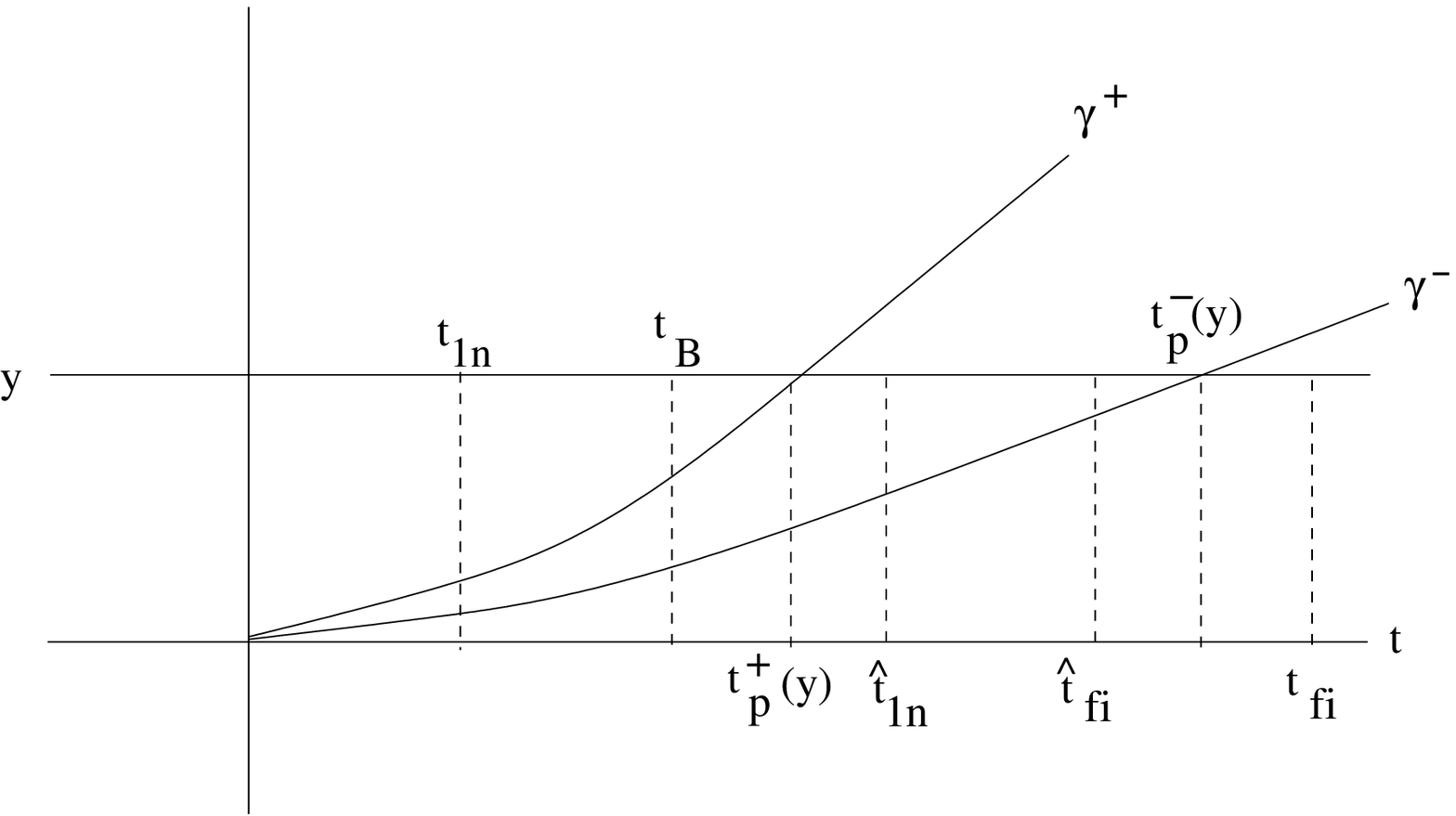}}
\par\nobreak
\centerline{Figure 3: Various times used in Theorem~\uniformbounda}

Of special interest is when $\ep=1/N$ and $t_n=1$ which implies that $n=N$. From equation~\qhat\ we see that $q_{\ep}(1)=1$ and $\gamma^{\pm}_{\ep}=b\pm2a$. The turning points are at values of $y$ such that $y=b+2a$ or $y=b-2a$. In the results below we  suppress the dependence upon $t$ since $t=1$.

\thm ftwoa. Suppose that $b-2a\le0$ and $t=1$. For $y\in \bar\bbc_+\setminus[b-2a,b+2a]$ 
$$
\eqalign{
&p_N(y,1/N)=\cr
&\kappa\kappa_1\left({y^2\over
(y-b)^2-4a^2}\right)^{1\over4}e^{N\int_0^1 \ln( z_1(u,y,1/N)+\sqrt{z_1(u,y,1/N)^2-1})du}(1+\phi_1(y,1/N)).} \eq\eno[asmext]
$$
The bound 
$$
|\phi_1(y,1/N)|\le \max(b/N^{1/\alpha},1/N^{2/\alpha},1/N), \eq\eno[ub]
$$
holds on compact subsets of $\bar\bbc_+\setminus[b-2a,b+2a]$. For $y\in L_1^+(0)$ given by Lemma~\anafreudpl\ and $N$ sufficiently large,
$$
\eqalign{
&{p_N(y,1/N)\over e^{NQ
(y,1/N)}}=\cr&
2\kappa\kappa_1\sqrt{\pi y}\left({N^{2/3}\rho_1(y,1/N)\over (y-b)^2-4a^2}\right)^{1/4}({\rm Ai}(N^{2/3}\rho_1(y,1/N))
+r_1(y,1/N)),}\eq\eno[cothreea]
$$
where $Q$ is given by equation~\repqygo,\ $\rho_1$ by \comdeqone,  $\kappa$ by \kap\ and $\kappa_1$ by \kapone. The error term $r_1$ satisfies  the bound
$$
|r_1(y,1/N)e^{{2\over3}N\Re(\rho_1^{3\over2}(y,1/N))}|\le d\max(|b|\ep^{1\over\alpha},\ep^{2\over\alpha},\ep),
\eq\eno[erroro]
$$
uniformly on compact subsets of $L_1^+(0)$. If $b-2a<0$ and $y\in L_2^+(0)$ given by Lemma~\anafreudpl\ then for $N$ sufficiently large,
$$
\eqalign{
&{(-1)^N p_N(y,1/N)\over
e^{NQ(y,1/N)}}=\cr&2\kappa\kappa_1\sqrt{\pi (-y)}\left({N^{2/3}\rho_2(y,1/N)\over(y-b)^2-4a^2}\right)^{1/2}({\rm Ai}(N^{2/3}\rho_2(y,1/N))+r_2(y,1/N)),}\eq\eno[cofour]$$
where $Q$ is given by equation~\repqylo\ and $\rho_2$ by \comdeqtwo. The error term $r_2$ satisfies the bound
$$
|r_2(y,1/N)e^{{2\over3}N\Re(\rho_2^{3\over2}(y,1/N))}|\le d\max(|b|\ep^{1\over\alpha},\ep^{2\over\alpha},\ep),
\eq\eno[errort]
$$
uniformly on compact subsets of $L_2^+(0)$.

\pf For $y\in\bar\bbc_+\setminus[b-2a,b+2a]$ as $t_n$ varies from $0$ to $1$ no turning points are encountered. Thus the result follows from Lemma~\ratio,\ equation~\asmexta\ of Theorem~\uniformbound\ with $t_n=1$ and the analytic properties of the solution. The uniformity of the error also follows from Theorem~\uniformbound. The result for the remaining regions follows in a similar manner.~\qed

Likewise for $b-2a>0$ from Theorem~\uniformbounda\ with $t_n=1$  we find,
\thm ftwob. Suppose that $b-2a>0$ and $t=1$. 
For $y\in \bar\bbc_+\setminus[0,b+2a]$ 
$$
\eqalign{
&p_N(y,1/N)=\cr
&\kappa\kappa_1\left({y^2\over
(y-b)^2-4a^2}\right)^{1\over4}e^{N\int_0^1 \ln( z_1(u,y,1/N)+\sqrt{z_1(u,y,1/N)^2-1})du}(1+\phi_1(y,N)),} \eq\eno[asmext]
$$
where $\phi_1$ satisfies equation~\ub\ uniformly
on compact subsets of $\bar\bbc_+\setminus[0,b+2a]$. For $y\in L_1^+(0)$ given by Lemma~\anafreudpla\ and $N$ sufficiently large,
$$
\eqalign{
&{p_N(y,1/N)\over e^{NQ
(y,1/N)}}=\cr&
2\kappa\kappa_1\sqrt{\pi y}\left({N^{2/3}\rho_1(y,1/N)\over (y-b)^2-4a^2}\right)^{1/4}({\rm Ai}(N^{2/3}\rho_1(y,1/N))
+r_1(y,1/N)),}\eq\eno[cothreeaa]$$
where $Q$ is given by equation~\repqygo,\ $\rho_1$ by \comdeqone, $\kappa$ by \kap\ and $\kapone$ by \kapone. The error term $r_1$ satisfies \erroro\ uniformly on compact subsets of $L_1^+(0)$. For $y\in(0,b+2a)$ 
$$\eqalign{
&{(-1)^N p_N(y,1/N)\over e^{NQ(y,1/N)}}\cr&\quad=2\kappa\kappa_1\sqrt{\pi y}\left({N^{2/3}\rho_2(y,1/N)\over(y-b)^2-4a^2}\right)^{1/4}\times\cr&\qquad\qquad
\Bigl[(\sin(N\Gamma_{1/N}(y))+R_1(y,1/N))({\rm Ai}(N^{2/3}\rho_2(y,1/N))+\hat r_1(y,1/N))\cr
&\qquad\qquad +2(\cos(N\Gamma_{1/N}(y))+
R_2(y,1/N))({\rm Bi}(N^{2/3}\rho_2(y,N))+\hat r_2(y,1/N))\Bigr],}\eq\eno[fouraa]$$
where uniformly on compact subintervals of $(0,b+2a)$ 
$$
\eqalign{&
|\hat r_1(y,1/N)e^{{2\over3}N\Re(\rho_2^{3\over2}(y,1/N))}|,\  
|\hat r_2(y,1/N)e^{-{2\over3}N\Re(\rho_2^{3\over2}(y,1/N))}|,\cr
&\quad |R_1(y,1/N)|,|R_2(y,1/N)|\le d
 \max(|b|/N^{1\over\alpha},1/N^{2\over\alpha},1/N).}\eq\eno[errorta] $$

\remk In the language of Deift et.\ al [\cite{DKMVZ}] and Biak et.\ al [\cite{BKMM}] the regions $(b+2a,\infty)$, $(b-2a,b+2a)$ and $(0, b-2a)$ are called bulk,  band and saturated region respectively.
The neighborhoods around $b+2a$ are called a band/void edge while the neighborhood of $b-2a$ is called a
band/saturated edge. Due to the error of the matching problem in the above technique the exponentially small error obtained by [\cite{BKMM}, Theorems 2.10 and 2.16] in the saturated region cannot be achieved.

We now examine the location of the zeros of the above orthogonal
polynomials. Denote
the zeros of the Airy function Ai by
$$
0>ai_1>ai_2 >\cdots.
$$
It is well known [\cite{O}] that the zeros of Ai lie in $(-\infty,0)$.
Denote the zeros of $p_N(y,1/N)$ in decreasing order $y_{1,N}>y_{2,N}>\cdots>y_{N,N}$.
We have, 
\thm zone. For fixed $k$,
$$
y_{k,N}= b+2a+{a^{1\over3}\over (N{d t_p^+\over dy}(b+2a,1/N))^{2\over3}}ai_k+O(\max{(1/N^{{2\over\alpha}+{1\over3}},|b|/N^{{1\over\alpha}+{1\over3}},1/N^{4\over3}})),\eq\eno[zeroone]
$$
and if $b-2a<0$
$$
y_{N-k,N}=b-2a-{a^{1\over3}\over (N{d t_p^-\over dy}(b-2a,1/N))^{2\over3}}ai_k+O(\max{(1/N^{{2\over\alpha}+{1\over3}},|b|/N^{{1\over\alpha}+{1\over3}},1/N^{4\over3}})),\ \eq\eno[zerotwo]
$$
Here $t_p^{\pm}=q^{-1}_{1/N}({y\over b\pm2a})$.

\pf From the relation between the equilibrium measure and $\rho_1$ given in Theorem~\freudtur. We see that for $t=1$ and fixed $N,\ \rho_1$ is an increasing function of $y$ and therefore has an inverse function. Equations~\cothreea\ or \cothreeaa\ and the arguments in [\cite{O}, p.~406--407] show that
$$
y_{k,N}=b+2a+{1\over N^{2\over3}{d\rho_1\over dy}(b+2a,1/N)}ai_k +O(\max{(1/N^{{2\over\alpha}+{1\over3}},|b|/N^{{1\over\alpha}+{1\over3}},1/N^{4\over3}})).
$$
Equation~\zeroone\ can now be obtained using \simzero\ and \tinv.
Likewise equation~\zerotwo\ is a consequence of \cofour\ and \errort. \qed

For $b-2a>0$ note that  $\hat q^{-1}$ is strictly increasing implies that for fixed $N, \Gamma_{1/N}(y)$ given by equation~\gammay\ is strictly increasing and therefore has an inverse. Using the notation above we have

\thm ztwo. Suppose that $b-2a>0$ then for fixed $k$,
$$
y_{k,N}= b+2a+{a^{1\over3}\over (N{d t_p^+\over dy}(b+2a,1/N))^{2\over3}}ai_k+O(\max{(1/N^{2\over\alpha},|b|/N^{1\over\alpha},1/N}))/N^{1/3}.\eq\eno[zeroonea]
$$
If $I\subset(0,b-2a)$ is a closed interval.  Then the zeros of $p_N(y,1/N)$ in $I$ are given by
$$
y=\Gamma^{-1}_{1/N}({2k+1\over \pi N}+o(1/N)).
$$

\pf  Equation~\zeroonea\ follows from \cothreeaa\ as above. To prove the second part note that in $I$, Ai
and Bi are nonzero. Furthermore in equation~\fouraa\ the term multiplying the $\sin$ is exponentially decreasing while the term multiplying the $\cos$ is exponentially increasing.  So that from \fouraa, a zero of $p_N(y,1/N)$, is given
by $\cos(N\Gamma_N(y))=-R_3(y,N)$, where $|R_3(y,N)|\le\max(1/N^{2\over\alpha},
|b|/N^{1\over\alpha},1/N)$.\qed

\advance\subsecno by 1\no=0\thmno=0
\beginsection VI. Model Coefficients

In order apply the above results to the special functions we are interested in 
we now consider in more detail special cases of recurrence
coefficients associated with $\hat q(t)=(t+s_1)^{1\over\alpha}, \alpha>0, s_1\ge0$ so that
$$
a(t,\ep)=a{(t/\ep+s_1)^{1\over\alpha}\over(1/\ep+s_1+1/2)^{1\over\alpha}}\quad{\rm and}\quad b(t,\ep)=b{(t/\ep+s_1+1/2)^{1\over\alpha}\over(1/\ep+s_1+1/2)^{1\over\alpha}}. \eq\eno[scoef]
$$
To make the connection with classical orthogonal polynomials the recurrence
coefficients given by equation~\ab\ can be further divided into the
following subcases,

\parindent=40pt
\item{case 1 } $b+2a>0, b-2a<0$\ (Asymmetric Freud,  Miexner-Pollaczek case),
\item{case 1a} $b=0$\ (symmetric Freud,  Hermite case),
\item{case 2 } $2a+b>0, b-2a=0$\ (Laguerre case),
\item{case 3 } $b+2a>0, b-2a>0$\ (Discrete Freud, Meixner case)
\parindent=20pt

\noindent The zero distribution for the above cases has been discussed by Kuijlaars and Van Assche (see [\cite{KVA}] and [\cite{KVAt}]).

In what follows we will always take $z^{\alpha}=e^{\alpha\ln z}$ where we
use the branch of the log that is analytic in a neighborhood of the
positive real axis and positive for large positive $z$. From Theorem~\thma\ we have the result,

\lem extfld.  With $\hat q(t)=(t+s_1)^{1\over\alpha}$ and $\ep$
sufficiently small we find,
$$
\eqalignno{
Q(y,\ep)&=\cases{A^+_1\ep(\hat
q({1\over\ep}+{1\over2}){y\over2a})^\alpha-\ep k(s_1,y,\ep)(1+r^+_1(y,\ep))& $0<x<\infty$\cr 
 A^-_1\ep(\hat q({1\over\ep}+{1\over2}){-y\over2a})^\alpha-\ep k(s_1,-y,\ep)(1+r^-_1(y,\ep))& $-\infty<x<0$}\quad {\rm for\ case\ 1} \cr
Q(y,\ep)&= A_2\ep(\hat q({1\over\ep}+{1\over2}){y\over2a})^\alpha-\ep
k(s_1,y,\ep)(1+r_2(y,\ep))\qquad\quad 0<x<\infty\qquad {\rm for\ case\ 2}\cr
Q(y,\ep)&= A_3\ep(\hat q({1\over\ep}+{1\over2}){y\over2a})^\alpha-\ep
k(s_1,y,\ep)(1+r_2(y,\ep))\qquad\quad 0<x<\infty\qquad {\rm for\ case\ 3}.}
$$
Here
$$
k(s_1,y,\ep)={1\over\alpha}\left(s_1+{1\over2}\right)+\ln\left({\hat q({1\over\ep}+{1\over2}) y\over a(s_1+{1\over2})^{1\over\alpha}}\right)^{s_1+{1\over2}},
$$
with
$$A_3=A^+_1=\alpha\int_0^{2a\over2a+b}u^{\alpha-1}\ln\left({1\over
u}-{b\over2a}+\sqrt{\left({1\over u}-{b\over2a}\right)^2-1}\,\right) du,$$\
$$A^-_1=\alpha\int_0^{2a\over2a-b}u^{\alpha-1}\ln\left({1\over
u}+{b\over2a}+\sqrt{\left({1\over u}+{b\over2a}\right)^2-1}\,\right) du,$$\
$$A_2=\alpha\int_0^{1\over2}u^{\alpha-1}\ln\left({1\over u}-1+\sqrt{{1\over
u}\left({1\over u}-2\right)}\,\right) du.$$ 
For $y>0$,  $r^+_1,\ r_2$ and $r_3$ have an analytic extensions 
to $\bbc\setminus(-\infty,0]$ and obey the bound
$$
|r^+_{1}(s_1,y,\ep)|\le C\max\{\ep^{2\over\alpha}, |b|\ep^{1\over\alpha}\},\qquad |r_i(s_1,y,\ep)|\le C \ep^{1\over\alpha} \quad i=2,3 \eq\eno[bbone]
$$ 
uniformly on compact subsets of $\bbc\setminus(-\infty,0]$. In the case of
1 with $y<0$, $r^-_1$ has an analytic extension to $\bbc\setminus[0,\infty)$ and obeys the bound
$$
|r^-_1(s_1,y,\ep)|\le C\max\{\ep^{2\over\alpha}, |b|\ep^{1\over\alpha}\}
$$
uniformly on compact subsets of $\bbc\setminus(-\infty,0]$.

\pf  From the second equation in \lnq\ we find for case 1 and $y>0$,
$$
\eqalign{
Q(y,\ep)=&A^+_1\ep
\left(\hat q\left({1\over\ep}+{1\over2}\right)
{y\over2a}\right)^\alpha\cr&
-\ep\alpha\hat q\left({1\over\ep}+{1\over2}\right)^{\alpha}
\int_0^{(s_1+1/2)^{1\over\alpha}\over\hat q({1\over\ep}+{1\over2})}
\ln\left({y\over2aw}-{b\over2a}+
\sqrt{\left({y\over2aw}-{b\over2a}\right)^2-1}\right)w^{\alpha-1}dw.}
$$
Since $y>0$ extracting $\ln({y\over aw})$ from the second term in the above
equation yields $k(s_1,y,\ep)$ and 
the remaining term can be expanded in a Taylor series to yield the bound
\bbone. 
The remaining cases can be derived in a similar manner. \qed

\remk When $b=0$, then
$A^+_1=A^+_1={\Gamma({\alpha\over2})\sqrt{\pi}\over2\Gamma({\alpha+1\over2})}$.

We also find that  Stirling's formula can be applied to $\kappa(n)$ in Lemma~\ratio\ Stirlings formula  to obtain,
$$
\kappa(n)={e^{(s_1+{1\over2})/\alpha}\Gamma(s_1+1)^{1\over\alpha}\over(2\pi)^{1\over2\alpha}(s_1+{1\over2})^{(s_1+{1\over2})/\alpha}}(1+O(\ep)).
$$
From formulas for $\rho$ given by equations \langer,\  \langero\ and \langert\ simplify. We find for case~1, 
$$
N^{2/3}\rho_1(1,y,1/N)=(N+s_1+1/2)^{2/3}\hat\rho_1(y),\ \eq\eno[rhoocaseone]
$$
where,
$$
\hat\rho_1(y)=\left\{\matrix{
({3\over2})^{2\over3}
\left(y\int_1^{y\over b+2a}{w^{\alpha-1} dw\over\sqrt{(y-bw)^2-4a^2 w^2}}-
\cosh^{-1}({y-b\over2a})\right)^{2/3} & b+2a\ge y\hfill \cr 
-({3\over2})^{2\over3}\left(\cos^{-1}({y-b\over2a})-
y\int^1_{y\over b+2a}{w^{\alpha-1}dw\over\sqrt{4a^2 w^2-(y-bw)^2}}\right)^{2/3}\cr
\hfill(b+2a)q_{1/N}(0)< y\le b+2a,}\right. \eq\eno[trhoocaseo]
$$
and
$$
N^{2/3}\rho_2(1,y,1/N)=(N+s_1+1/2)^{2/3}\hat\rho_2(y),\ \eq\eno[rhotcaseone]
$$
with,
$$
\hat\rho_2(y)=\left\{\matrix{({3\over2})^{2\over3}\left((-y)\int_1^{y\over b-2a}{w^{\alpha-1} dw\over\sqrt{(bw-y)^2-4a^2 w^2}}-\cosh^{-1}({b-y\over2a})\right)^{2/3} 
& y\le b-2a\hfill \cr 
-({3\over2})^{2\over3}\left(\cos^{-1}({b-y\over2a})
+y\int^1_{y\over b-2a}{w^{\alpha-1}dw\over\sqrt{4a^2 w^2-(bw-y)^2}}\right)^{2/3}
\hfill\cr\hfill b-2a \le y < (b-2a)q_{1/N}(0).}\right. \eq\eno[trhotcaseo]
$$
For case 2,
$$
N^{2/3}\rho_1(1,y,1/N)=(N+s_1+1/2)^{2/3}\hat\rho_1(y),\ \eq\eno[rhoocasetwo]
$$
where,
$$
\hat\rho_1(y)=\left\{\matrix{
({3\over2})^{2\over3}\left(y\int_1^{y\over 4a}{w^{\alpha-1} dw\over\sqrt{y(y-4aw) }}-\cosh^{-1}({y\over2a}-1)\right)^{2/3} 
&  4a\le y\hfill \cr 
-({3\over2})^{2\over3}\left(\cos^{-1}(1-{y\over2a})-y\int^1_{y\over 4a}{w^{\alpha-1}dw\over\sqrt{y(4aw-y)}}\right)^{2/3} &4aq_{1/N}(0)< y\le 4a.}\right. \quad\eq\eno[trhocaset]
$$
For case 3 and $N$ large enough so that $(b+2a)q_{1/N}(0)<b-2a$ we have
$$
N^{2/3}\rho_1(1,y,1/N)=(N+s_1+1/2)^{2/3}\hat\rho_1(y),\ \eq\eno[rhoocasethree]
$$
where,
$$
\hat\rho_1(y)=\left\{\matrix{
({3\over2})^{2\over3}\left(y\int_1^{y\over b+2a}{w^{\alpha-1} dw\over\sqrt{(y-bw)^2-4a^2 w^2}}-\cosh^{-1}({y-b\over2a})\right)^{2/3} 
& b+2a\le y \hfill\cr 
-({3\over2})^{2\over3}\left(\cos^{-1}({b-y\over2a})-y\int^1_{y\over b+2a}{w^{\alpha-1}dw\over\sqrt{4a^2 w^2-(y-bw)^2}}\right)^{2/3} 
& b-2a< y\le b+2a,}\right. \quad\eq\eno[trhoocaseth]
$$
and
$$
N^{2/3}\rho_2(1,y,1/N)=(N+s_1+1/2)^{2/3}\hat\rho_2(y),\ \eq\eno[rhotcasethree]
$$
with,
$$
\hat\rho_2(y)=\left\{\matrix{
-({3\over2})^{2\over3}\left(y\int^1_{y\over b-2a}{w^{\alpha-1}dw\over\sqrt{4a^2 w^2-(bw-y)^2}}-\cos^{-1}({b-y\over2a})\right)^{2/3} 
& b-2a\le y< b+2a\hfill \cr
({3\over2})^{2\over3}\left(\cosh^{-1}({b-y\over2a})-y\int^1_{y\over b-2a}{w^{\alpha-1} dw\over\sqrt{(bw-y)^2-4a^2 w^2}}\right)^{2/3}  \cr
\hfill (b-2a)q_{1/N}(0)< y \le b-2a.}\right.\quad\eq\eno[trhotcaseth]
$$

For case 3 the contraining measure has a particularly simple density [\cite{KVAt}] (see equation~\eqqmeathree) given by
$$
{d\sigma\over dy}={q(N+1/2)^{\alpha}\over N} c_{\alpha}y^{\alpha-1},
\quad  y>(b+2a)q_{1\over N}(0), \eq\eno[cons]
$$
where
$$
c_{\alpha}={\alpha\over\pi}\int_{b-2a}^{b+2a}{u^{-\alpha}du\over\sqrt{4a^2-(u-b)^2}}.\ \eq\eno[ca]
$$
Also
$$
N\Gamma_{q/N}(y)=(N+s_1+1/2)\pi c_{\alpha}y^{\alpha}-\pi s_1 \eq\eno[ga]
$$
\remk The above formulas can be extended into the complex plane
with the use of Theorem~\freudtur\ which also gives the relation between the equilibrium measure and $\rho$.

The above results allow us to recast  Theorems~\ftwoa\ and \ftwob.

\thm ftwoaa. Suppose $a(t,\ep)$ and $b(t,\ep)$ are given by equation~\scoef\ with $b-2a\le0$ . Let $a_1(t,\ep), b_1(t,\ep)$ satisfy equations~\abep\ and \abepw\ and set,
$$
\tilde\kappa={\Gamma(s_1+1)^{1\over\alpha} a^{s_1+1/2}\over(2\pi)^{1\over2\alpha}(N+s_1+1/2)^{1\over2\alpha}}.\eq\eno[tkap]
$$
For $y\in \bar\bbc_+\setminus[b-2a,b+2a]$ 
$$
\eqalign{
&((N+s_1+1/2)^{1\over\alpha} y)^{s_1}
p_N(y,1/N)=\cr&\tilde\kappa\kappa_1({1\over
(y-b)^2-4a^2})^{1\over4}e^{\alpha(N+s_1+1/2)\int_0^1 u^{\alpha-1} \ln(\tilde z_1(u,y)+\sqrt{(\tilde z_1(u,y)^2-1})du}(1+\phi_1(y,1/N)).} \eq\eno[asmextaa]
$$
where $\tilde z_1(u,y)={y\over2au}-{b\over2a}$.
The bound 
$$
|\phi_1(y,1/N)|\le \max(b/N^{1/\alpha},1/N^{2/\alpha},1/N), \eq\eno[ubaa]
$$
holds on compact subsets of $\bar\bbc_+\setminus[b-2a,b+2a]$. For $y\in L_1^+(0)$ given by Lemma~\anafreudpl\ and $N$ sufficiently large,
$$
\eqalign{
&{((N+s_1+1/2)^{1\over\alpha}y)^{s_1}p_N(y,1/N)\over e^{(N+s_1+1/2)A({y\over2a})^{\alpha}}}=\cr&
{2\sqrt{\pi}\tilde \kappa\kappa_1\over(N+s_1+1/2)^{{1\over2\alpha}-{1\over6}}}\left({\hat\rho_1(y)\over(y-b)^2-4a^2}\right)^{1/4}({\rm Ai}((N+s_1+1/2)^{2/3}\hat\rho_1(y))
+r_1(y,1/N))}.\eq\eno[cothreeaa]$$
where $\hat\rho_1$ is the extension of \trhoocaseo\  for $b-2a<0$ or \trhocaset\  if $b-2a=0$ and $A=A_1^+$ for $b-2a<0$ or $A_2$ for $b-2a=0$. The error term $r_1$ satisfies,
$$
|r_1(y,1/N)e^{{2\over3}(N+s_1+1/2)\hat\rho_1^{3\over2}(y,1/N)}|\le d\max(|b|\ep^{1\over\alpha},\ep^{2\over\alpha},\ep),
\eq\eno[erroroa]
$$
uniformly on compact subsets of $L_1^+(0)$. If $b-2a<0$ and $y\in L_2^+(0)$ given by Lemma~\anafreudpl\ then for $N$ sufficiently large,
$$
\eqalign{
&{(-1)^N((N+s_1+1/2)^{1\over\alpha}(-y))^{s_1}p_N(y,1/N)\over e^{(N+s_1+1/2)A_1^-(-{y\over2a})^{\alpha}}}=\cr&
{2\sqrt{\pi}\tilde \kappa\kappa_1\over(N+s_1+1/2)^{{1\over2\alpha}-{1\over6}}}\left({\hat\rho_2(y)\over(y+b)^2-4a^2}\right)^{1/4}({\rm Ai}((N+s_1+1/2)^{2/3}\hat\rho_2(y))
+r_2(y,1/N)).}\eq\eno[cofouraa]$$
where $\tilde\rho_2$ is the extension of \trhotcaseo. The error term $r_2$ satisfies the bound,
$$
|r_2(y,1/N)e^{{2\over3}(N+s_1+1/2)\hat\rho_2^{3\over2}(y,1/N)}|\le d\max(|b|\ep^{1\over\alpha},\ep^{2\over\alpha},\ep),
\eq\eno[errort]
$$
uniformly on compact subsets of $L_2^+(0)$.

Likewise for $b-2a>0$  we find,
\thm ftwob. Suppose $a(t,\ep)$ and $b(t,\ep)$ are given by equation~\scoef\ with $b-2a>0$ . Let $a_1(t,\ep), b_1(t,\ep)$ satisfy equations~\abep\ and \abepw. For $y\in \bar\bbc_+\setminus[0,b+2a]$, 
$$
\eqalign{
&((N+s_1+1/2)^{1\over\alpha} y)^{s_1}
p_N(y,1/N)=\cr&\tilde\kappa\kappa_1({1\over
(y-b)^2-4a^2})^{1\over4}e^{\alpha(N+s_1+1/2)\int_0^1 w^{\alpha-1} \ln(\tilde z_1(u,y)+\sqrt{(\tilde z_1(u,y)^2-1})du}(1+\phi_1(y,1/N)).} \eq\eno[asmextaa]
$$
where $\tilde z_1(u,y)={y\over2au}-{b\over2a}$ and $\tilde\kappa$ is given by \tkap. The bound, 
$$
|\phi_1(y,1/N)|\le \max(b/N^{1/\alpha},1/N^{2/\alpha},1/N), \eq\eno[ubaat]
$$
holds on compact subsets of $\bar\bbc_+\setminus[0,b+2a]$.
For $y\in L_1^+(0)$ given by Lemma~\anafreudpla\ and $N$ sufficiently large,
$$
\eqalign{
&{((N+s_1+1/2)^{1\over\alpha}y)^{s_1}p_N(y,1/N)\over e^{(N+s_1+1/2)({y\over2a})^{\alpha}}}=\cr&
{2\sqrt{\pi}\tilde \kappa\kappa_1\over(N+s_1+1/2)^{{1\over2\alpha}-{1\over6}}}\left({\hat\rho_1(y)\over(y-b)^2-4a^2}\right)^{1/4}({\rm Ai}((N+s_1+1/2)^{2/3}\hat\rho_1(y))
+r_1(1,y,1/N)).}\eq\eno[cothreeaa]$$
where $\hat\rho_1$ is the extension of \trhoocaseth. The error term $r_1$ satisfies \ubaat\ uniformly on compact subsets of $L_1^+(0)$. For $y\in(0,b+2a)$ 
$$
\eqalign{
&{((N+s_1+1/2)^{1\over\alpha}y)^{s_1}p_N(y,1/N)\over e^{(N+s_1+1/2)({y\over2a})^{\alpha}}} = 
(-1)^N{2\sqrt{\pi}\tilde \kappa\kappa_1\over(N+s_1+1/2)^{{1\over2\alpha}-{1\over6}}}\left({\hat\rho_2(y)\over4a^2-(y-b)^2}\right)^{1/4}\cr
&\quad \Bigl[(\sin\pi((N+s_1+1/2) y^{\alpha}c_{\alpha}-s_1)r
+R_1(y,1/N))\cr
&\qquad\qquad\times ({\rm Ai}((N+s_1+1/2)^{2/3}\hat\rho_2(y)) +r_1(y,1/N))\cr
&\quad +(\cos\pi((N+s_1+1/2) y^{\alpha}c_{\alpha}-s_1)+R_2(y,1/N))\cr
&\qquad\qquad\times(
{\rm Bi}((N+s_1+1/2)^{2/3}\hat\rho_2(y))+ r_2(y,1/N))\Bigr],}\eq\eno[cofiveaa]$$
where uniformly on compact subintervals of $(0,b+2a)$ 
$$
\eqalign{&
|\hat r_1(y,1/N)e^{{2\over3}(N+s_1+1/2)\hat\rho_2^{3\over2}(y,1/N)}|,\  |\hat r_2(y,1/N)e^{-{2\over3}(N+s_1+1/2)\hat\rho_2^{3\over2}(y,1/N)}|,\cr&|R_1(y,1/N)|,\|R_2(y,1/N)|\le d
 \max(|b|/N^{1\over\alpha},1/N^{2\over\alpha},1/N).}\eq\eno[errorta] $$

\remk In the above cases ${dt_p^+\over dy}(b+2a,1/N)={\alpha(N+s_1+1/2)\over N(b+2a)}$ so that the formulas for the zero's given in Theorems~\zone\ and \ztwo\   simplify.

\advance\subsecno by 1\no=0\thmno=0\exmno=0
\beginsection VII. Applications

\exm exone. Hermite polynomials, $b=0$

We consider well known classes of orthonormal polynomials associated with the various  cases given in the previous section. 
We begin with the orthonormal Hermite polynomials $\{\hat H_n(x)\}$ 
which have 
$$w(x)=e^{-x^2}$$ 
as their weight function
(\cite{O}, \cite{S}). In this case $b_n=0$, $n\ge0$, $a_n=\sqrt{n\over2}$,
$ n\ge1$ and  $\hat q_{n}=\sqrt{n}$ so that $a(t,1/N) = \sqrt{Nt/2\over N+1/2}$. In Theorems~\cthmnextone\ and \thminit\
we take $a_1(n)=a(n)$,
$b_1(n)=0$ so that $\kappa_1=1$ and $A_1=1$. Since $p_0(y)=1$, $\hat H_n={p_n\over\pi^{1/4}}$. In order to compare the previous results with the literature we make the scaling $y\to\sqrt{2}y$, $\tilde\rho_1(y)={1\over(\sqrt{2})^{2\over3}}\hat\rho_1(y)$ and $\lam_N=\sqrt{2N+1}$ so that $x=\lam_N y$.
Thus
$$
\tilde\rho_1(y)=\left\{\matrix{\left({3\over 4}\left(y\sqrt{y^2-1}-\cosh^{-1}y\right)\right)^{2/3} & \hbox{\rm for } 1\le y\hfill\cr
\left({3\over 4}\left(\cos^{-1}y-y\sqrt{1-y^2}\right)\right)^{2/3} 
&\hbox{\rm for }q_{1\over N}(0)< y\le1.}\right.
$$
Since $s_1=0$, $b=0$, $\alpha=2$ and $A_1^{\pm}=1$ we find uniformly on compact subsets of $\bbc\setminus[-1,1]$
$${\hat H_N(\lam_N y)\over e^{(\lam_N y)^2/2}} = 
{\left(y+\sqrt{y^2-1}\right)^{\lambda^2_N/2}\over
\sqrt{2\lam_N\pi}(y^2-1)^{1/4}}
e^{-{\lambda^2_N\over 2}y\sqrt{y^2-1}}
\left(1+O(1/N)\right).\eq\eno[hoasym]$$
Also  on   $L^+_1(0)$
$${\hat H_N(\lam_N y)\over e^{(\lam_N y)^2/2}} = 
{\sqrt 2\over \lam_N^{1/6}}
\left(\tilde\rho_1(y)\over\sqrt{y^2-1}\right)^{1/4}\left({\rm Ai}\left(
\lam_N^{4/3}\tilde\rho_1(y)\right)+r_1(y,1/N)\right), \eq\eno[hro]$$
where uniformly on compact subsets of $L_1^+(0)$,
$$
|e^{{2\over3}\lambda_N^2\tilde\rho^{3\over2}_1(y)}r_1(y,1/N)|<d/N.
$$
On $L^+_2(0)$
$${\hat H_N(\lam_N y)\over e^{(\lam_N y)^2/2}} = (-1)^N
{\sqrt 2\over \lam_N^{1/6}}
\left(\tilde\rho_1(-y)\over\sqrt{y^2-1}\right)^{1/4}\left({\rm Ai}\left(
\lam_N^{4/3}\tilde\rho_1(-y)\right)+r_2(y,1/N)\right).\ \eq\eno[hrt]$$
where uniformly on compact subsets of $L^+_1(0)$,
$$
|e^{{2\over3}\lambda_N^2\tilde\rho^{3\over2}_1(-y)}r_2(y,1/N)|<d/N.
$$
Formula~\hoasym\ was first derived from the point of view of difference equations by [\cite{VG}] and formula~\hro\ by  [\cite{GBVA}]. For $y>0$ this formula is  found in Olver [\cite{O}, p.~493] (see also Szego [\cite{S}, Thm~8.22b]).  
From Theorem~\zone\ we find that the zeros of $\hat H_N(\lam_N y)$ are given by,
$$
y_{k,N}=1+{ai_k\over2^{1\over2}\lam_N^{4\over3}}+O(1/N^{4\over3}),\ y_{N-k,N}=-y_{k,N}.
$$
(See also Szego [\cite{S}, p.~132].)

\exm extwo.  Meixner-Pollaczek polynomials, $b+2a>0$, $b-2a <0$

The  Meixner-Pollaczek (Meixner polynomials  of the second kind) $\hat M_n(x,\delta,\eta)$ are orthonormal with respect to the weight function 
$$
w(x)=e^{x\tan^{-1}\delta}\Bigl|\Gamma\left({\eta+ix\over2}\right)\Bigr|^2.
$$
Since $p_0(x)=1$ we find that [\cite{KS}] 
$$
\hat M_n(x,\delta,\eta)={p_n(x)\over\sqrt{2\pi\Gamma(\eta)}({a\over2})^{\eta\over2}}.
$$
The coefficients in the recurrence formula for the 
orthonormal Meixner-Pollaczek polynomials  
[\cite{C}], 
$\hat M_n(x:\delta,\eta)$ are given by,
$$
a_1(n)=\sqrt{\delta^2+1}\sqrt{n(n+\eta-1)}, \eta>0, \delta\ge0 ,
$$
and
$$
b_1(n)=\left(n+{\eta\over2}\right)2\delta
$$
In this case we will use a comparison system whose coefficients are given by,
$$
a(n)=\sqrt{\delta^2+1}\left(n+{\eta-1\over2}\right),\quad
 \hbox{ and } b(n)=b_1(n),
$$ 
so that
$$
\kappa_1={\sqrt{\Gamma(\eta)}\over\Gamma({\eta+1\over2})}.
$$
With $\hat q(t)=t+{\eta-1\over2}$ we find $s_1={\eta-1\over2}$, $a(t,\ep)=a{q({t\over\ep})\over q({1\over\ep}+{1\over2})}$,
$b(t,\ep)=b{\hat q({t\over\ep}+{1\over2})\over\hat q({1\over\ep}+{1\over2})}=b_1(t,\ep)$ and $a_1(t,\ep)=a{\sqrt{{t\over\ep}({t\over\ep}+\eta-1)}\over\hat q({1\over\ep}+{1\over2})}$ where $a=\sqrt{\delta^2+1}$ and $b=2\delta$. For $t$ strictly greater than zero $|a_1(t,\ep)-a(t,\ep)|=O(\ep^2)$ and
$\sup_{t\in[0,t_{fi}]}|a^2_1(t,\ep)-a^2(t,\ep)|=O(\ep^2)$. Thus equations~\abep\ and \abepw\ are satisfied. With  the scalings $y\to2y$, $\tilde \rho=({1\over2})^{2\over3}\hat\rho$, $\lam_N=2N+\eta$ and $x=\lam_N y$
we find,
$$
\eqalign{
\int_0^1 {du\over\sqrt{y^2-2\delta yu-u^2}}&=i\ln
\left({y(i\delta+1)\over(\delta y+1)i+\sqrt{y^2-2\delta y-1}}\right)\cr
&=2\tan^{-1}\left({1\over y+\sqrt{y^2-2\delta y-1}}\right),}\ \eq\eno[alt]
$$
so that,
$$
\eqalign{
&\int_0^1\ln(\tilde z_1(y,u)+\sqrt{\tilde z_1(y,u)^2-1})du\cr
&=\cosh^{-1}\left({y-\delta\over a}\right)+
iy\ln\left({y(i\delta+1)\over(\delta y+1)i+\sqrt{y^2-2\delta y-1}}\right).} 
$$
Also
$$
 \tilde\rho_1(y)=\left\{\matrix{
({3\over4}y\cos^{-1}({\delta\over a}+{1\over ay})-{3\over4}\cosh^{-1}({y-\delta\over a}))^{2/3} 
& \hbox{for } 1\le {y\over\delta+\sqrt{\delta^2+1}}\hfill\cr 
-({3\over4}\cos^{-1}({y-\delta\over a})-{3\over4}y\cosh^{-1}({\delta\over a}+{2\over ay}))^{2/3} 
&\hbox{for } q_{1/N}(0)<{y\over\delta+\sqrt{\delta^2+1}}\le 1}\right.
$$
which can be extended to $L_1^+(0)$. We also find that $A_1^+=a\cot^{-1}(\delta)$. In a similar manner
$A_1^-=a(\pi-\cot^{-1}(\delta))$ and
$$\tilde\rho_2(y)=\left\{\matrix{
({3\over4}(-y)\cos^{-1}(-{\delta\over a}-{1\over ay})-{3\over4}\cosh^{-1}({\delta-y\over a}))^{2/3} 
&\hbox{for }{y\over\delta-\sqrt{\delta^2+1}}\le 1\hfill\cr 
-({3\over4}\cos^{-1}({\delta-y\over a})-{3\over4}(-y)\cosh^{-1}(-{\delta\over a}-{1\over ay}))^{2/3} 
&\hbox{for } q_{1/N}(0)< {y\over\delta-\sqrt{\delta^2+1}}\le 1,}\right.
$$
which can be extended to $L^+_2(0)$. Thus uniformly on compact subsets of $\bbc\setminus[\delta-a,\delta+a]$,
$$
\eqalign{
&(\lam_N y)^{\eta-1\over2}\hat M_N(\lam_N y:\delta,\eta){a^{\lam_N\over2}\over(y-\delta+\sqrt{y^2-2\delta y-1})^{\lam_N\over2}}\cr&={2^{\eta-1\over2}\over\pi(\lam_N)^{1\over2}(y^2-2\delta y-1)^{1\over4}}\left({y(i\delta+1)\over(\delta y+1)i+\sqrt{y^2-2\delta y-4}}\right)^{iy\lam_N\over2}(1+0(1/N)).}\eq\eno[mpoutasym]
$$
On $L^+_1(0)$ 
$$
\eqalign{
&{(\lam_N y)^{\eta-1\over2}\hat M_N(\lam_N y:\delta,\eta)\over
e^{\cot^{-1}(\delta)\lam_N y/2}}\cr&=
{2^{\eta\over2}\over\sqrt{\pi}\lam_N^{1/3}}\left({\tilde\rho_1(y)\over 
y^2-2\delta y-1}\right)^{1/4}(
{\rm Ai}(\lam_N^{2\over3}\tilde\rho_1(y))+r_1(y,1/N)),}\ \eq\eno[mprhoo]$$
where uniformly on compact subsets of $L_+^1(0)$,
$$
|e^{{2\over3}\lambda_N\tilde\rho^{3\over2}_1(y)}r_1(y,1/N)|<d/N.
$$
On $L^+_2(0)$ we have,
$$
\eqalign{
&{(\lam_N(-y))^{\eta-1\over 2}\hat M_N(\lam_N y:\delta,\eta)\over
e^{-(\pi-\cot^{-1}(\delta))\lam_N y/2}}\cr&={2^{\eta\over2}\over\sqrt{\pi}\lam_N^{1/3}}\left({\tilde\rho_1(y)\over 
y^2-2\delta y-1}\right)^{1/4}
({\rm Ai}(\lam_N^{2/3}\tilde\rho_2(y))+r_2(y,1/N)),}\ \eq\eno[mprhot]$$
where uniformly on compact subsets of $L^+_2(0)$,
$$
|e^{{2\over3}\lambda_N^2\tilde\rho^{3\over2}_2(y)}r_2(y,1/N)|<d/N.
$$
From Theorem~\zone\ we have for $k$ fixed,
$$
y_{k,N}=\delta+\sqrt{\delta^2+1}+
\left({\delta+\sqrt{\delta^2+1}\over\lam_N}\right)^{2\over3}
\left({\sqrt{\delta^2+1}\over2}\right)^{1\over3}
ai_k+O(1/N^{4\over3}),\ \eq\eno[mpzo]
$$
and
$$
y_{N-k,N}=\delta-\sqrt{\delta^2+1}-
\left({\delta-\sqrt{\delta^2+1}\over\lam_N}\right)^{2\over3}
\left({\sqrt{\delta^2+1}\over2}\right)^{1\over3}ai_k+O(1/N^{4\over3}).\ \eq\eno[mpzt]
$$
Equation~\mpoutasym\ was first obtained by Geronimo and Van Assche [\cite{GVA}] with a slightly different scaling. With the use of equation~\alt\ this is similar to the asymptotics obtained by Li and Wong [\cite{LW}] equations~(6.9) and (6.10)  with the scaling $\lam_N y=2N\tilde y$. Equations~\mprhoo\ and \mprhot\ agree  with
equations~(6.23) and (6.24) of [\cite{LW}] when the above scaling is taken into account. The location of the zeros given by equations~\mpzo\ and \mpzt\ should be compared with equations~(7.17) and (7.18) of [\cite{LW}] (see also Chen and Ismail~[\cite{CI}]).

\exm exthree.  Laguerre polynomials,  $b-2a=0$

The orthonormal Laguerre polynomials $\hat L^{\alpha}_n(x)$ have weight function 
$$w(x)=x^{\alpha}e^{-x},\quad x\ge0,\quad  \alpha>-1,
$$ 
with recurrence coefficients,
$$
a_1(n)=\sqrt{n(n+\alpha)},\ {\rm and}\ b_1(n)=2\left(n+{\alpha+1\over 2}\right)
$$
Thus $\hat L^{\alpha}_n(x)={p_n(x)\over\sqrt{\Gamma(\alpha+1)}}$. As a comparison system we will use,
$$
a(n)=n+{\alpha\over2}, b(n)=b_n.
$$
With $\hat q(t)=t+{\alpha\over2}$ we find $s_1={\alpha\over2}$, $a(t,\ep)={q({t\over\ep})\over q({1\over\ep}+{1\over2})}$,
$b(t,\ep)=b{\hat q({t\over\ep}+{1\over2})\over\hat q({1\over\ep}+{1\over2})}=b_1(t,\ep)$ and $a_1(t,\ep)={\sqrt{{t\over\ep}({t\over\ep}+\alpha)}\over\hat q({1\over\ep}+{1\over2})}$ where $b=2$. For $t$ strictly greater than zero $|a_1(t,\ep)-a(t,\ep)|=O(\ep^2)$ and
$\sup_{t\in[0,t_{fi}]}|a^2_1(t,\ep)-a^2(t,\ep)|=O(\ep^2)$. Thus equations~\abep\ and \abepw\ are satisfied. To move the oscillatory interval to $[0,1]$ we scale $y\to4y$, and set $\hat\rho=4^{-{2\over3}}\tilde\rho$,  $\lam_N=4N+2\alpha+2$ and $x=\lam_N y$. In this case $A_2=1$,
$$
\kappa_1={\sqrt{\Gamma(\alpha+1)}\over\Gamma({\alpha\over2}+1)},
$$
and
$$
 \tilde\rho_1(y)=\left\{\matrix{({3\over4}(\sqrt{y^2-y}-{1\over2}\cosh^{-1}(2y-1)))^{2/3} & {\rm for}\quad 1\le y\cr -({3\over4}({1\over2}\cos^{-1}(2y-1)-\sqrt{y-y^2}))^{2/3} &\quad\quad {\rm for}\quad 0<y\le1,}\right.
$$
which can be extended to $L^+_1(0)$.
Thus uniformly on compact subsets of $\bbc\setminus[0,1]$
$$
{(\lam_N y)^{\alpha\over2}\hat L^{\alpha}_N\over e^{\lam_Ny\over2}}={1\over\sqrt{2\pi\lam_N}(y^2-y)^{1\over4}}\left({2y-1+2\sqrt{y^2-y}\over e^{2\sqrt{y^2-y}}}\right)^{\lam_N\over4}(1+O(1/N)). \ \eq\eno[loasym]
$$
On  $L^+_1(0)$ we have
$$
{(\lam_N y)^{\alpha\over2}\hat L_N^{\alpha}(\lam_N y)\over e^{\lam_N
y\over2}}={\sqrt{2}\over\lam_N^{1\over3}}\left({\tilde\rho_1(y)\over y^2-y}\right)^{1\over4}({\rm Ai}(\lam_N^{2\over3}\tilde\rho_1(y))+r_1(y,1/N)),\ \eq\eno[lro]
$$
where uniformly on compact subsets of $L^+_1(0)$,
$$
|e^{{2\over3}\lambda_N\tilde\rho^{3\over2}_1(y)}r_1(y,1/N)|<d/N.
$$
Starting from the recurrence formula equation~\loasym\ with a different scaling was first given in [\cite{GVA}]. 
Equation~\loasym\ can be found  in Szego [\cite{S}, Thm~8.22.8b] for $y>1$. See also Frenzen and Wong [\cite{FW}].  We have for the zeros of $\hat L_N^{\alpha}(\lam_N y)$,
$$
y_{k,N}=1+ {4^{1\over3}\over\lam_N^{2\over3}}ai_k+O(1/N^{4\over3}).
$$
(See  [\cite{S}, p.~132].)

\exm exfour.  Meixner polynomials,  $b-2a>0$

For the orthonormal Meixner polynomials $\hat m_n(x:\beta,c)$ have an atomic measure with masses
$$
w(x)={(\beta)_x\over x!}c^x, \quad x=0,1,\ldots,
$$
and recurrence coefficients
$$
a_1(n)={\sqrt{c}\over1-c}\sqrt{n(n+\beta-1)},\hbox{ and }  
b_1(n)={1+c\over1-c}\left(n+{c\over1+c}\beta\right), \quad 0<c<1,\quad 0<\beta.
$$
Thus 
$$\hat m_n(x:\beta,c)={p_n(x)\over(1-c)^{\beta\over2}}.
$$
We apply the scaling $x\to \hat q_{\ep} y-\lam_1$, where $\lam_1={\beta\over2}$ and $\hat q_{\ep}$ is given below.
The shifted Meixner polynomials $\tilde m_n(\tilde x:\beta,c)$ satisfy the recurrence formula 
$$\tilde a_1(n+1)\tilde m_{n+1}(\tilde x)+(\tilde b_1(n) -\tilde x
)\tilde m_{n}(\tilde x)+\tilde a_1(n)\tilde
m_{n-1}(\tilde x)=0
$$
where $\tilde a_1(n)$ are as above and $\hat b_1(n)={1+c\over1-c}(n+{\beta\over2})$.

The comparison system is
$$
a(n)={\sqrt{c}\over1-c}\left(n+{\beta-1\over2}\right), \quad
\hbox{ and } b(n)=\tilde b_1(n).
$$
Thus 
$$\kappa_1={\Gamma(\beta)^{1/2}\over\Gamma({\beta+1\over2})}.
$$
With $\hat q(t)=t+s_1, s_1={\beta-1\over2}$  
$a(t,\ep)=a{q({t\over\ep})\over q({1\over\ep}+{1\over2})}$,
$b(t,\ep)=b{\hat q({t\over\ep}+{1\over2})\over\hat q({1\over\ep}+{1\over2})}=\tilde b_1(t,\ep)$ and $a_1(t,\ep)=a{\sqrt{{t\over\ep}({t\over\ep}+\beta-1)}\over\hat q({1\over\ep}+{1\over2})}$ where $a={\sqrt{c}\over1-c}$ and $b={1+c\over1-c}$ we see that ${b\over2a}>1$. For $t$ strictly greater than zero $|a_1(t,\ep)-a(t,\ep)|=O(\ep^2)$ and
$\sup_{t\in[0,t_{fi}]}|a^2_1(t,\ep)-a^2(t,\ep)|=O(\ep^2)$ so equations~\abep\ and \abepw\ are satisfied. In this case
$$
\hat\rho_1(y)=\left\{\matrix{
({3\over2})^{2\over3}(y\cosh^{-1}({b\over 2a}-{1\over2ay})-\cosh^{-1}({y-b\over 2a}))^{2/3} & \hbox{for } b+2a\le y\hfill \cr 
-({3\over2})^{2\over3}(\cos^{-1}({y-b\over2a})-y\cos^{-1}({b\over2a}-{1\over2ay}))^{2/3} &\hbox{for } b-2a<y\le b+2a,}\right. 
$$
and
$$
\hat\rho_2(y)=\left\{\matrix{
-({3\over2})^{2\over3}(y\cos^{-1}({1\over2ay}-{b\over 2a})-\cos^{-1}({b-y\over2a}))^{2/3} & \hbox{for } b-2a\le y<b+2a\cr 
({3\over2})^{2\over3}(\cosh^{-1}({b-y\over 2a})-y\cosh^{-1}({1\over2ay}-{b\over2a}))^{2/3} &\hbox{for } 0<y\le b-2a.\hfill}\right. 
$$
We also find that  $A_3=2a\ln({b+1\over2a})$ and from \ca\ $c_1=1$.
With $\lam_N=N+\beta/2$, we find uniformly on compact subsets of $\bbc\setminus[0,b+2a]$
$$
\eqalign{
&(\lam_N y)^{\beta-1\over2}{\hat m_N(\lam_N y-{\beta\over2}:\beta,c)\over({b+1\over2a})^{\lam_N y}}\cr&={\Gamma(\beta)^{1\over2} c^{\beta\over4}((y-b+\sqrt{y^2-2by+1})/2a)^{\lam_N}\over\sqrt{2\pi\lam_N}(y^2-2by+1)^{1\over4}}\left({2ay\over  by-1+\sqrt{y^2-2by+1}}\right)^{\lam_N y}(1+O(1/N)).} \eq\eno[moasym]
$$
On $(b-2a,\infty)$,
$$
{(\lam_N y)^{\beta-1\over2}\hat m_N(\lam_N y-{\beta\over2}:\beta,c)\over ({b+1\over2a})^{\lam_N y}}={\sqrt{2}\Gamma(\beta)^{1\over2} c^{\beta\over4}\over(\lam_N)^{1\over3}}\left({\hat\rho_1(y)\over y^2-2by+1}\right)^{1\over4}({\rm Ai}(\lam_N^{2\over3}\hat\rho_1(y))+r_1(y,1/N)),
$$
where uniformly on closed intervals of $(b-2a,\infty)$,
$$
|e^{{2\over3}\lambda_N\hat\rho^{3\over2}_1(y)}r_1(y,1/N)|<d/N.
$$
On $(0,b+2a)$,
$$
\eqalign{
&{(\lam_N y)^{\beta-1\over2}\hat m_N(\lam_N y-{\beta\over2}:\beta,c)\over ({b+1\over2a})^{\lam_N y}}=(-1)^N{\sqrt{2}\Gamma(\beta)^{1\over2} c^{\beta\over4}\over(\lam_N)^{1\over3}}\left({\hat\rho_1(y)\over y^2-2by+1}\right)^{1\over4}\cr
&\Biggl[\cos\left(\pi\left(\lam_N y-{\beta\over2}\right)
+R_1(y,1/N)\right)({\rm Ai}((\lam_N^{2/3}\hat\rho_2(y))+r_1(y,1/N))\cr
&+\left(\sin\pi\left(\lam_N y-{\beta\over2}\right)+R_2(y,1/N)\right)(
{\rm Bi}(\lam_N^{2/3}\hat\rho_2(y))+ r_2(y,1/N))\Biggr],}
$$
where uniformly on closed intervals of $(0,b+2a)$,
$$
|e^{{2\over3}\lambda_N\hat\rho^{3\over2}_1(y)}r_1(y,1/N)|,\ 
|e^{-{2\over3}\lambda_N\hat\rho^{3\over2}_2(y)}r_2(y,1/N)|,\ |R_1(y,1/N)|,\ |R_2(y,1/N)|<d/N.
$$
From Theorem~\ztwo\ we find
$$
y_{k,N}={1+c+2\sqrt{c}\over1-c}+ {\sqrt{c}^{1\over3}\over1-c}{1+c+2\sqrt{c}\over\lam_N^{2\over3}}ai_k+O(1/N^{4\over3}).
$$
and 
$$
y_{N-k,k}={(2k-1)\pi+\beta\over2\lam_N}.
$$

\exm exfive.  Continuous dual Hahn polynomials $b-2a=0$

In order to keep the exposition contained we will assume the parameters $a_1, b_1,$ and $c_1$ are positive and $a_1+b_1+c_1>3/2$. In this case
the normalized continuous daul Hahn polynomials $\hat h_n(x; a_1,b_1,c_1)$ are orthonormal with respect to the weight [\cite{KS}],
$$
w(x)={1\over2\pi}\left|\Gamma(a_1+ix)\Gamma(b_1+ix)\Gamma(c_1+ix)\over\Gamma(2ix)\right|^2,\ x>0.
$$
If
$$
A_n=(n+a_1+b_1)(n+a_1+c_1)\ {\rm and}\ C_n=n(n+b_1+c_1-1),
$$
then the recurrence coefficients are
$$
b_1(n)=A_n+C_n-a_1^2= 2
\left(n^2+(a_1+b_1+c_1-1/2)n+{a_1b_1+a_1c_1+c_1b_1\over2}\right),
$$
and
$$
a^2_1(n)=A_{n-1}C_n=n(n+a_1+b_1-1)(n+a_1+c_1-1)(n+c_1+b_1-1).
$$
In this case
$$
\hat h_n(x; a_1,b_1,c_1)={p_n(x)\over\sqrt{\Gamma(a_1+b_1)\Gamma(a_1+c_1)\Gamma(b_1+c_1)}}.
$$
We chose the comparison system with recurrence coefficients,
$$
a(n)=\hat q(n)\ {\rm and}\ b(n)=2\hat q(n+1/2),
$$
where
$$
\hat q(n)=(n+(a_1+b_1+c_1-3/2)/2)^2.
$$
Thus
$$
\kappa_1={\sqrt{\Gamma(a_1+b_1)\Gamma(a_1+c_1)\Gamma(b_1+c_1)}\over\Gamma({2a_1+2b_1+2c_1+1\over4})^2}.
$$
With $\hat q(t)$ as above $s_1={2a_1+2b_1+2c_1-3\over4}$, $A_2={\pi\over\sqrt{2}}$, $a(t,\ep)={q({t\over\ep},\ep)\over q({1\over\ep}+{1\over2})}$, $a_1(t,\ep)={a_1({t\over\ep})\over q({1\over\ep}+{1\over2})}$, $b(t,\ep)=2{q({t\over\ep},\ep)\over q({1\over\ep}+{1\over2})}$ and $b_1(t,\ep)={b_1({t\over\ep})\over q({1\over\ep}+{1\over2})}$ we find 
for $t$ strictly greater than zero $|a_1(t,\ep)-a(t,\ep)|=O(\ep^2)$ and $\sup_{t\in[0,t_{fi}]}|a_1^2(t,\ep)-a^2(t,\ep)|=O(\ep^2)=\sup_{t\in[0,t_{fi}]}|b_1(t,\ep)-b(t,\ep)|$ so equations~\abep\ and \abepw\ are satisfied. We make the scaling $y\to 4 y$  and set $\lambda_N=(2N+a_1+b_1+c_1-1/2)^2$. For $y\in C\setminus[0,1]$, we find that
$$
\int_0^1{dw\over\sqrt{wy-w^2}}=
\left(\pi-\cos^{-1}\left({2\over y}-1\right)\right)=
{1\over i}\ln\left(1-{2\over y}+i2\sqrt{{1\over y}\left(1-{1\over y}\right)}\right),
$$
so that
$$
\eqalign{
&\int_0^1 u^{-{1\over2}}\ln\left({2y\over u}-1
+\sqrt{\left({2y\over u}-1\right)^2-1}\,\right)\cr
&=2\ln\left(2y-1+2\sqrt{y(y-1)}\,\right)+{2\sqrt{y}\over i}
\ln\left(1-{2\over y}+i2\sqrt{{1\over y}\left(1-{1\over y}\right)}\right).}\eq\eno[dhe]
$$
Also 
$$
 \hat\rho_1(y)=\left\{\matrix{
{3\over2}^{2\over3}(\sqrt{y}\cos^{-1}({2\over y}-1)-\cosh^{-1}(2y-1))^{2/3} 
&\hbox{for } 1\le y\hfill\cr 
-{3\over2}^{2\over3}(\cos^{-1}(1-2y)-\sqrt{y}\cosh^{-1}({2\over y}-1))^{2/3} 
&\hbox{for } q_{1\over N}(0)<y\le1,}\right.  \eq\eno[dhrho]
$$
which can be extended to $L_1^+(0)$.
Thus uniformly on compact subsets of $\bbc\setminus[0,1]$,
$$
\eqalign{
&(\lambda_N y)^{({2a_1+2b_1+2c_1-3\over4})}\hat h_n(\lambda_N y;a_1,b_1,c_1)=\cr
&{\left(2y-1+2\sqrt{y(y-1)}\,\right)^{\lambda_N^{1/2}\over2}
\left(1-{2\over y}+i2\sqrt{{1\over y}(1-{1\over y}})\right)^{\sqrt{\lambda_N y}\over 2i}\over\sqrt{2}\pi\lambda_N^{1\over2}(y(y-1))^{1\over4}}(1+O(1/N)).}
$$
For $y\in L^+_1(0)$ we have,
$$
\eqalign{
&{(\lambda_N y)^{({2a_1+2b_1+2c_1-3\over4})}\hat h_n(\lambda_N y;a_1,b_1,c_1)\over e^{\pi({\lambda_N y\over2})^{1\over2}}}=\cr&\sqrt{2\over\pi}{1\over\lambda_N^{5\over12}}\left({\tilde\rho_1(y)\over y(y-1)}\right)^{1\over4}({\rm Ai}(\lambda_N^{1\over3}\tilde\rho_1(y))+r_1(y,1/N)),
}
$$ 
where $\tilde\rho_1={1\over2^{2/3}}\hat\rho_1$.
Uniformly on compact subsets of $L_1^+(0)$
$$
|e^{{2\over3}\lambda_N^{1\over2}\tilde\rho^{3\over2}_1(y)} r_1(y,1/N)|\le d/N.
$$
From Theorem~\zone\ we find
$$
y_{k,N}=1+{2^{2\over3}ai_k\over\lambda_N^{1\over3}}+O(1/N^{4\over3}).
$$

\exm exsix.  Wilson polynomials $b-2a=0$

The Wilson polynomials are the most general hypergeometric polynomials in the Askey scheme which do not have a purely discreet orthogonality measure. We denote the orthonormal Wilson polynomials by $\hat W_n(x^2:a_1,b_1,c_1,d_1)$ and for simplicity we will assume that $\Re(a_1,b_1,c_1,d_1)>0$, non-real parameters occur in conjugate pairs and $a_1+b_1+c_1+d_1>1$. These polynomials are orthonormal [\cite{KS}] with respect to the weight
$$
w(x)=\left|{\Gamma(a_1+ix)\Gamma(b_1+ix)\Gamma(c_1+ix)\Gamma(d_1+ix)\over\Gamma(2ix)}\right|^2, \ x>0.
$$
With
$$
A_n={(n+a_1+b_1+c_1+d_1-1)(n+a_1+b_1)(n+a_1+c_1)(n+a_1+d_1)\over(2n+a_1+b_1+c+1+d_1-1)(2n+a_1+b_1+c_1+d_1)},
$$
and
$$C_n={n(n+b_1+c_1-1)(n+b_1+d_1-1)(n+c_1+d_1-1)\over(2n+a_1+b_1+c_1+d_1-1)(2n+a_1+b_1+c_1+d_1-2)},
$$
we have
$$
a_1(n)=\sqrt{A_{n-1}C_{n}},\ {\rm and}\ b_1(n)=A_n+C_n-a_1^2.
$$
In this case,
$$
\hat W(x^2:a_1,b_1,c_1,d_1)=\sqrt{\Gamma(a_1+b_1+c_1+d_1)\over\Gamma(a_1+b_1)\Gamma(a_1+c_1)\cdots\Gamma(c_1+d_1)}p_n(x).
$$
We chose  the comparison system to have recurrence coefficients
$$
a(n)={1\over4}\hat q(n)\quad\hbox{ and }\quad b(n)={1\over2}\hat q(n+1/2),
$$
where
$$
q(n)=\left(n+{a_1+b_1+c_1+d_1-2\over2}\right)^2.
$$ 
This comparison system is very similar to the one used for the continuous dual Hahn polynomials so those
results can be used here. With $\hat q(t)$ as above $s_1={a_1+b_1+c_1+d_1-2\over2}$, $A_2={\pi\over\sqrt{2}}$, $a(t,\ep)={{1\over4}q({t\over\ep},\ep)\over q({1\over\ep}+{1\over2})}$, $a_1(t,\ep)={a_1({t\over\ep})\over q({1\over\ep}+{1\over2})}$, $b(t,\ep)={1\over2}{q({t\over\ep},\ep)\over q({1\over\ep}+{1\over2})}$ and $b_1(t,\ep)={b_1({t\over\ep})\over q({1\over\ep}+{1\over2})}$ we find 
for $t$ strictly greater than zero $|a_1(t,\ep)-a(t,\ep)|=O(\ep^2)$ and $\sup_{t\in[0,t_{fi}]}|a_1^2(t,\ep)-a^2(t,\ep)|=O(\ep^2)=\sup_{t\in[0,t_{fi}]}|b_1(t,\ep)-b(t,\ep)|$ so equations~\abep\ and \abepw\ are satisfied. We set $\lambda_N=(N+{a_1+b_1+c_1+d_1-1\over2})^2$.
Using equation~\dhe\ we find that uniformly on compact subsets of $\bbc\setminus[0,1]$,
$$
\eqalign{
&(\lambda_N y)^{({a_1+b_1+c_1+d_1-2\over2})}\hat W_n((\lambda_N y)^2;a_1,b_1,c_1,d_1)=\cr&{(2y-1+2\sqrt{y(y-1)})^{\lambda_N^{1/2}}
(1-{2\over y}+i2\sqrt{{1\over y}(1-{1\over y}}))^{\sqrt{\lambda_N y}\over  i}\over2^{3\over2}\pi^2\lambda_N^{1\over2}(y(y-1))^{1\over4}}(1+O(1/N)),}
$$
while for $y\in L^+_1(0)$ equation~\dhrho\ gives,
$$
\eqalign{
&{(\lambda_N y)^{({a_1+b_1+c_1+d_1-2\over2})}\hat W_n((\lambda_N y)^2;a_1,b_1,c_1,d_1)\over e^{\pi\sqrt{\lambda_N y}}}=\cr&{1\over\sqrt{2\pi^3}\lambda_N^{5\over12}}\left({\hat\rho_1(y)\over y(y-1)}\right)^{1\over4}({\rm Ai}(\lambda_N^{1\over3}\hat\rho_1(y))+r_1(y,1/N)),
}
$$ 
where uniformly on compact subsets of $L_1^+(0)$,
$$
|e^{{2\over3}N\hat\rho^{3\over2}_1(y)} r_1(y, 1/N)|\le d/N.
$$
From Theorem~\zone\ we find
$$
y_{k,N}=1+{ai_k\over\lambda_N^{1\over3}}+O(1/N^{4\over3}).
$$

\no=0

\beginsection Appendix A. 

In this appendix we sketch a proof of Theorem~\cthmnextone. We begin by considering solutions to the differential equation
$$
{d^2\chi\over d\rho^2}=\ep^{-2}\rho \chi.\ \eq\ano[diffairy]
$$
We assume that $\rho$ is an infinitely differentiable function of $t>0$  The above differential equation
shows that the Taylor expansion for $\chi(\ep^{-{2\over3}}\rho(t\pm\ep))$ can be written interms of
$\chi(\ep^{-{2\over3}}\rho(t))$ and $\chi(\ep^{-{2\over3}}\rho(t))$. We examine this expansion by setting,
$$
\tilde\rho_{\pm}=\pm{\rho(t\pm\ep)-\rho(t)\over\ep}=\sum_{i=0}^n(\pm\ep)^i{\rho^{(i+1)}(t)\over (i+1)!} + R_n(t)
$$
where $R_n$ is the remainder term in the Taylor expansion.
Wang and Wong [\cite{WW}] essentially show,

\lemapp Ao. Suppose $\chi$ is any solution of \diffairy\ then,
$$\eqalign{
\chi(\ep^{-{2\over3}}\rho(t\pm\ep))&=\chi(\ep^{-{2\over3}}\rho(t)\pm\ep^{1\over3}\tilde\rho_{\pm})\cr&=\chi(\ep^{-{2\over3}}\rho(t))X_1(\rho,\tilde\rho_{\pm},\pm\ep)\pm\ep^{1\over3}\chi'(\ep^{-{2\over3}}\rho(t))X_2(\rho,\tilde\rho_{\pm},\pm\ep),}\eq\ano[airyexp]
$$
and
$$\eqalign{
\chi'(\ep^{-{2\over3}}\rho(t)&\pm\ep^{1\over3}\tilde\rho_{\pm})=\cr&\pm\ep^{-{1\over3}}\chi(\ep^{-{2\over3}}\rho(t))X_3(\rho,\tilde\rho_{\pm},\pm\ep)+\chi'(\ep^{-{2\over3}}\rho(t))X_4(\rho,\tilde\rho_{\pm},\pm\ep),}\eq\ano[airypexp]
$$
where
$$X_3={\partial\over\partial\tilde\rho}X_1,\eq\ano[xft]
$$
and
$$X_4={\partial\over\partial\tilde\rho}X_2.\eq\ano[xff].
$$
If
$$
X_i(\rho,\tilde\rho_{\pm},\pm\ep)=\sum_{n=0}^{\infty}(\pm\ep)^n
X_{i,n}(\rho,\tilde\rho_{\pm})\ i=1,2, \eq\ano[xexp]
$$
then for $\rho>0$,
$$
X_{1,0}(\rho,\tilde\rho)=\cosh\sqrt{\rho}\tilde\rho, \
X_{2,0}(\rho,\tilde\rho)={\sinh\sqrt{\rho}\tilde\rho\over\sqrt\rho},
\eq\ano[xo]
$$
and
$$
\eqalign{
X_{i,k}(\rho,\tilde\rho_{\pm})&=\pm{1\over\sqrt{\rho}}\int_0^{\tilde\rho_{\pm}}sX_{i,k-1}(\rho,s)\sinh\sqrt{\rho}(\tilde\rho_{\pm}-s)ds\cr&={\pm\tilde\rho^2_{\pm}\over\sqrt{\rho}}\int_0^1 sX_{i,k-1}(\rho,s\tilde\rho_{\pm})\sinh\sqrt{\rho}\tilde\rho_{\pm}(1-s)ds.}\eq\ano[xhigher]
$$
Above if $\rho<0$ then $\sqrt{\rho}$ is replaced by $i\sqrt{-\rho}$.

If \airyexp\ is differentiated twice with respect to $\tilde\rho_{\pm}$, then the Airy differential equation, and  the conditions
$$
X_1(\rho,0,\pm\ep)=1,\qquad {\partial X_1(\rho,0,\pm\ep)\over\tilde\partial\tilde\rho_{\pm}}=0,
$$
and
$$
X_2(\rho,0,\pm\ep)=0,\qquad  {\partial X_2(\rho,0,\pm\ep)\over\tilde\partial\tilde\rho_{\pm}}=1
$$
yield upon equating powers of $\ep$ the following differential equations  [\cite{WW}],
$$
{\partial^2 X_{1,0}\over\partial\tilde\rho^2_{\pm}}=\rho X_{1,0},  \qquad X_{1,0}(\rho,0)=1,\qquad {\partial X_{1,0}(\rho,0)\over\partial\tilde\rho_{\pm}}=0,
$$
$$
{\partial^2 X_{1,k}\over\partial\tilde\rho^2_{\pm}}=\rho X_{1,k}\pm\tilde\rho_{\pm} X_{1, k-1}\qquad X_{1,k}(\rho,0)=0,\qquad  {\partial X_{1,k}(\rho,0)\over\tilde\partial\rho_{\pm}}=0,
$$
and
$$
{\partial^2 X_{2,0}\over\partial\tilde\rho^2_{\pm}}=\rho X_{2,0},  \qquad X_{2,0}(\rho,0)=0,\qquad {\partial X_{1,0}(\rho,0)\over\partial\tilde\rho_{\pm}}=1,
$$
$$
{\partial^2 X_{2,k}\over\partial\tilde\rho^2_{\pm}}=\rho X_{2,k}\pm\tilde\rho_{\pm} X_{2, k-1}\qquad X_{2,k}(\rho,0)=0,\qquad  {\partial X_{2,k}(\rho,0)\over\tilde\partial\rho_{\pm}}=0.
$$
Integrating these equations yields \xo\ and \xhigher.

Using induction Wang and Wong essentially show that,
$$
|X_{1,k}(\rho,\tilde\rho)|\le
 3^k\left({1\over3}\right)_k{|\tilde\rho|^{3k}\over(3k)!}e^{|\Re(\sqrt{\rho}\tilde\rho)|},\eq\ano[boundxone]
$$
and
$$
|X_{2,k}(\rho,\tilde\rho)|\le
 3^k\left({2\over3}\right)_k{|\tilde\rho|^{3k+1}\over(3k+1)!}e^{|\Re(\sqrt{\rho}\tilde\rho)|}.\eq\ano[boundxtwo]
$$
Substituting the formula for $\tilde\rho_{\pm}$ into $X_{i,n}(\rho,\tilde\rho_{\pm})$
yields
$$
X_{i,n}(\rho,\tilde\rho_{\pm})=\sum_{j=0}^m(\pm\ep)^j X_{i,n,j}(\rho,\rho')+\tilde R_m,\eq\ano[xinj]
$$

Following [\cite{WW}] and [\cite{GBVA}] we now look for solutions of the equation 
$$
a(t+\ep)\psi(t+\ep)+(b(t,\ep)-y)\psi(t,\ep)+a(t,\ep)\psi(t-\ep)=0. \eq\ano[edif]
$$
Write,
$$
a(t+\ep)=\sum_{n=0}^{\infty}{a^{(n)}(t+{\ep\over2})\over n!}
\left({\ep\over2}\right)^n=\sum_{n=0}^{\infty}a_n(t)
\left({\ep\over2}\right)^n, \eq\ano[ap]
$$
$$
a(t)=\sum_{n=0}^{\infty}{a^{(n)}(t+{\ep\over2})\over n!}
\left(-{\ep\over2}\right)^n=\sum_{n=0}^{\infty}a_n(t)
\left(-{\ep\over2}\right)^n, \eq\ano[apo]
$$
and
$$
b(t)=b_0(t)+\sum_{n=2}^{\infty}b_n(t)\ep^n,\ \eq\ano[bi]
$$
The assumption on the form of $b(t)$ allows us to seek solutions of the form
$$
\tilde\psi(t)=\chi(\ep^{-{2\over3}}\rho(t))C(t, \ep)+\ep^{4\over3}\chi'(\ep^{-{2\over3}}\rho(t))D(t, \ep),
\ \eq\ano[pcd]$$
where
$$
C(t,\ep)=\sum_{s=0}^{\infty} C_{s}(t)\ep^{s}\ {\rm and}\ D(t,\ep)=\sum_{s=0}^{\infty} D_{s}(t)\ep^{s}.
\eq\ano[cd]
$$

Substitution of equations~\pcd, \airyexp\ and \airypexp\ into equation~\edif\ then  equating coefficients of $\chi$ and $\chi'$ 
yields
$$
\eqalign{
&a(t+\ep)(X_1(\rho,\tilde\rho_+,\ep)C(t+\ep)+
 \ep X_3(\rho,\tilde\rho_+,\ep)D(t+\ep))+(b(t)-y)C(t)\cr& a(t)(X_1(\rho,\tilde\rho_-,-\ep)C(t-\ep)-
 \ep X_3(\rho,\tilde\rho_-,-\ep)D(t-\ep)),}\ \eq\ano[cdo]
$$
and
$$
\eqalign{
&a(t+\ep)(X_2(\rho,\tilde\rho_+,\ep)C(t+\ep)+
 \ep X_4(\rho,\tilde\rho_+,\ep)D(t+\ep))+\ep (b(t)-y)D(t)\cr& a(t)(-X_2(\rho,\tilde\rho_-,-\ep)C(t-\ep)+\ep X_4(\rho,\tilde\rho_-,-\ep)D(t-\ep)).}\ \eq\ano[cdt]
$$

Substitute the expansions for $C$, $D$, $a$, $b$ into equations~
\cdo\ and \cdt\ then collect powers of $\ep$ and use equation~\xhigher.
For $\ep^0$ we find from \cdo
$$
(2a_0 X_{1,0,0}+b_0-y)C_0=0, \eq\ano[xxo]
$$
Thus  \xo\ shows
$$
X_{1,0,0}=\cosh(\sqrt{\rho}\rho')={y-b_0\over2 a_0}.\ \eq\ano[xooo]
$$
For the coefficient of $\ep$ in \cdt\ we find
$$
2a_0[(X_{2,0,1}+X_{2,1,0})C_0+X_{2,0,0}C_0']+a_1 X_{2,0,0}C_0+(2a_0 X_{4,0,0}+(b_0-y))D_0=0.
$$
Equations~\xff\ and \xxo\ show that the term multiplying $D_0$ is equal to zero and \xo\ gives,
$$
X_{2,0,0}={\sinh\sqrt{\rho}\rho'\over\sqrt{\rho}},\ X_{2,1,0}= {\rho''\over2}\cosh\sqrt{\rho}\rho', \eq\ano[xtoo]
$$
and
$$X_{2,0,1}={(\rho')^2\over4\rho}\cosh\sqrt{\rho}\rho'-{\rho'\over4\rho^{3/2}}\sinh\sqrt{\rho}\rho' .\eq\ano[xtow]
$$
Thus
$$
{C_0'\over C_0}=-{a_1\over 2a_0}-
{\cosh\sqrt{\rho}\rho'\over2\sinh\sqrt{\rho}\rho'}
\left({{\rho'}^2\over2\sqrt{\rho}}+\sqrt{\rho}\rho''\right)+{\rho'\over4\rho},
$$
which has $C_0(t)=g(t)$ as a solution. To continue on we utilize \xft\  to obtain
$$
X_{3,0,0}=\sqrt{\rho}\sinh\sqrt{\rho}\rho',\ X_{3,0,1}={\rho\rho''\over2}\cosh\sqrt{\rho}\rho' ,
$$
and
$$
X_{3,1,0}={\rho'\over4\sqrt{\rho}}\sinh\sqrt{\rho}\rho'+{(\rho')^2\over4}\cosh\sqrt{\rho}\rho'.
$$
If we now look at the coefficient of $\ep^n$  and use the results just obtained we find by induction that,
$$
\eqalign{
2a_0(X_{3,0,1}+X_{3,1,0})D_{n-2}+&a_1 X_{3,0,0}D_{n-2}+2 a_0 X_{3,0,0}D_{n-2}'\cr&=g_n(C_0,\dots, C_{n-2},D_0,\dots, D_{n-4}),\ n\ge2,}
$$
which can be recast as
$$
\eqalign{
&{d\over dt}\left[\left(\rho4a_0^2\sinh^2(\sqrt{\rho}\rho')\right)^{1/4} D_{n-2}\right]\cr&=\left({1\over \rho4a^2_0\sinh^2(\sqrt{\rho}\rho')}\right)^{1/4}g_n
(C_0,C_2,\dots, C_{n-2},D_0,\dots, D_{n-4}),\quad n\ge2.} \eq\ano[dnn]
$$
Likewise
$$
\eqalign{
&{d\over dt}\left[\left({a_0^2\sinh^2(\sqrt{\rho}\rho')\over\rho}\right)^{1/4} C_{n-2}\right]\cr
&=\left({\rho\over a^2_0\sinh^2(\sqrt{\rho}\rho')}\right)^{3/4}
f_n(C_0,C_2,\dots, C_{n-3},D_0,\dots, D_{n-3}),\quad n\ge2}\eq\ano[cnn]
$$
where $f_n$ and $g_n$ are linear in the variables in their arguments and their higher derivatives. In the above equations we take the convention that $D_n=0=C_n$ if $n$ is negative. 

\thmapp thtwo. 
Suppose \ab\ and ic)--iiic) hold, $b_0=b(t,\ep)$ and $b_i(t,\ep)=0, i>0$ 
in equation~\bi. In equations 
\cpsione--\cpsithree\ let $\rho=\rho_1=(t_p^+-t)(\zeta_1)^{3/2}$ and let $L_1(\ep)$
be given as in Lemmas~\anafreudpl\ or \anafreudpla\ and
$L_1^+(\ep)=L_1(\ep)\cap \bar C_+$. If $y\in L^+_1(0)$ then there exists an $\ep_1$ such that $y\in L^+_1(\ep),\ 0\le\ep\le \ep_1$. Let $\chi$ be any entire function solution of \diffairy\  and  $\psi=g\chi$. For each $\ep,\ 0<\ep\le\ep_1$ and all $t: t\in[t_{in},t_{fi}]$ 
$$
a(t+\ep,\ep)\psi(t+\ep,y,\ep)+(b(t,\ep)-y)\psi(t,y,\ep)+a(t,\ep)\psi(t-\ep,
y,\ep)=\beta(t,y,\ep),
\eq\ano[ppsi]
$$
where 
$$
|\beta(t,y,\ep)|<c(y,\ep)\ep^2 \sup_{u\in[t-\ep,t+\ep]}
\left[|\chi(\ep^{-{2\over3}}\rho(u,y,\ep))|+\ep^{1\over3}|\chi'(\ep^{-{2\over3}}\rho(u,y,\ep))|\right]. \eq\ano[ppsier]
$$
If $K$ is a compact set in $L^+_1(0)$ then there exists an $\ep_K$ such that $c(y,\ep)$ is uniformly bounded  on $K\times[0,\ep_K]$. Furthermore for fixed $(t,\ep)$, $\beta\in H(K\cap \bbc_+)$ and  $\beta(t,\cdot,\cdot)\in C(K\times(0,\ep_K])$. If $b-2a<0$ let $\rho=\rho_2=(t_p^--t)(\zeta_2)^{3/2}$  or if $b-2a>0$ let $\rho=\rho_2=(t-t_p^-)(\zeta_2)^{3/2}$ in equations \cpsione--\cpsithree. For both cases let  $L_2(\ep)$ be given as in Lemmas~\anafreudpl\ or \anafreudpla\ and $L_2^+(\ep)=L_2(\ep)\cap\bar C_+$ . If $y\in L^+_2(0)$ then there exists an $\ep_1$ such that $y\in L^+_2(\ep),\ 0\le\ep\le \ep_1$. Let $\chi$ be any entire function solution of \diffairy\  and  $\psi=g\chi$. For each $\ep,\ 0<\ep\le\ep_1$ and all $t: t\in[t_{in},t_{fi}]$ 
$$
a(t+\ep,\ep)\psi(t+\ep,y,\ep)+
(y-b(t,\ep))\psi(t,y,\ep)+a(t,\ep)\psi(t-\ep,y,\ep)=\beta(t,y,\ep),
$$
where the error $\beta$ satisfies \ppsier. If $K$ is a compact set in $L^+_2(0)$ then there exists an $\ep_K$ such that $c(y,\ep)$ is uniformly bounded  on $K\times[0,\ep_K]$. Furthermore for fixed $(t,\ep)$, $\beta\in H(K\cap \bbc_+)$ and  $\beta(t,\cdot,\cdot)\in C(K\times(0,\ep_K])$. 

\pf  We first take $y$ real and choose $\ep_1\le\ep_0$ so that $t_{in}-\ep_1>0$.  
From equations~\xo--\xinj, \xooo--\xtow, Theorem~\freudtur\ and the properties of $q$ we find 
$$
|X_{i,0}(\rho,\tilde\rho_{\pm},\pm\ep)-(X_{i,0,0}\pm\ep X_{i,0,1})|<c(y)\ep^2,
\quad i=1,2 \eq\ano[xonezero]
$$
$$
|X_{i,1}(\rho,\tilde\rho_{\pm},\pm\ep)-X_{i,1,0}|<c(y)\ep,
\quad i=1,2 \eq\ano[xoneone]
$$
and
$$
|X_{1,k}|< 3^k\left({1\over3}\right)_k{c(y)^k\over(3k)!}, \quad
 |X_{2,k}|< 3^k\left({2\over3}\right)_k{c(y)^k\over(3k)!}, \eq\ano[bser]
$$
where $c(y)$ is independent of 
$(t,\ep)\in[t_{in},t_{fi}]\times[0,\ep_1]$. This implies from \airyexp\ that
$$
\eqalign{
\chi(\ep^{-{2\over3}}\rho(t\pm\ep))&=\chi
\left(\ep^{-{2\over3}}\rho(t)\right)
\sum_{l=0}^1\sum_{j=0}^{1-l}(\pm\ep)^{l+j} X_{1,l,j}\cr
&\quad \pm\ep^{1\over3} \chi'(\ep^{-{2\over3}}\rho(t))\sum_{l=0}^1\sum_{j=0}^{1-1}(\pm\ep)^{l+j} X_{2,l,j}
+e_1(t,y,\ep).
}
$$
With the above estimates we find that,
$$
|e_1(t,y,\ep)|<\ep^2 c(y)\sup_{u\in[t-\ep,t+\ep]}
\left[|\chi(\ep^{-{2\over3}}\rho(u,y,\ep))|+\ep^{1\over3}|\chi'(\ep^{-{2\over3}}\rho(u,y,\ep))|\right].
$$
With the choice of $C_0=g(t)$,  the terms of 
$\ep^i$, $i=0,\dots, 1$ vanish in equations~\cdo\ and \cdt. The remaining terms times $\chi$ or $\chi'$ when added together give $\beta$. For $y\in L^+_1(0)$ differentiation with respect to $t$ of \qgrone\ and  
Theorem~\freudtur\ show that $\cosh(\sqrt{\rho}\rho')={y-b_0\over a_0}$. Thus for fixed $(t,\ep)$ Morera's Theorem implies the above equations for $X_{i,k}$ can be analytically continued to $L^+_1(0)$. Theorem~\freudtur\ also shows that for fixed $(y,\ep)$ in $L^+_1(0)\times[0,\ep_1],\ \psi_1$ and $\psi_2$ satisfy equation~\ppsi\ with $\beta$ satisfying the bounds and smoothness given above. The other cases can be argued in a similar manner.
\qed

\remk Integration of Equations \dnn\ and \cnn\ as in [\cite{O}, p.~409] allows the extension of the above results. However since this is not needed for the applications considered above.

We now have,

\thmapp ththree. 
Suppose \ab\ and ic)--iiic) hold. Let $a_1(t,\ep), b_1(t,\ep)\in C([0,\infty)\times[0,\ep_0])$
satisfy \abep\ 
with $b_1(t,\ep)$ real and $a_1(t,\ep)$ strictly positive on every
compact subset of $(0,\infty]\times[0,\ep_0]$. In equations 
\cpsione--\cpsithree\ let $\rho=\rho_1=(t_p^+-t)(\zeta_1)^{3/2}$ and let $L_1(\ep)$
be given as in Lemmas~\anafreudpl\ or \anafreudpla\ and
$L_1^+(\ep)=L_1(\ep)\cap \bar C_+$. Suppose $K\in L^+_1(0),\ K$ compact then there exists an $\ep_K$ such that for each $(y,\ep)\in K\times(0,\ep_K]$ and all $n: n\ep\in[t_{in},t_{fi}]$ there exist solutions $f_{i}\ i=1,2$ of
equation~\fourteen\ such that 
$$
f_i(n)=\psi_i(n)+r_i(n), \eq\ano[aieqone]
$$
where
$$
|{r_i(n)\over u^{(i)}(n)}|=|{f_i(n)-\psi_i(n)\over u^{(i)}(n)}|\le d(K)\ep
,\quad i=1,2. \quad\eq\ano[cineqone]
$$ 
Furthermore for fixed $\ep\ {r_i(n)\over u^{(i)}(n)}\in H(K\cap C_+)$ and ${r_i(n)\over u^{(i)}(n)}\in C(K\times[0,\ep_K])$. If $b-2a<0$ let $\rho=\rho_2=(t_p^--t)(\zeta_2)^{3/2}$ or if $b-2a>0$ let $\rho=\rho_2=(t-t_p^-)(\zeta_2)^{3/2}$ in equations \cpsione--\cpsithree. For both cases let  $L_2(\ep)$ be given as in 
Lemmas~\anafreudpl\ or \anafreudpla\ and $L_2^+(\ep)=L_2(\ep)\cap\bar C_+$.  
Suppose $K\subset L^+_2(0)$, $K$ compact, then there exists an $\ep_K$ such that for each $(y,\ep)\in K\times(0,\ep_K]$ and all $n: n\ep\in[t_{in},t_{fi}]$ there exists solutions $f_{i}$, $i=1,2$ of
equation~\fourteen\ such that 
$$
(-1)^n f_i(n)=\psi_i(n)+r_i(n), \eq\ano[aieqtwo]
$$
where
$$
|{r_i(n)\over u^{(i)}(n)}|=|{(-1)^nf_i(n)-\psi_i(n)\over u^{(i)}(n)}|\le d(K)\ep
,\quad i=1,2. \eq\ano[cineqtwo]
$$
Furthermore for fixed $\ep, {r_i(n)\over u^{(i)}(n)}\in H(K\cap C_+)$ and ${r_i(n)\over u^{(i)}(n)}\in C(K\times[0,\ep_K])$.

\pf The proof follows closely that of Theorem 4.4 in [\cite{GBVA}]. Given $K$ a compact subset of $L_1^+(0)$ there exists and $\ep_K$ such that $K\subset L_1^+(\ep)$ for $\ep\in[0,\ep_K]$. From Theorem~\ththree, equations~\cpsione\ and \cpsitwo\ we find for $j=1,2$ that $\psi_j$ and $u^{(j)}$ satisfy equation~\ppsi\ with error $\beta_j$ and $\beta^{(j)}$ respectively. Also Lemmas~\anafreudpl\ or \anafreudpla\ tells us that for fixed $y$ and $\ep$ $\Re\rho_1^{3\over2}(t,y,\ep)$  is a decreasing function of $t$. With $t_n = \ep n$, 
$\sigma_j = {f(n)-\psi_j (n)\over u^{(j)}(n)}$, $\hat\psi_j (n)={\psi_j (n)\over u^{(j)}(n)}$, 
$\hat \beta_j={\beta_j(n)\over u^{(j)}(n)}$, $\hat \beta^{(j)}={\beta^{(j)}(n)\over u^{(j)}(n)}$, and $\Delta(w(n))=w(n)-w_1(n)$, we find using  \fourteen\ that 
$$\eqalign{u^{(2)}(n-1) \, (\sigma_2 (n)-\sigma_2 (n-1))-&u^{(2)}(n+1){a((n+1)\ep,\ep)\over a(n\ep,\ep)}(\sigma_2 (n+1)-\sigma_2 (n))\cr& 
=h_2(n)+q_2(n)} \eq\ano[fifteen]$$ 
and
$$\eqalign{u^{(1)}(n+1)(\sigma_1 (n+1)-\sigma_1 (n))-&{a(n\ep,\ep)\over a((n+1)\ep,\ep)}u^{(1)}(n-1) \,  
(\sigma_1 (n)-\sigma_1 (n-1))\cr& 
=h_1(n)+q_1(n),}\eq\ano[lower]$$
where 
$$\eqalign{ 
h_j(n)& = (-1)^{j}\big[(\Delta({y-b(n\ep,\ep)\over a((n+2-j)\ep,\ep)})+{\hat\beta^{(j)}(n\ep)\over a((n+2-j)\ep,\ep)})u^{(j)}(n)\sigma_j(n)\cr
&\quad +\Delta({a((n+j-1)\ep,\ep)\over a((n+2-j)\ep,\ep)})u^{(j)}(n+2j-3)\sigma_j(n+2j-3)\big],}$$ 
and 
$$\eqalign{ 
q_j(n)& = (-1)^{j}\big[\Delta({y-b(n\ep,\ep)\over a((n+2-j)\ep,\ep)})u^{(j)}(n)\hat\psi_j(n)\cr
&\quad +\Delta({a((n+j-1)\ep,\ep)\over a((n+2-j)\ep,\ep)})u^{(j)}(n+2j-3)\hat\psi_j(n+2j-3)\cr
&\quad +\hat\beta_j(n\ep){u^{(j)}(n)\over a((n+2-j)\ep,\ep)}.}$$ 
Selecting a solution $f_2$ such that $\sigma_2(n_{fi})=0=\sigma_2(n_{fi}-1)$ where $n_{fi}$ is the largest integer such that $n_{fi}\ep\le t_{fi}$ yields  
$$\sigma_2(n)=\sum^{n_{fi}-1}_{i=n+1} G_2(n,i){h_2(i)+q_2(i)\over u^{(2)}(i-1)}$$ 
where 
$$G_2(n,i)=-\sum^{i-1}_{k=n}{a(i\ep,\ep)u^{(2)}(i-1)u^{(2)}(i)\over a((k+1)\ep,\ep)u^{(2)}(k)u^{(2)}(k+1)}.$$ 
The above formula for $\sigma$ can be recast as 
$$\sigma_2(n)=\sum^{n_{fi}-1}_{i=n+1}\tilde G_2(n,i){q_2(i)\over u^{(2)}(i-1)}+\sum^{n_{fi}}_{i=n+1}K_2(n,i)\sigma_2(i) ,\eq\ano[sione]$$ 
where   
$$\eqalign{ 
K_2(n,i)&=\tilde G_2(n,i)\left(\Delta
\left({y-b(i\ep,\ep)\over a(i\ep,\ep)}\right)+{\hat\beta^{(2)}(i\ep)\over a(i\ep,\ep)}\right){u^{(2)}(i)\over u^{(2)}(i-1)}\cr
&\quad + \tilde G_2(n,i-1)\Delta\left({a((i)\ep,\ep)\over a((i-1)\ep,\ep)}\right){u^{(2)}(i)\over u^{(2)}(i-2)}.}$$ 
Here $ \tilde G_2(n,i)= G_2(n,i)$ for $i\le n_{fi}-1$ and zero otherwise. Note that $G_2(n,i)=0$ for $i\le n$. 
For $j=1$ we apply a similar analysis to equation~\lower\ with
the conditions $\sigma_1(n_{in})=0=\sigma_1(n_{in}+1)$ where $n_{in}$ is the smallest integer such that $n_{in}\ep\ge t_{in}$ yields
$$\sigma_1(n)=\sum^{n-1}_{i=n_1+1}G_1(n,i){q_1(i)\over u^{(1)}(i+1)}+\sum^{n-1}_{i=n_1}K_1(n,i)\sigma_1(i).\eq\ano[sitwo]$$ 
Here $\tilde G_1(n,i)=G_1(n,i)$ for $i\ge n_1+1$ and zero otherwise, 
$$G_1(n,i)=\sum^{n-1}_{k=i} {a((i+1)\ep,\ep)u^{(1)}(i+1)u^{(1)}(i)\over a((k+1)\ep,\ep) u^{(1)}(k)u^{(1)}(k+1)},$$ 
and 
$$\eqalign{ 
K_1(n,i)&=-\tilde G_1(n,i)
\left(\Delta\left({y-b(i\ep,\ep)\over a((i+1)\ep,\ep)}\right)+{\hat\beta^{(1)}(i\ep)\over a((i+1)\ep,\ep)}\right){u^{(1)}(i)\over u^{(1)}(i+1)}\cr
&\quad+\tilde G_1(n,i+1)\Delta\left({a((i+1)\ep,\ep)\over a(i+2)\ep,\ep
)}\right){u^{(1)}(i)\over u^{(1)}(i+2)}.}$$ 

Let
$$ 
\tilde E(t) = |e^{{2\over3}{1\over\ep}\rho(t,y,\ep)^{3\over2}}|. \eq\ano[tweight] 
$$ 
We interrupt the proof of Theorem A3 to show,

\lemapp lemafour.  
Suppose that \ab\ and ic)-- iiic) hold. Let $K$ be a compact subset of $L_1^+(0)$. Then there are constants $d(K)$ and $\ep_K$ so that for all $(n,i) :[n\ep,i\ep]\subset[t_{in},t_{fi}],\ep\in[0,\ep_K]$ the following inequality holds, 
$$ 
|\tilde G_1(n,i)|\le \tilde G_1(i)= d(K)\ep^{-{1\over3}}|\hat u^{(1)}(i)\hat u^{(1)}(i+1)|, \eq\ano[tgreenone] 
$$ 
and 
$$ 
|\tilde G_2(n,i)|\le \tilde G_2(i)= d(K)\ep^{-{1\over3}}|\hat u^{(2)}(i)\hat u^{(2)}(i-1)|. \eq\ano[tgreentwo] 
$$ 
The above inequalities imply that
$$\sum_{j=n_{in}+1}^{n_{fi}}\tilde G_j(i)<d(K)\ep^{-1},\qquad j = 1,2. \eq\ano[cgreenot]$$
 
\pf  For fixed $y$ and $\ep$ we write (see [\cite{GBVA}, eqs~(3.45) and (3.46)]) 
as,
$$\eqalign{ 
&{1\over a(k+1)u^{(2)}(k)u^{(2)}(k+1)}\cr&={1\over a(k+1)}{1\over u^{(1)}(k+1)u^{(2)}(k)-u^{(1)}(k)u^{(2)}(k+1)}
\left({u^{(1)}(k+1)\over u^{(2)}(k+1)}-{u^{(1)}(k)\over u^{(2)}(k)}\right)\cr&={\beta^{(1)}(k+1)u^{(1)}(k+1)-\beta^{(2)}(k+1)u^{(1)}(k+1)\over D_k}\left({u^{(1)}(k+1)\over u^{(2)}(k+1)}-{u^{(1)}(k)\over u^{(2)}(k)}\right),}
$$
where Theorem~\thtwo\ has been used to obtain the last equality. Here
$$
D_k=a(k+1)(u^{(1)}(k+1)u^{(2)}(k)-u^{(1)}(k)u^{(2)}(k+1)).
$$
 Summation by parts yields
$$\eqalign{ 
&\sum_{k=n}^{i-1}{1\over a(k+1)u^{(2)}(k)u^{(2)}(k+1)}\cr&= 
\sum_{k=n}^{i-2}\left({u^{(1)}(k+1)\over u^{(2)}(k+1)}-
{u^{(1)}(j)\over u^{(2)}(j)}\right)
{\beta^{(1)}(k+1)u^{(2)}(k+1)-\beta^{(2)}(k+1)u^{(1)}(k+1)\over D_k}\cr& 
\quad +\left({u^{(1)}(i)\over u^{(2)}(i)}-{u^{(1)}(n)\over u^{(2)}(n)}\right)
{1\over a(i)}{1\over u^{(1)}(i)u^{(2)}(i-1)-u^{(1)}(i-1)u^{(2)}(i)}.}\eq\ano[intpartthree] 
$$ 
Since $L_1^+(0)\in S_0\cup S_1$ we find from the properties of $\rm{Ai}_{i},\ i=0,1$ and $\tilde w^{(1)}$, ([\cite{O}, p.~238 and p.~418] and [\cite{GBVA}]), the nonvanishing of $g$ and the monotonicity of $\Re\rho^{3\over2}$ that for $m\le i$,
$$
\left|{u^{(1)}(m)\over u^{(2)}(m)}\right|\le d(K)\tilde E^{-1}(i)^2.
$$
Also Theorem~\ththree\ implies that
$$|\beta^{(2)}(k+1)u^{(1)}(k+1)|,|\beta^{(1)}(k+1)u^{(2)}(k+1)|<d(K)\ep^2.
$$
Lemma~\Ao, the Wronskian for ${\rm Ai}_1$ and $\tilde w^{(1)}$ and the nonvanishing of $g$ and $a(t,\ep)$ 
show that
$$
{1\over |D_k|}\le d(K)\ep^{-{1\over3}}.
$$
In the above inequalities $d(K)$ is independent of $\ep\in[0,\ep_K]$, $n:n\ep\in[t_{in},t_{fi}]$ and $y\in K$. Utilizing these inequalities the definition of $\hat u^{(i)},\ i=1,2$ and the fact that there are at most $n_{fi}~\ep$ terms in the summation gives equation~\tgreenone. This bound shows that
$$
\sum_{i=n_{in}+1}^{n_{fi}}\tilde G_2(i)\le d(K)\ep^{-{1\over3}}\sum_{i=n_{in}}^{n_{fi}}|\hat u^{(2)}(i)\hat u^{(2)}(i-1)|.
$$
The above sum can be rewritten as,
$$\eqalign{ 
\sum_{i=n_{in}+1}^{n_{fi}}|\hat u^{(2)}(i-1)\hat u^{(2)}(i)|&\le d_1(K)+\sum_{i=n_{in}+1}^{n_{fi}-1}|\hat u^{(2)}(i-1)\hat u^{(2)}(i)|\cr
&\le 2d_1(K)+\sum_{(i:\Re(\rho_1(i)^{3\over2})\le0,\rho_1(i)\ne0)}|\hat u^{(2)}(i-1)\hat u^{(2)}(i)|\cr
&\quad +\sum_{(i:\Re(\rho_1(i)^{3\over2})>0)}|\hat u^{(2)}(i-1)\hat u^{(2)}(i)|.} 
$$
where 
$$d_1(K) =\max_{i\in[n_{in}+1,n_{fi}],y\in K,\ep\in[0,\ep_K]}|\hat u^{(2)}(i-1,y,\ep)\hat u^{(2)}(i,y,\ep)|.
$$
With 
$$
d_2(K)=\sup_{t\in[t_{in}+\ep,t_{fi}],y\in K,\ep\in[0,\ep_K]}|t-t_p^+(y,\ep)|^{1/2}|\hat u^{(2)}(t-\ep,y,\ep)\hat u^{(2)}(t,y,\ep)|,
$$
which from the asymptotic properites of $u^{(j)},\ j=1,2$ and Theorem~\freudtur\ is finite. We find using the integral test that,
$$
\eqalign{
\sum_{(i:\Re(\rho_1(i)^{3\over2})>0)}|\hat u^{(2)}(i-1)\hat u^{(2)}(i)|&\le \sum_{(i:\Re(\rho_1(i)^{3\over2})>0)}{d_2(K)\ep^{1\over3}\over|t_i-t_p^+(y,\ep)|^{1/2}}\cr
& \le\ep^{-{2\over3}} d_2(K)\int_{t_{in}}^{t_{fi}}{dt\over|t-t_p^+(y,\ep)|^{1\over2}}\le\ep^{-{2\over3}} d(K).}
$$
A similar argument bounds the remaining sum which gives the result
for $j=2$. The argument for $j=1$ is similar. \qed

We now complete the proof of Theorem~\ththree. The bounds above, the nonvanishing, and asymptotic properties of $u^{(2)}$ when used in equation~\sione\ imply
$$
|\sigma_2(n)|\le d_3(K)\ep+d_4(K)\ep^2\sum_{i=n+1}^{n_{fi}-1}\tilde G_2(i)|\sigma_2(i)|.
$$
The discrete version of Gronwall's inequality and the fact 
that ${r_2{n}\over u^{(2)}(n)}=\sigma_2(n)$ 
gives the bound \cineqone\ for $i=2$. The uniformitiy of the bound on $K$ as well as the smoothness assertions follow from Theorems~\freudtur\ and \thtwo. The result for $i=1$ follows in a similar manner.  
The proof for $y\in L_2^+(0)$ follows as above. \qed

\beginsection References

\frenchspacing

\myinput{references}

\end

%% file: gen-dfn.tex
\def\IP{{\rm I\kern -1.6pt{\rm P}}}
\def\IC{\hbox{\rm C\kern-.58em{\raise.53ex\hbox{$\scriptscriptstyle|$}}
     \kern-.55em{\raise.53ex\hbox{$\scriptscriptstyle|$}} }}
\def\IN{\hbox{I\kern-.2em\hbox{N}}}
\def\IR{\hbox{\rm I\kern-.2em\hbox{\rm R}}}
\def\sIR{{\sl\hbox{I\kern-.2em\hbox{R}}}}
\def\ZZ{{{\rm Z}\kern-.28em{\rm Z}}}

\def\remk{\bigskip\noindent{\bf Remark.~~}}
\def\pf{\bigskip\noindent{\bf Proof.~~}}

\def\frac#1/#2{\leavevmode
   \kern.1em\raise .5ex\hbox{\the\scriptfont0 #1}%
\kern-.1em /%
\kern-.15em\lower.25ex\hbox{\the\scriptfont0 #2}}
\def\fracc#1/#2{\leavevmode
   \kern.1em\raise .5ex\hbox{$\scriptstyle #1$}%
\kern-.1em /%
\kern-.15em\lower.25ex\hbox{$\scriptstyle #2$}}

\def\sfrac#1/#2{\leavevmode
   \kern.1em\raise .5ex\hbox{\the\scriptscriptfont0 #1}%
\kern-.1em /%
\kern-.15em\lower.25ex\hbox{\the\scriptscriptfont0 #2}}

\outer\def\beginsection#1\par{\vskip0pt plus.3\vsize
   \vskip0pt plus-.3\vsize\bigskip\vskip\parskip
   \message{#1}\leftline{\bf#1}\nobreak\smallskip}

\outer\def\breaksection#1\par{\vskip0pt plus.3\vsize
   \vskip0pt plus-.3\vsize\bigskip\vskip\parskip
   \message{#1}{\noindent\obeylines\bf#1}\nobreak\smallskip}

\outer\def\ctrsection#1\par{\vskip0pt plus.3\vsize
   \vskip0pt plus-.3\vsize\bigskip\vskip\parskip
   \message{#1}\centerline{\bf#1}\nobreak\smallskip}

\outer\def\subsection#1\par{\vskip0pt plus.3\vsize
   \vskip0pt plus-.3\vsize\bigskip\vskip\parskip
   \message{#1}\leftline{\it#1}\nobreak\smallskip}

\outer\def\centersection#1\par{\vskip0pt plus.3\vsize
   \vskip0pt plus-.3\vsize\bigskip\vskip\parskip
   \message{#1}\centerline{\bf#1}\nobreak\smallskip}

\def\sqr#1#2{{\vcenter{\vbox{\hrule height.#2pt
\hbox{\vrule width.#2pt height #1pt \kern#1pt
\vrule width.#2pt}
\hrule height.#2pt}}}}
\def\square{\mathchoice\sqr56\sqr56\sqr{2.1}3\sqr{1.5}3}
\def\qed{\hfill$\square$}

\def\pmb#1{\setbox0=\hbox{$#1$}%
\kern-.025em\copy0\kern-\wd0
\kern.05em\copy0\kern-\wd0
\kern-.025em\raise.0433em\box0 }

\def\pmbit#1{\setbox0=\hbox{\it#1}%
\kern-.025em\copy0\kern-\wd0
\kern.05em\copy0\kern-\wd0
\kern-.025em\raise.0433em\box0 }

\def\pmbb#1{\setbox0=\hbox{$\scriptstyle#1$}%
\kern-.025em\copy0\kern-\wd0
\kern.05em\copy0\kern-\wd0
\kern-.025em\raise.0433em\box0 }

\newcount\no
\newcount\subsecno
\newcount\thmno
\newcount\exmno
\newcount\appno
\global\subsecno=1\no=0\thmno=0\exmno=0\appno=0

\def\ano[#1]{\global\advance\no by 1   
\expandafter\xdef\csname#1\endcsname{{\rm(A.\the\no)}}
{{\rm(A.\the\no)}}}

\def\eqn[#1]{\global\advance\no by 1  
\expandafter\xdef\csname#1\endcsname{(\the\subsecno.\the\no)}
{(\the\subsecno.\the\no)}}           

\def\leqn[#1]{\global\advance\no by 1  
\expandafter\xdef\csname#1\endcsname{(\the\no)}
{(\the\no)}}           

\def\eno[#1]{\global\advance\no by 1   
\expandafter\xdef\csname#1\endcsname{{\rm(\the\no)}}
{{\rm(\the\no)}}}                            
\let\eq=\eqno

\def\sno[#1]{\global\advance\no by 1   
\expandafter\xdef\csname#1\endcsname{{\rm(\the\subsecno.\the\no)}}
{{\rm(\the\subsecno.\the\no)}}}           

\def\lbl[#1]{\global\advance\no by 1  
\expandafter\xdef\csname#1\endcsname{(\the\no\rm}
{(\the\no}\rm}

\def\slbl[#1]{\global\advance\no by 1   
\expandafter\xdef\csname#1\endcsname{(\the\subsecno.\the\no\rm}
{(\the\subsecno.\the\no}\rm}

\def\thm#1. #2\par{\medbreak\global\advance\thmno by 1
\expandafter\xdef\csname#1\endcsname{{\rm\the\subsecno.\the\thmno}}
\noindent{\bf Theorem~\the\subsecno.\the\thmno.\ }{\sl#2}\par\medbreak}

\def\thmn#1. #2\par{\medbreak\global\advance\thmno by 1
\expandafter\xdef\csname#1\endcsname{{\rm\the\thmno}}
\noindent{\bf Theorem~\the\thmno.\ }{\sl#2}\par\medbreak}

\def\exm#1. #2\par{\medbreak\global\advance\exmno by 1
\expandafter\xdef\csname#1\endcsname{{\rm\the\exmno}}
\noindent{\bf Example~\the\exmno.\ }{\rm#2}\par\medbreak}

\def\lemapp#1. #2\par{\medbreak\global\advance\appno by 1
\expandafter\xdef\csname#1\endcsname{{\rm A\the\appno}}
\noindent{\bf Lemma~A\the\appno.\ }{\sl#2}\par\medbreak}

\def\thmapp#1. #2\par{\medbreak\global\advance\appno by 1
\expandafter\xdef\csname#1\endcsname{{\rm A\the\appno}}
\noindent{\bf Theorem~A\the\appno.\ }{\sl#2}\par\medbreak}

\def\prop#1. #2\par{\medbreak\global\advance\thmno by 1
\expandafter\xdef\csname#1\endcsname{{\rm\the\subsecno.\the\thmno}}
\noindent{\bf Proposition~\the\subsecno.\the\thmno.\ }{\sl#2}\par\medbreak}

\def\lem#1. #2\par{\medbreak\global\advance\thmno by 1
\expandafter\xdef\csname#1\endcsname{{\rm\the\subsecno.\the\thmno}}
\noindent{\bf Lemma~\the\subsecno.\the\thmno.\ }{\sl#2}\par\medbreak}

\def\coro#1. #2\par{\medbreak\global\advance\thmno by 1
\expandafter\xdef\csname#1\endcsname{{\rm\the\subsecno.\the\thmno}}
\noindent{\bf Corollary~\the\subsecno.\the\thmno.\ }{\sl#2}\par\medbreak}

\def\dfn#1. #2\par{\medbreak\global\advance\thmno by 1
\expandafter\xdef\csname#1\endcsname{{\rm\the\subsecno.\the\thmno}}
\noindent{\bf Definition~\the\subsecno.\the\thmno.\ }{\rm#2}\par\medbreak}

\def\clm#1. #2\par{\medbreak\global\advance\thmno by 1
\expandafter\xdef\csname#1\endcsname{{\rm\the\subsecno.\the\thmno}}
\noindent{\bf Claim~\the\subsecno.\the\thmno.\ }{\rm#2}\par\medbreak}

\def\rem#1. #2\par{\medbreak\global\advance\thmno by 1
\expandafter\xdef\csname#1\endcsname{{\rm\the\subsecno.\the\thmno}}
\noindent{\bf Remark~\the\subsecno.\the\thmno.\ }{\rm#2}\par\medbreak}

\def\exam#1. #2\par{\medbreak\global\advance\thmno by 1
\expandafter\xdef\csname#1\endcsname{{\rm\the\subsecno.\the\thmno}}
\noindent{\bf Example~\the\subsecno.\the\thmno.\ }{\rm#2}\par\medbreak}

\def\fact#1. #2\par{\medbreak\global\advance\thmno by 1
\expandafter\xdef\csname#1\endcsname{{\rm\the\subsecno.\the\thmno}}
\noindent{\bf Fact~\the\subsecno.\the\thmno.\ }{\sl#2}\par\medbreak}
\def\fac#1. #2\par{\medbreak\global\advance\thmno by 1
\expandafter\xdef\csname#1\endcsname{{\rm\the\thmno}}
\noindent{\bf Fact~\the\thmno.\ }{\sl#2}\par\medbreak}

\newcount\refnum
\refnum = 0
\def\ref{\baselineskip=12pt\frenchspacing
     \bigskip\global\advance\refnum by 1 \item{\the\refnum .}}
\newcount\brefnum
\brefnum = 0
\def\bref{\baselineskip=12pt\frenchspacing
                    \bigskip\global\advance\brefnum by 1 \item{[\the\brefnum]}}

\font\teneusm=eusm10
\font\seveneusm=eusm7
\font\fiveeusm=eusm5
\newfam\eusmfam
\textfont\eusmfam=\teneusm
\scriptfont\eusmfam=\seveneusm
\scriptscriptfont\eusmfam=\fiveeusm

\font\teneurm=eurm10
\font\seveneurm=eurm7
\font\fiveeurm=eurm5
\newfam\eurmfam
\textfont\eurmfam=\teneurm
\scriptfont\eurmfam=\seveneurm
\scriptscriptfont\eurmfam=\fiveeurm

\font\ninerm=cmr9
\font\eightrm=cmr8
\font\sixrm=cmr6
 
\font\ninei=cmmi9
\font\eighti=cmmi8
\font\sixi=cmmi6
\skewchar\ninei='177 \skewchar\eighti='177 \skewchar\sixi='177
 
\font\ninesy=cmsy9
\font\eightsy=cmsy8
\font\sixsy=cmsy6
\skewchar\ninesy='60 \skewchar\eightsy='60 \skewchar\sixsy='60

\font\ninebf=cmbx9
\font\eightbf=cmbx8
\font\sixbf=cmbx6
 
\font\ninett=cmtt9
\font\eighttt=cmtt8
\font\sixtt=cmtt8 
\hyphenchar\tentt=-1 
\hyphenchar\ninett=-1
\hyphenchar\eighttt=-1
 
\font\ninesl=cmsl9
\font\eightsl=cmsl8
\font\sixsl=cmsl8 

\font\nineit=cmti9
\font\eightit=cmti8
\font\sixit=cmti8
 
\newskip\ttglue
 
\catcode`@=11 
 
\def\ninepoint{\def\rm{\fam0\ninerm}%
  \textfont0=\ninerm \scriptfont0=\sixrm \scriptscriptfont0=\fiverm
  \textfont1=\ninei \scriptfont1=\sixi \scriptscriptfont1=\fivei
  \textfont2=\ninesy \scriptfont2=\sixsy \scriptscriptfont2=\fivesy
  \textfont3=\tenex \scriptfont3=\tenex \scriptscriptfont3=\tenex
  \def\it{\fam\itfam\nineit}%
  \textfont\itfam=\nineit
  \def\sl{\fam\slfam\ninesl}%
  \textfont\slfam=\ninesl
  \def\bf{\fam\bffam\ninebf}%
  \textfont\bffam=\ninebf \scriptfont\bffam=\sixbf
   \scriptscriptfont\bffam=\fivebf
  \def\tt{\fam\ttfam\ninett}%
  \textfont\ttfam=\ninett
  \tt \ttglue=.5em plus.25em minus.15em
  \normalbaselineskip=11pt
  \def\MF{{\manual hijk}\-{\manual lmnj}}%
  \let\sc=\sevenrm
  \let\big=\ninebig
  \setbox\strutbox=\hbox{\vrule height8pt depth3pt width\z@}%
  \normalbaselines\rm}
 
\def\eightpoint{\def\rm{\fam0\eightrm}%
  \textfont0=\eightrm \scriptfont0=\sixrm \scriptscriptfont0=\fiverm
  \textfont1=\eighti \scriptfont1=\sixi \scriptscriptfont1=\fivei
  \textfont2=\eightsy \scriptfont2=\sixsy \scriptscriptfont2=\fivesy
  \textfont3=\tenex \scriptfont3=\tenex \scriptscriptfont3=\tenex
  \def\it{\fam\itfam\eightit}%
  \textfont\itfam=\eightit
  \def\sl{\fam\slfam\eightsl}%
  \textfont\slfam=\eightsl
  \def\bf{\fam\bffam\eightbf}%
  \textfont\bffam=\eightbf \scriptfont\bffam=\sixbf
   \scriptscriptfont\bffam=\fivebf
  \def\tt{\fam\ttfam\eighttt}%
  \textfont\ttfam=\eighttt
  \tt \ttglue=.5em plus.25em minus.15em
  \normalbaselineskip=9pt
  \def\MF{{\manual opqr}\-{\manual stuq}}%
  \let\sc=\sixrm
  \let\big=\eightbig
  \setbox\strutbox=\hbox{\vrule height7pt depth2pt width\z@}%
  \normalbaselines\rm}
 
\def\sixpoint{\def\rm{\fam0\sixrm}%
  \textfont0=\sixrm \scriptfont0=\sixrm \scriptscriptfont0=\fiverm
  \textfont1=\sixi \scriptfont1=\sixi \scriptscriptfont1=\fivei
  \textfont2=\sixsy \scriptfont2=\sixsy \scriptscriptfont2=\fivesy
  \textfont3=\tenex \scriptfont3=\tenex \scriptscriptfont3=\tenex
  \def\it{\fam\itfam\sixit}%
  \textfont\itfam=\sixit
  \def\sl{\fam\slfam\sixsl}%
  \textfont\slfam=\sixsl
  \def\bf{\fam\bffam\sixbf}%
  \textfont\bffam=\sixbf \scriptfont\bffam=\sixbf
   \scriptscriptfont\bffam=\fivebf
  \def\tt{\fam\ttfam\sixtt}%
  \textfont\ttfam=\sixtt
  \tt \ttglue=.5em plus.25em minus.15em
  \normalbaselineskip=9pt
  \def\MF{{\manual opqr}\-{\manual stuq}}%
  \let\sc=\sixrm
  \let\big=\sixbig
  \setbox\strutbox=\hbox{\vrule height7pt depth2pt width\z@}%
  \normalbaselines\rm}
\def\leftmatrix#1{\begingroup \m@th
  \setbox0=\vbox{\def\cr{\crcr\noalign{\kern2pt\global\let\cr=\endline}}
      \ialign{\hfill$##$\kern2pt\kern\p@renwd&\thinspace\hfil$##$
            &&\quad\hfil$##$\crcr
                   \omit\strut\hfil\crcr\noalign{\kern-\baselineskip}
                   #1\crcr\omit\strut\cr}}
   \setbox2=\vbox{\unvcopy0\global\setbox1=\lastbox}
\setbox2 = \hbox{\unhbox1\unskip\global\setbox1=\lastbox}
\setbox2=\hbox{$\kern\wd1\kern-\p@renwd\kern-\wd1
   \global\setbox1=\vbox{\box1\kern2pt}
  \vcenter{\kern-\ht1\unvbox0\kern-\baselineskip}\,$}
\null\;\vbox{\kern\ht1\box2}\endgroup}
\def\rightmatrix#1{\begingroup \m@th
  \setbox0=\vbox{\def\cr{\crcr\noalign{\kern2pt\global\let\cr=\endline}}
      \ialign{\hfil$##$\kern2pt\kern\p@renwd&\thinspace\hfil$##$
            &&\quad\hfil$##$\crcr
                   \omit\strut\hfil\crcr\noalign{\kern-\baselineskip}
                   #1\crcr\omit\strut\cr}}
   \setbox2=\vbox{\unvcopy0\global\setbox1=\lastbox}
\setbox2 = \hbox{\unhbox1\unskip\global\setbox1=\lastbox}
\setbox2=\hbox{$\kern\wd1\kern-\p@renwd\kern-\wd1
   \global\setbox1=\vbox{\box1\kern2pt}
  \vcenter{\kern-\ht1\unvbox0\kern-\baselineskip}\,$}
\null\;\vbox{\kern\ht1\box2}\endgroup}